\documentclass[preprint,review,3p]{elsarticle}

\usepackage{graphicx}
\usepackage{subfigure}
\usepackage{float}
\usepackage{amsmath}

\usepackage{amssymb}

\biboptions{sort&compress}
\journal{Computers $\&$ Fluids}

\begin{document}

\begin{frontmatter}



\title{Stationary two-dimensional turbulence statistics \\ using a Markovian forcing scheme}


\author{Omer San\corref{cor1}}
\ead{omersan@vt.edu}
\cortext[cor1]{Corresponding author.}
\author{and Anne E. Staples}

\address{Department of Engineering Science and Mechanics\\ Virginia Tech, Blacksburg, VA 24061, USA}

\begin{abstract}
In this study we investigate the statistics of two-dimensional stationary turbulence using a Markovian forcing scheme, which correlates the forcing process in the current time step to the previous time step according to a defined memory coefficient. In addition to the Markovian forcing mechanism, the hyperviscous dissipation mechanism for small scales and the Ekman friction type of linear damping mechanism for the large scales are included in the model. We examine the effects of various dissipation and forcing parameters on the turbulence statistics in both wave space and physical space. Our analysis includes the effects of the effective forcing scale, the bandwidth of the forcing, the memory correlation coefficient, and the forcing amplitude, along with the large scale friction and small scale dissipation coefficients. Scaling exponents of structure functions and energy spectra are calculated, and the role of the parameters associated with the Markovian forcing is discussed. We found that the scaling exponents are approximately invariant and show a universal behavior for the various forms of forcing schemes used. We found, however, that the final states strongly depend on the large scale friction mechanism considered. When the large scale friction mechanism is included in the model with a high friction coefficient, we demonstrate that the behavior is no longer universal. Our analysis also shows that the second-order vorticity structure function has an asymptotic scaling exponent for larger dissipation. Additionally, we confirmed that vorticity behaves as a passive scalar when the dissipation mechanism becomes less effective. Finally, although turbulence is not believed to have a separation of time scales in the dynamics of the velocity field, we conjectured that a separation of time scales exists in the dynamics of the energy spectrum.\\
\end{abstract}

\begin{keyword}
Two-dimensional turbulence  \sep turbulence statistics \sep Markovian forcing scheme \sep Ekman friction \sep forced-dissipative nonlinear systems \sep Fourier-Galerkin pseudospectral method


\end{keyword}

\end{frontmatter}


\section{Introduction}
\label{sec:intro}
The phenomenology of turbulence was described by Richardson \cite{richardson1920supply} and quantified in a scaling theory by Kolmogorov \cite{kolmogorov1941dissipation}. It is believed that the turbulent flow phenomena are describable through the three-dimensional (3D) Navier-Stokes equations \cite{kaneda2003energy}. Two-dimensional (2D) turbulence, to the first approximation, is a reduced dimensional version of 3D turbulence, where the flow is constrained to two dimensions. In reality, 2D turbulence is never realized in nature or in the laboratory, both of which have some degree of three-dimensionality \cite{boffetta2012two}. Nevertheless, many aspects of idealized 2D turbulence appear to be relevant for physical systems in geophysics, astronomy and plasma physics \cite{rosa1998characterization}. One of the most important reasons for studying two-dimensional turbulence is to improve our understanding of geophysical flows in the atmosphere and ocean \cite{lilly1971numerical,charney1971geostrophic,herring1974decay,mcwilliams1984emergence,maltrud1991energy,lindborg1999can,ferziger2002numerical,san2011approximate}, in which stratification and rotation suppress vertical motions in the thin layers of fluid.

From a theoretical perspective, 2D turbulence is not simply a reduced dimensional version of 3D turbulence because a completely different phenomenology arises from new conservation laws in two dimensions \cite{boffetta2012two}. In fact, two-dimensional turbulence behaves in a profoundly different way from three-dimensional turbulence due to different energy cascade behavior, and follows the Kraichnan-Batchelor-Leith (KBL) theory  \citep{kraichnan1967inertial,batchelor1969computation,leith1971atmospheric}. In three-dimensional turbulence, energy is transferred forward, from large scales to smaller scales, via the vortex stretching and tilting mechanism. In two dimensions that mechanism is absent, and it turns out that under most forcing and dissipation conditions energy will be transferred from smaller scales to larger scales. This is largely because of another quadratic invariant, the potential enstrophy, defined as the integral of the square of the potential vorticity. Despite the apparent simplicity of dealing with two rather than three spatial dimensions, two-dimensional turbulence is possibly richer in its dynamics due to its conservation properties, such as its inverse energy and forward enstrophy cascading mechanisms, which three-dimensional turbulence does not possess.

\begin{figure}[!h]
\centering
\includegraphics[width=0.65\textwidth]{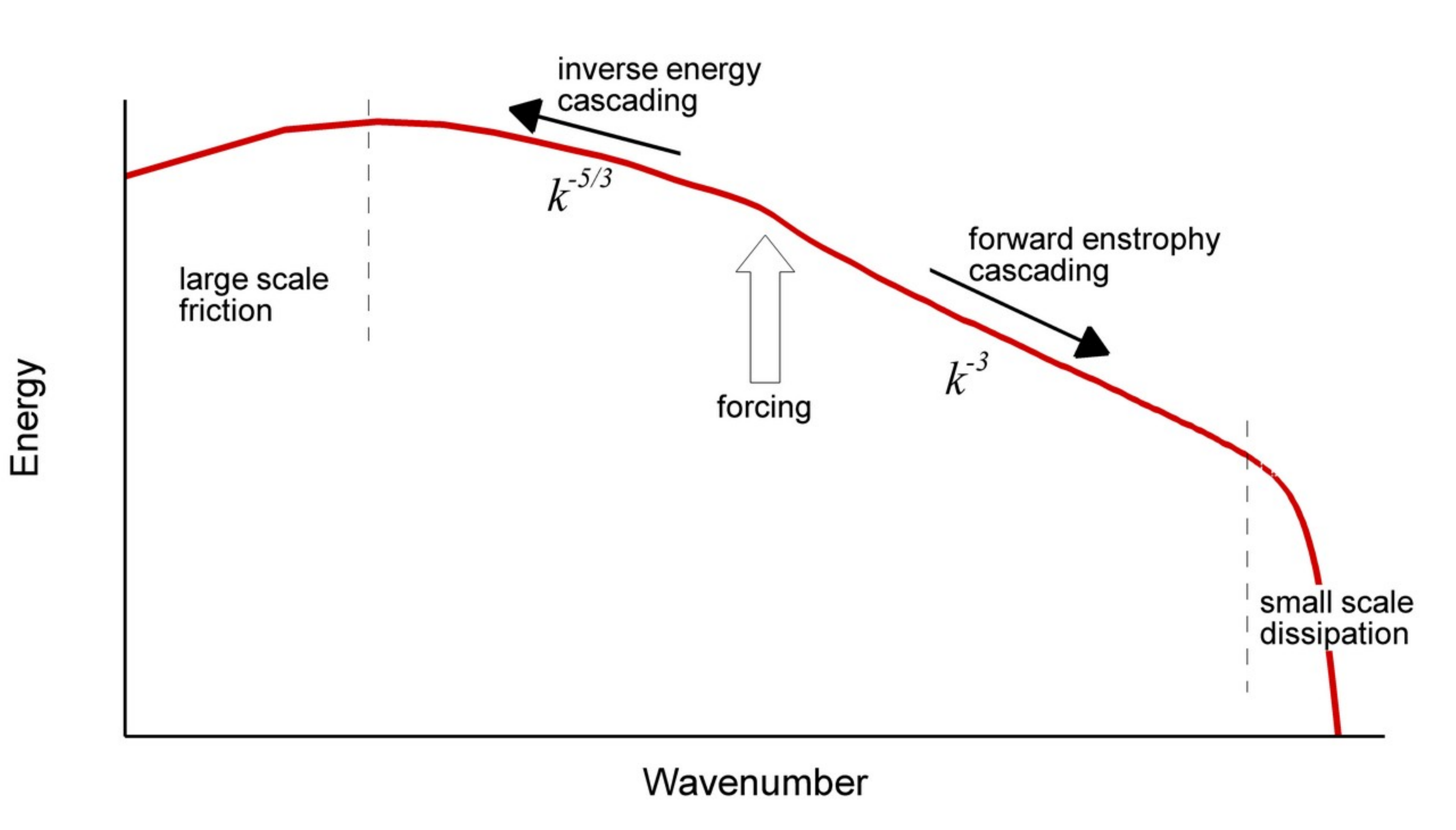}
\caption{A qualitative picture of the energy spectrum which shows double cascade scalings of stationary two-dimensional turbulence.}
\label{fig:def}
\end{figure}

The physics of two-dimensional turbulence have been elucidated substantially during the past decades by theoretical models, intensive numerical investigations, and dedicated soap film experiments \cite{goldburg1997experiments,bruneau2009influence}. Danilov and Gurarie \cite{danilov2000quasi}, Kellay and Goldburg \cite{kellay2002two}, and Tabeling \cite{tabeling2002two} reviewed both theoretical and experimental two-dimensional turbulence studies. More recent reviews on two-dimensional turbulence are also provided by Clercx and van Heijst \cite{clercx2009two} and Boffetta and Ecke \cite{boffetta2012two}. Recent studies in two-dimensional turbulence, both forced (stationary) turbulence \citep{lindborg2000kinetic,danilov2001forced,tran2002constraints,boffetta2010evidence,bracco2010reynolds,vallgren2011enstrophy} and unforced (decaying) turbulence \citep{yin2004easily,lindborg2010testing,kuznetsov2010sharp,fox2010freely} provide high resolution computational confirmation of the KBL theory. The conjecture in KBL theory is that enstrophy, not energy, cascades to the small scales, and the energy, on the contrary, cascades to the large scales \citep{weiss1991dynamics}. Therefore, in two-dimensional turbulence there are two inertial ranges, one for the forward cascade of enstrophy and one for the inverse cascade of energy. In this dual cascading phenomenon, the relative locations of the Kolmogorov and Kraichnan scalings in 2D turbulence energy spectra depend on the forcing scale \citep{tran2003dual}. According to Kolmogorov theory, in the energy cascade range, the only parameters of practical importance would be energy injection rate $\epsilon$ and wave number $k$. Dimensional reasoning states that the energy density (i.e., energy spectrum) is $E(k) \sim \epsilon^{2/3} k^{-5/3}$. In KBL theory, a similar argument gives that $E(k) \sim \eta^{2/3} k^{-3}$ in the forward enstrophy cascade range; where $\eta$ is the enstrophy injection rate (i.e., energy injection rate is related to the enstrophy injection rate by $\epsilon = \eta/k_f^2$ in which $k_f$ is the energy injection scale). These predictions for dual cascade given by KBL theory are illustrated in Figure~\ref{fig:def}.

The dual cascading conjecture, along with the dissipation mechanisms, challenges the universality of the scalings in stationary two-dimensional turbulence. In general, the dissipation mechanism contains an energy sink on large-scales and an enstrophy sink on small-scales \cite{maltrud1991energy,borue1993spectral,smith1993bose,smith1994finite,legras1988high,danilov2001forced,haugen2004inertial}.
The presence of these two sinks is necessary to reach a stationary regime if two different fluxes are assumed to flow in opposite directions from the forcing zone as hypothesized in two-dimensional turbulence. In other words, due to inverse energy cascading in two-dimensional turbulence, the stationary final state can only be obtained if there exist a dissipation mechanism in large scales.
Therefore, large-scale dissipation mechanisms have been routinely employed in numerical simulations of stationary two-dimensional turbulence to absorb energy at large scales. Blackbourn and Tran \cite{blackbourn2011effects} have recently studied the effects of friction on forced two-dimensional Navier-Stokes turbulence.

Two-dimensional turbulence models are useful for understanding large scale motions of forced-dissipative systems occurring in the atmosphere and oceans. The frictional and forcing effects discussed in this study are crucial to understanding the dynamics of these systems. The statistics of such a forced-dissipative two-dimensional turbulence system are investigated here for a wide range of physical considerations. One of our major goals is to analyze the universality of the scaling laws for these statistics in both wave space and physical space by considering the energy or enstrophy spectra and velocity or vorticity structure functions. In order to minimize the effects of dissipation at intermediate scales of the spectrum, following many two-dimensional turbulence studies (e.g., \cite{maltrud1991energy,borue1993spectral,smith1993bose,smith1994finite,legras1988high,danilov2001forced,haugen2004inertial}), we use high-order Laplacian for viscosity (sometimes called hyperviscosity) which separates sharply the inertial and dissipative ranges. To model the forcing and energy injection mechanism, we use a Markovian process, which correlates the forcing structure in the current time step to the previous one according to a defined memory coefficient. The effects of the amplitude of the forcing, the effective forcing scale, and the bandwidth of the forcing are considered here in addition to the memory coefficient parameter that measures the stochastic behavior of the forcing mechanism in the system. We integrate the hyperviscous Navier-Stokes equations with a pseudospectral method using the third-order Runge-Kutta scheme for nonlinear terms and the second-order Crank-Nicolson scheme for the linear dissipation terms. We should also note that our hyperviscous model reduces to the classical Navier-Stokes turbulence model in the limiting case.

The paper is organized as follows: the mathematical formulation of the forced-dissipative two-dimensional turbulence is given in Section~\ref{sec:model}. The numerical methods are briefly presented in Section~\ref{sec:numerics}. The results for two-dimensional isotropic homogeneous stationary turbulence are provided in Section~\ref{sec:results}. This section also explores the effects of large scale friction and small scale dissipation mechanisms, and systematically analyzes the relevant parameters of the Markovian forcing scheme. The turbulence statistics in the stationary regime and their scaling exponents are shown in this section as well. Final conclusions and comments are drawn in Section~\ref{sec:conclusions}.

\section{Mathematical model}
\label{sec:model}
The governing equation for two-dimensional incompressible flows can be written in its vorticity formulation in the following form \cite{danilov2000quasi}
\begin{equation}
\frac{\partial \omega}{\partial t} + J = D + F
\label{eq:ge}
\end{equation}
where $\omega$ is the vorticity which is a scalar quantity in two-dimensional flows. Here, $J$ is the nonlinear Jacobian which symbolizes nonlinear interactions, $D$ represents the dissipation mechanism, and $F$ is for forcing. The Jacobian term in Eq.~(\ref{eq:ge}) is defined as
\begin{equation}
J = \textbf{u} . \nabla \omega = \frac{\partial \psi}{\partial y}\frac{\partial \omega}{\partial x} - \frac{\partial \psi}{\partial x}\frac{\partial \omega}{\partial y}
\label{eq:jac}
\end{equation}
using the definition of the velocity stream function
\begin{equation}
u = \frac{\partial \psi}{\partial y}; \quad v = - \frac{\partial \psi}{\partial x}
\label{eq:vel}
\end{equation}
where $u$ and $v$ are components of the two-dimensional velocity vector field $\textbf{u}$. The dissipation term in classical Navier-Stokes equation is given as $D =\upsilon \nabla^2\omega$, where $\nabla^2$ is the Laplacian. In this study, we use a generalized form of the dissipation $D =\upsilon \nabla^{2p}\omega$ in place of the regular viscosity dissipation. The purpose of using high order Laplacian, which is called as hyperviscosity, is to eliminate as much as possible the effects of viscosity at the intermediate scales, thus extending turbulence inertial scales. On the other hand, in order to sink energy at the large scales we use Ekman type friction to be able reach a statistically steady state. In the case of periodic boundary conditions the applied forcing mechanism would result in an unbounded growth of the total kinetic energy if we would not include a large scale friction mechanism. The Ekman friction terminology is usually used in the contexts of the rotating flows. We can also interpret this large scale friction term as Rayleigh friction for stratified flows, or Hartman friction for magnetohydrodynamic flows. Finally, the dissipation mechanism in our study is modeled by the following generalized form
\begin{equation}
D = (-1)^{p+1} \upsilon \nabla^{2p} \omega - \lambda \omega
\label{eq:nuNS}
\end{equation}
in which it reduces to the classical Ekman-Navier-Stokes equation for $p=1$. The kinematic relationship between vorticity and stream function according is given as
\begin{equation}
\frac{\partial^2 \psi}{\partial x^2} + \frac{\partial^2 \psi}{\partial y^2} = -\omega.
\label{eq:ke}
\end{equation}

From a computational point of view, this formulation has several advantages over the primitive variable formulation. It eliminates pressure from the Navier-Stokes equations and hence has no corresponding odd-even decoupling between the pressure and velocity components, as well as projection inaccuracies usually observed in fractional step approaches \cite{brown2001accurate}. The vorticity-stream function formulation also automatically satisfies the divergence-free condition and allows one to reduce the number of equations to be solved.

\section{Numerical methods}
\label{sec:numerics}
Fourier series expansion based methods are often used for solving problems with periodic boundary conditions. One of the most accurate methods for solving the Navier-Stokes equations in periodic domains is the pseudospectral method, which exploits fast Fourier transform (FFT) algorithms, resulting in spectral accuracy \cite{press1992numerical,moin2001fundamentals}. By transforming Eq.~(\ref{eq:ge}) to Fourier space the governing equation becomes
\begin{equation}
\frac{\partial \hat{\omega}_{\textbf{k}}}{\partial t} + J_\textbf{k} = D_\textbf{k} + F_\textbf{k}
\label{eq:gef}
\end{equation}
where a hat over the variable represents the Fourier coefficients of the corresponding variable in the wave space $\textbf{k} = (k_x, k_y)$. The relationship between vorticity and stream function in the Fourier space becomes
\begin{equation}
(-k_x^2 - k_y^2) \hat{\psi}_{\textbf{k}} = -\hat{\omega}_{\textbf{k}}.
\label{eq:kef}
\end{equation}
The nonlinear Jacobian in Fourier space is
\begin{equation}
J_\textbf{k} = (\textbf{\emph{i}} k_y  \hat{\psi}_{\textbf{k}}) \circ (\textbf{\emph{i}} k_x  \hat{\omega}_{\textbf{k}}) - (\textbf{\emph{i}} k_x  \hat{\psi}_{\textbf{k}}) \circ (\textbf{\emph{i}} k_y  \hat{\omega}_{\textbf{k}})
\label{eq:convl}
\end{equation}
where $\emph{i}$ is the complex unit number (i.e., $\emph{i}^2 = -1 $). The convolution sum in the nonlinear Jacobian term is computed in the spatial
domain using the convolution theorem. In the pseudospectral method, the convolution sum of these nonlinear terms is actually computed in the physical domain and Fast Fourier transforms are used to go back and forth between Fourier wave space and physical space. In the present study, we use the standard Fourier-Galerkin pseudospectral method in a periodic square box of a length $2\pi$ \cite{canuto1988spectral,san2012high}. The dissipation terms in the Fourier space becomes
\begin{equation}
D_\textbf{k} = -[\upsilon k^{2p} + \lambda]\hat{\omega}_{\textbf{k}} = -[\nu(\frac{k}{k_d})^{2p} + \lambda]\hat{\omega}_{\textbf{k}}
\label{eq:disp}
\end{equation}
where $k=\sqrt{k_x^2 + k_y^2}$. We redefine the small scale viscosity coefficient by using the expression of $\nu = \upsilon k_d^{2p}$ where $k_d$ is the effective dissipation wave number, which we set in our study as $k_d = 0.96 N/2$, where $N^2=512^2$ is the resolution of the problem. Note that the power $p=1$ corresponds the regular constant property Navier-Stokes equations for Reynolds number of $Re=k_d^{2}/\nu$. In our model, we use a Markovian forcing scheme, which correlates the forcing in the current time step to the previous one according to a defined memory coefficient. The forcing is localized within narrow spectral range $(k_f-\sigma, k_f+\sigma)$ in the vicinity of the forcing wave number $k_f$. The bandwidth of the effective forcing is determined by the $\sigma$ variable. The Markovian forcing process at the current time step becomes
\begin{equation}
F_\textbf{k}^n = f_0 (1-\rho^2)^{1/2}e^{\emph{i}\zeta} + \rho f^{n-1}_{k}
\label{eq:disp}
\end{equation}
where $f_0$ is the forcing amplitude, $\rho$ is the memory correlation coefficient, and the $\zeta$ is the uniformly distributed phase on interval [$0,2\pi$]. In this study, the semi-discrete vorticity transport equation in Fourier space, Eq.~(\ref{eq:gef}), is solved by a combination of the third-order Runge-Kutta and second-order Crank-Nicolson schemes in a periodic square. The dissipation terms are treated by an implicit Crank-Nicolson scheme and nonlinear Jacobian and forcing terms are treated by the explicit Runge-Kutta scheme. Therefore, starting with the value of the Fourier coefficients of vorticity, $\hat{\omega}^{n}_{\textbf{k}}$, at the current time step, the time marching algorithm for computing the vorticity at the next time step, $\hat{\omega}^{n+1}_{\textbf{k}}$, consists of the following three substeps:
\begin{eqnarray}
\hat{\omega}^{(1)}_{\textbf{k}} &=& \alpha_1 \hat{\omega}^{n}_{\textbf{k}} + \beta_1 (-J_\textbf{k}^{n} + F_\textbf{k}^n) \nonumber \\
\hat{\omega}^{(2)}_{\textbf{k}} &=& \alpha_2 \hat{\omega}^{(1)}_{\textbf{k}} + \beta_2 (-J_\textbf{k}^{(1)} + F_\textbf{k}^n) - \gamma_2 (-J_\textbf{k}^{n} + F_\textbf{k}^n)  \nonumber \\
\hat{\omega}^{n+1}_{\textbf{k}} &=& \alpha_3 \hat{\omega}^{(2)}_{\textbf{k}} + \beta_3 (-J_\textbf{k}^{(2)} + F_\textbf{k}^n) - \gamma_3 (-J_\textbf{k}^{(1)} + F_\textbf{k}^n).
\label{eq:TVDRK}
\end{eqnarray}
where the coefficients are
\begin{eqnarray}
\alpha_1 &=& \frac{1 + \frac{4}{15}\alpha \Delta t}{1 - \frac{4}{15}\alpha \Delta t}; \quad
\beta_1  = \frac{\frac{8}{15} \Delta t}{1 - \frac{4}{15}\alpha \Delta t}; \quad \alpha = -[\nu (\frac{k}{k_d})^{2p} + \lambda] \nonumber \\
\alpha_2 &=& \frac{1 + \frac{1}{15}\alpha \Delta t}{1 - \frac{1}{15}\alpha \Delta t}; \quad
\beta_2  = \frac{\frac{5}{12} \Delta t}{1 - \frac{1}{15}\alpha \Delta t}; \quad  \gamma_2 = \frac{\frac{17}{60} \Delta t}{1 - \frac{1}{15}\alpha \Delta t} \nonumber \\
\alpha_3 &=& \frac{1 + \frac{1}{6}\alpha \Delta t}{1 - \frac{1}{6}\alpha \Delta t}; \quad \
\beta_3  = \frac{\frac{3}{4} \Delta t}{1 - \frac{1}{6}\alpha \Delta t}; \quad \ \gamma_3 = \frac{\frac{5}{12} \Delta t}{1 - \frac{1}{6}\alpha \Delta t}
\label{eq:coeff}
\end{eqnarray}
in which $\Delta t$ is the time step.

\section{Results}
\label{sec:results}
In this section we present numerical results for homogeneous isotropic stationary turbulence for various physical parameters which determine the forcing and dissipation mechanisms. The spatial resolution is fixed $N^2=512^2$ Fourier modes, the time step size is $\Delta t =0.005$, and the solution is advanced forward in time in Fourier space. For all the simulations we start from a rest state, integrate the model until a statistically steady state is obtained, and continue for enough time to compute turbulence statistics. The boundary conditions are $2 \pi$-periodic in both directions. The domain-integrated kinetic energy is tracked for each simulation to measure the energy level of the system, which is quantified by the following integral
\begin{equation}
E(t)
= \frac{1}{2} \iint \left(\frac{\partial \psi}{\partial x}\right)^2
+ \left(\frac{\partial \psi}{\partial y} \right)^2 dx \, dy ,
\label{eq:totE}
\end{equation}
in which we can estimate when the system reaches quasi-stationary regime. Averaging over time is not possible unless a statistically steady state is established by a large scale friction mechanism. The mean values are obtained by ensemble averaging the data between time $t=50$ and $t=100$ throughout the study.

In order to examine the characteristics of two-dimensional stationary turbulence, we first define two statistical measures; one is the energy spectrum in wave space, and the other is the structure function in physical space. The energy spectrum is defined as
\begin{equation}
\hat{E}(\textbf{k},t)=\frac{1}{2}k^2|\hat{\psi}(\textbf{k},t)|^2
\label{eq:esp}
\end{equation}
and the angle averaged energy spectrum is
\begin{equation}
E(k,t)= \sum_{k\leq|\acute{\small\textbf{k}}|\leq k+1} \hat{E}(\acute{\textbf{k}},t).
\label{eq:Aesp}
\end{equation}
In the forward enstrophy cascade (i.e., in the inertial scale between the forcing scale and the small scale dissipation), it is known from the KBL theory that the energy spectrum in the inertial range approaches the classical $k^{-3}$ scaling in the inviscid limit. On the other hand, in the inverse energy cascade (i.e., the inertial scale which includes the scales larger than the forcing scale), the theory predicts that the energy spectrum scales by $k^{-5/3}$, since the scaling arguments leading to this Kolmogorov scaling nowhere assume a specified direction of the cascade.

The statistics of two-dimensional turbulent flow can be further investigated considering powers of velocity or vorticity differences in the physical space. A commonly used statistical quantity in two-dimensional turbulence is the second-order vorticity structure function which is defined as
\begin{equation}
\langle \delta \omega (r)^{2}\rangle = \langle |\omega(\textbf{x} + \textbf{r}) - \omega(\textbf{x})|^{2} \rangle
\label{eq:str}
\end{equation}
with $r=|\textbf{r}|$ being the spatial separation.  Assuming that the system is homogeneous
and isotropic, the structure functions depend on $r$ only. The classical $k^{-5/3}$ scaling of energy spectrum in the inverse energy cascade relates to a $r^{2/3}$ scaling of second-order velocity structure function. In the enstrophy cascade range, where Kraichnan $k^{-3}$ scaling appears, the corresponding $r^2$ scaling is equal to the upper bound. Hence, steeper energy spectra always results in $r^2$ scaling \citep{kramer2011structure}. Vorticity transport equation in two-dimension is formally identical to passive scalar transport equation that describes the transport of a scalar quantity $\theta$ passively advected by velocity field \citep{benzi1990power,celani2005large,falkovich2005anomalous}. In the context of passive scalar, according to the theory \citep{benzi1990power,paret1999vorticity,bracco2010reynolds}, the second-order vorticity structure function in the inviscid limit is proportional with $\langle \delta \omega (r)^{2}\rangle \sim r^{d}$ for a given energy spectrum $E(k)\sim k^{-3-d}$, for $0<d<2$.

\begin{figure*}[!t]
\centering
\mbox{
\subfigure[]{\includegraphics[width=0.33\textwidth]{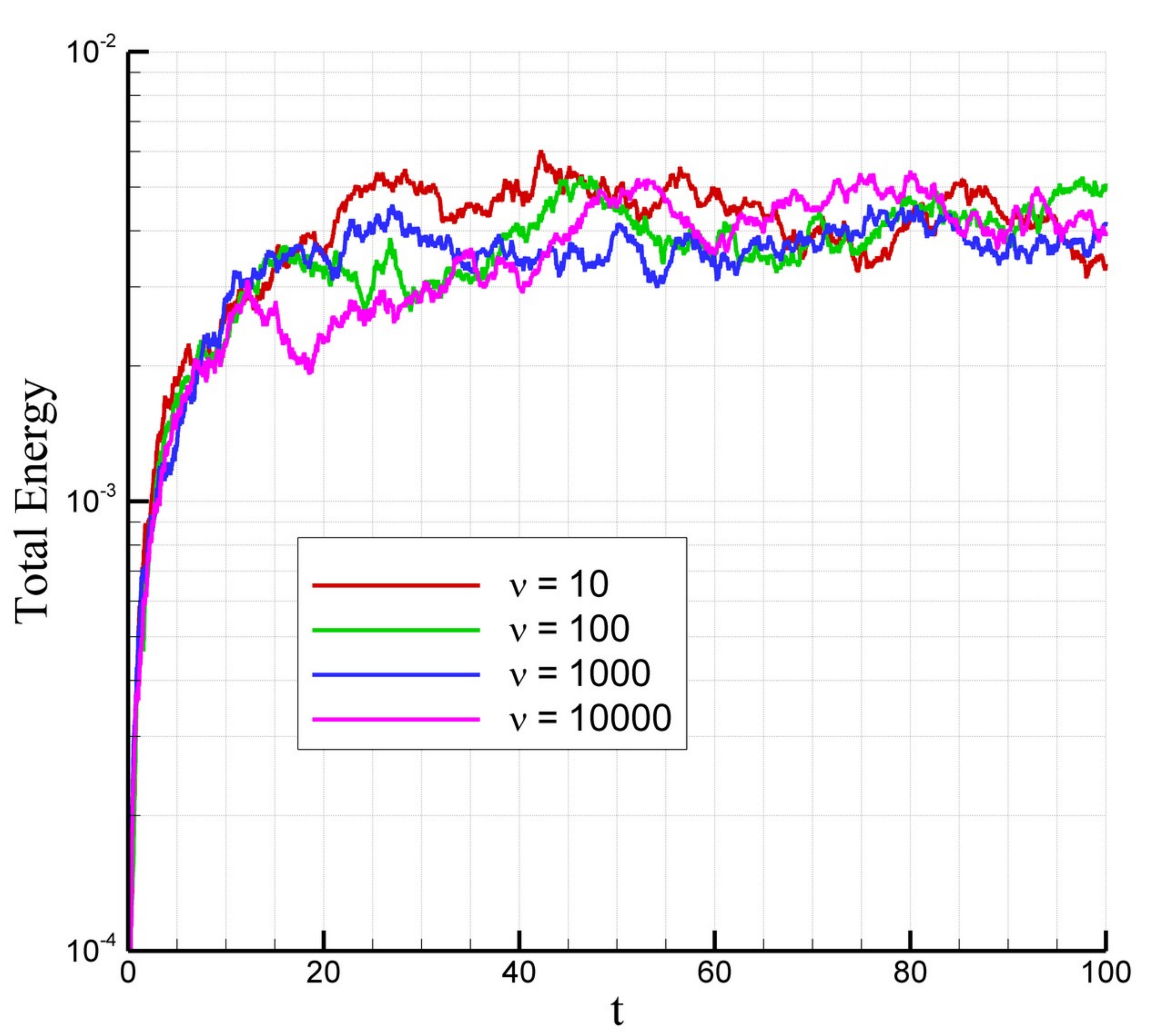}}
\subfigure[]{\includegraphics[width=0.33\textwidth]{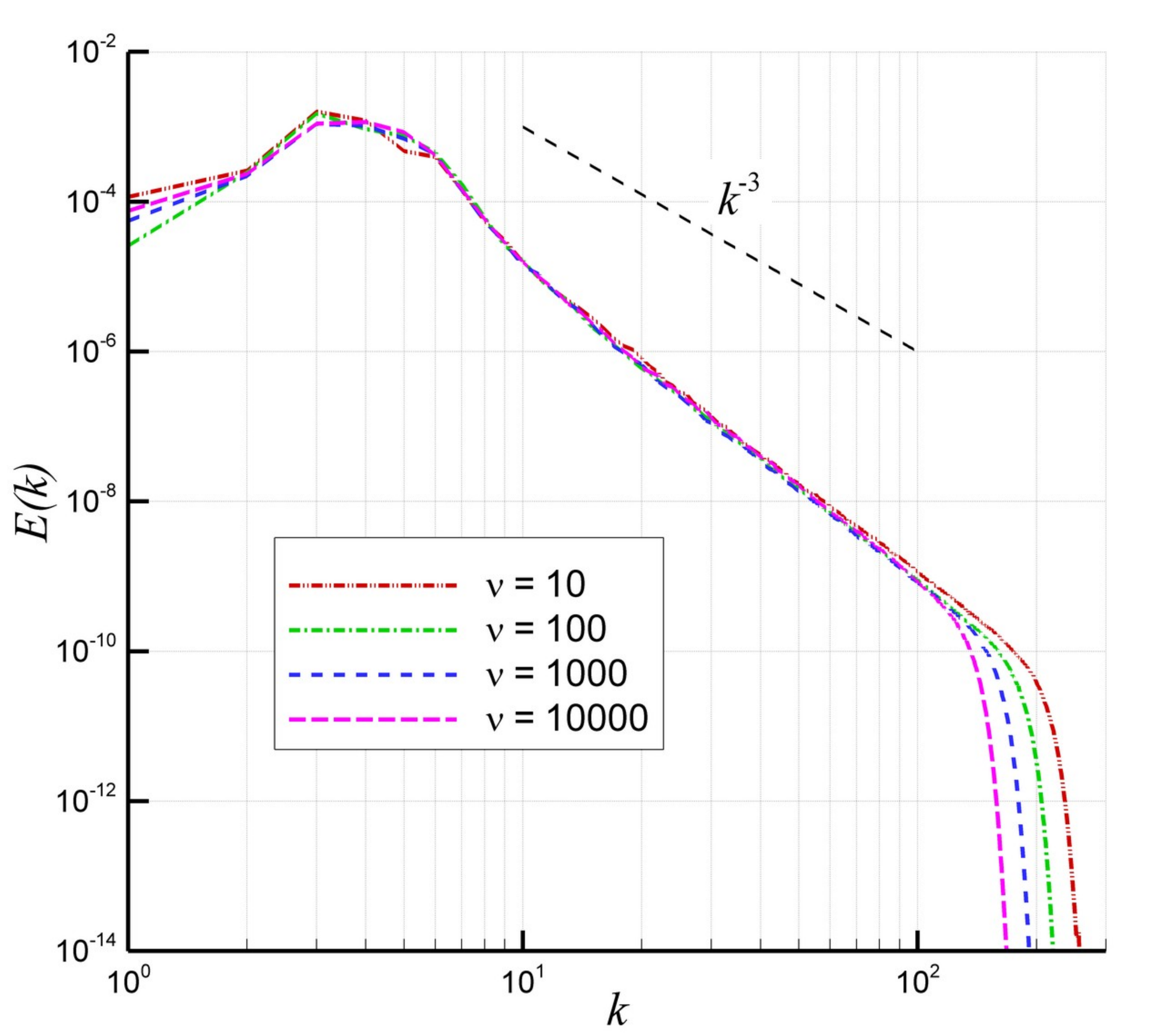}}
\subfigure[]{\includegraphics[width=0.33\textwidth]{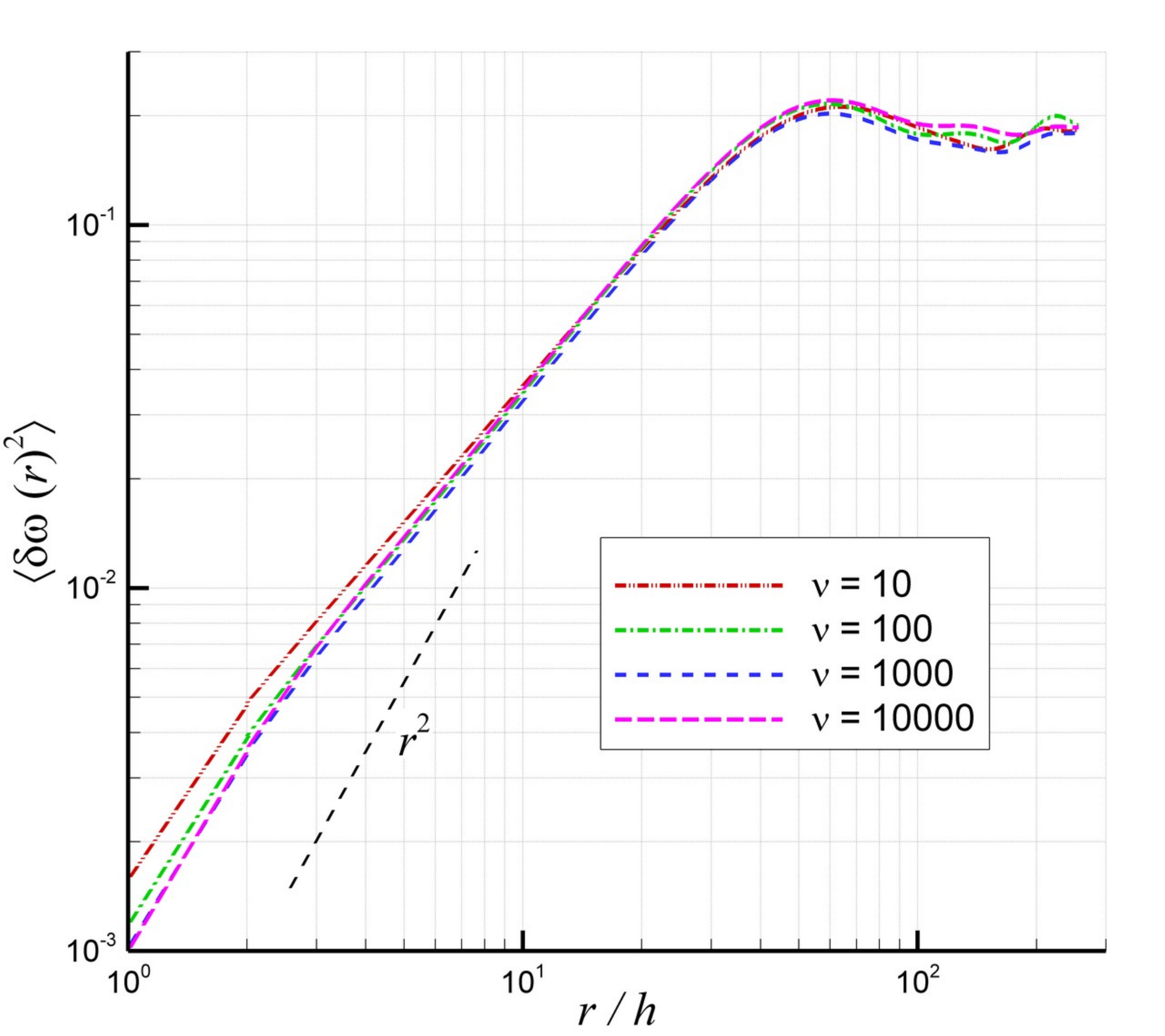}}
}
\caption{The effects of the small scale dissipation coefficient on the statistics ($\lambda=0.05$, $p=8$ and $k_f=5$); (a) time series of total energy, (b) angle averaged energy spectra, and (c) second-order vorticity structure functions.}
\label{fig:stat_nu_k5}
\end{figure*}

\begin{figure*}[!t]
\centering
\mbox{
\subfigure[$\nu=10$]{\includegraphics[width=0.245\textwidth]{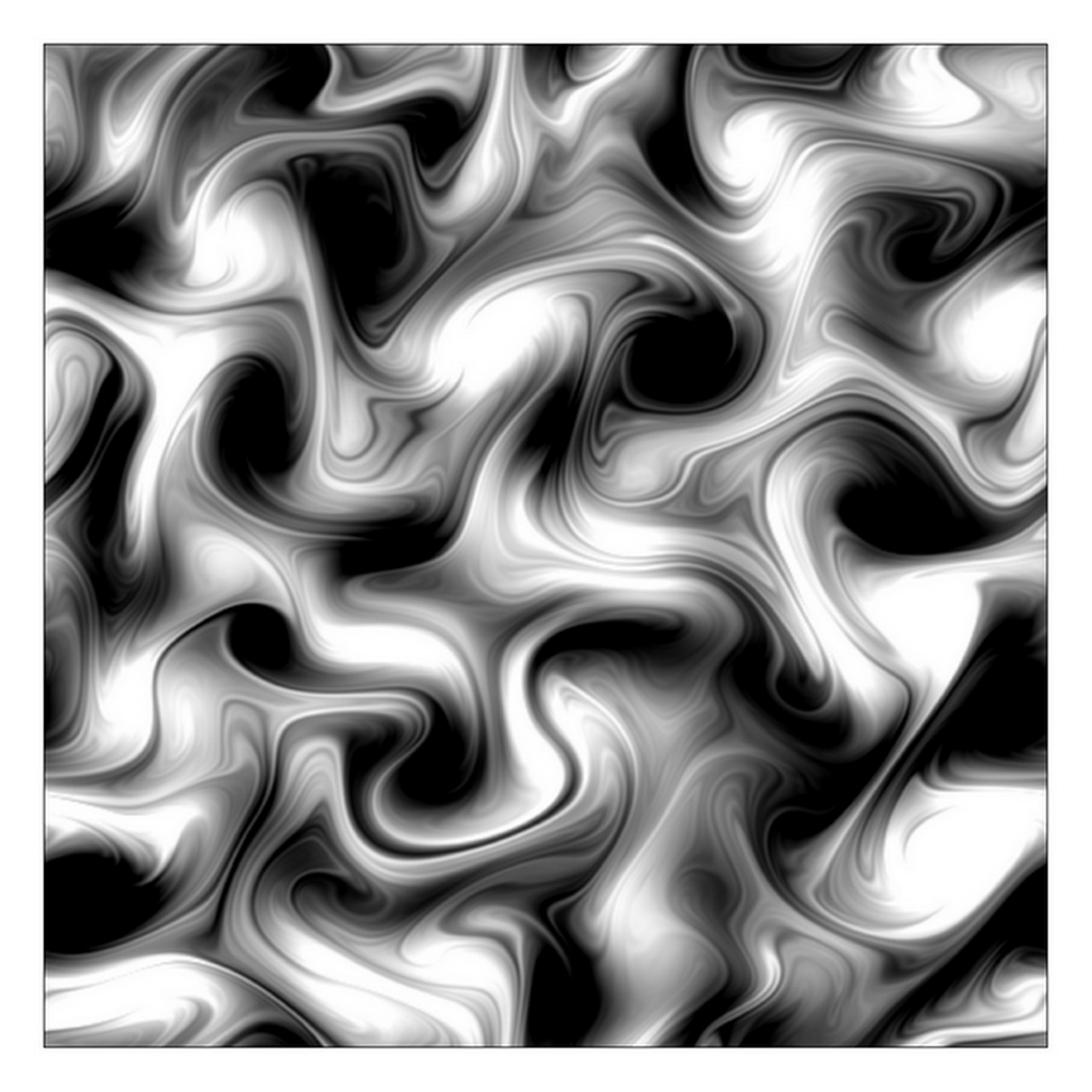}}
\subfigure[$\nu=100$]{\includegraphics[width=0.245\textwidth]{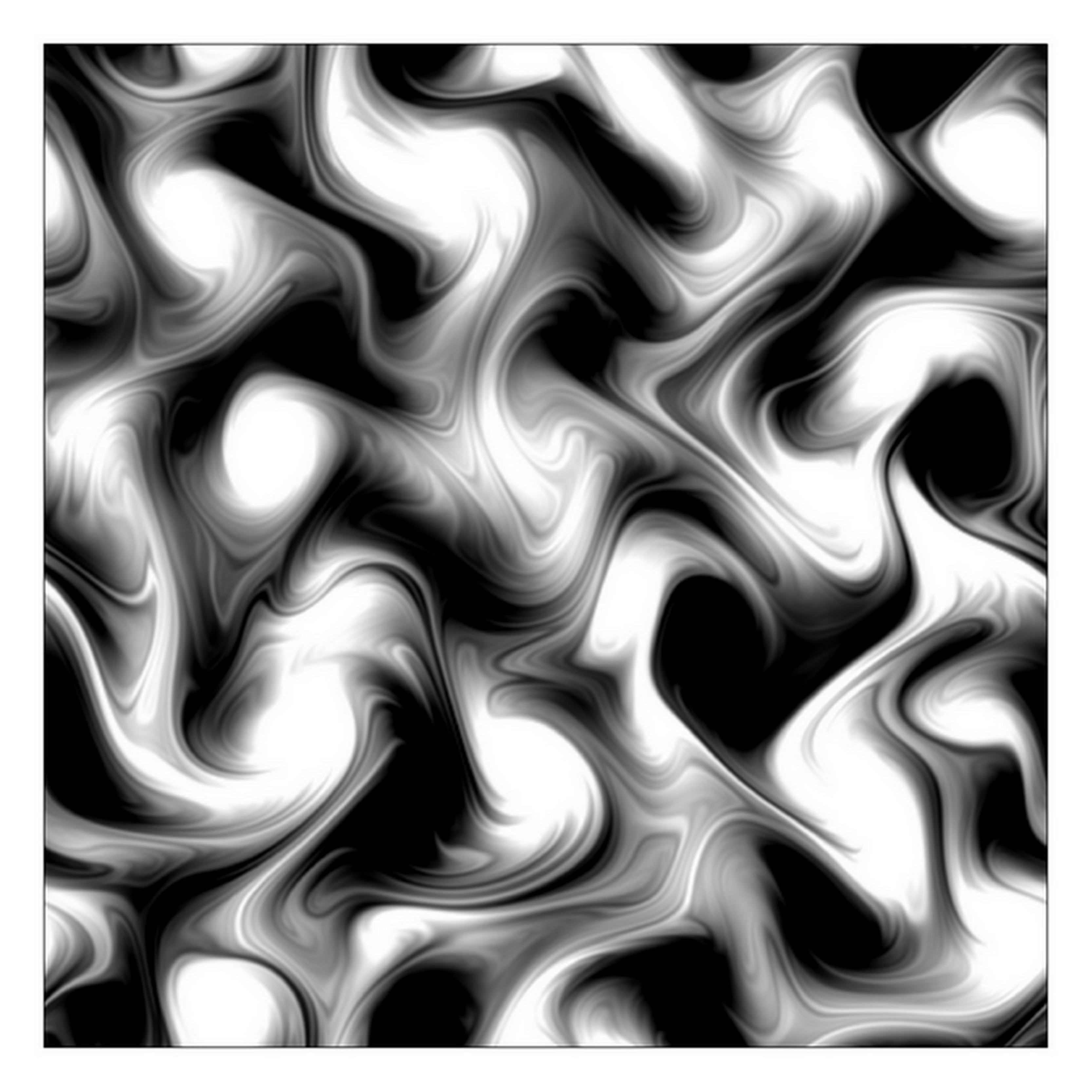}}
\subfigure[$\nu=1000$]{\includegraphics[width=0.245\textwidth]{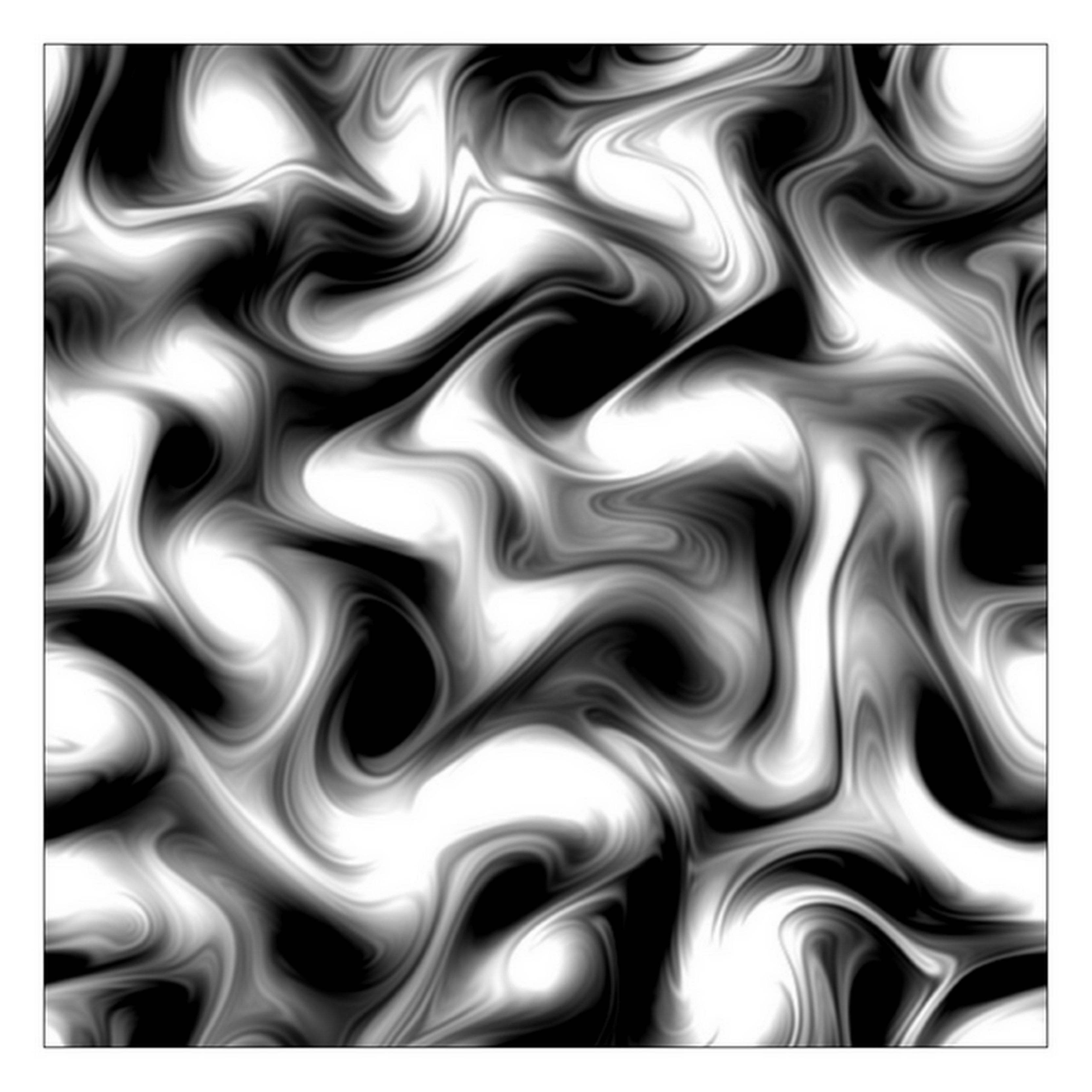}}
\subfigure[$\nu=10000$]{\includegraphics[width=0.245\textwidth]{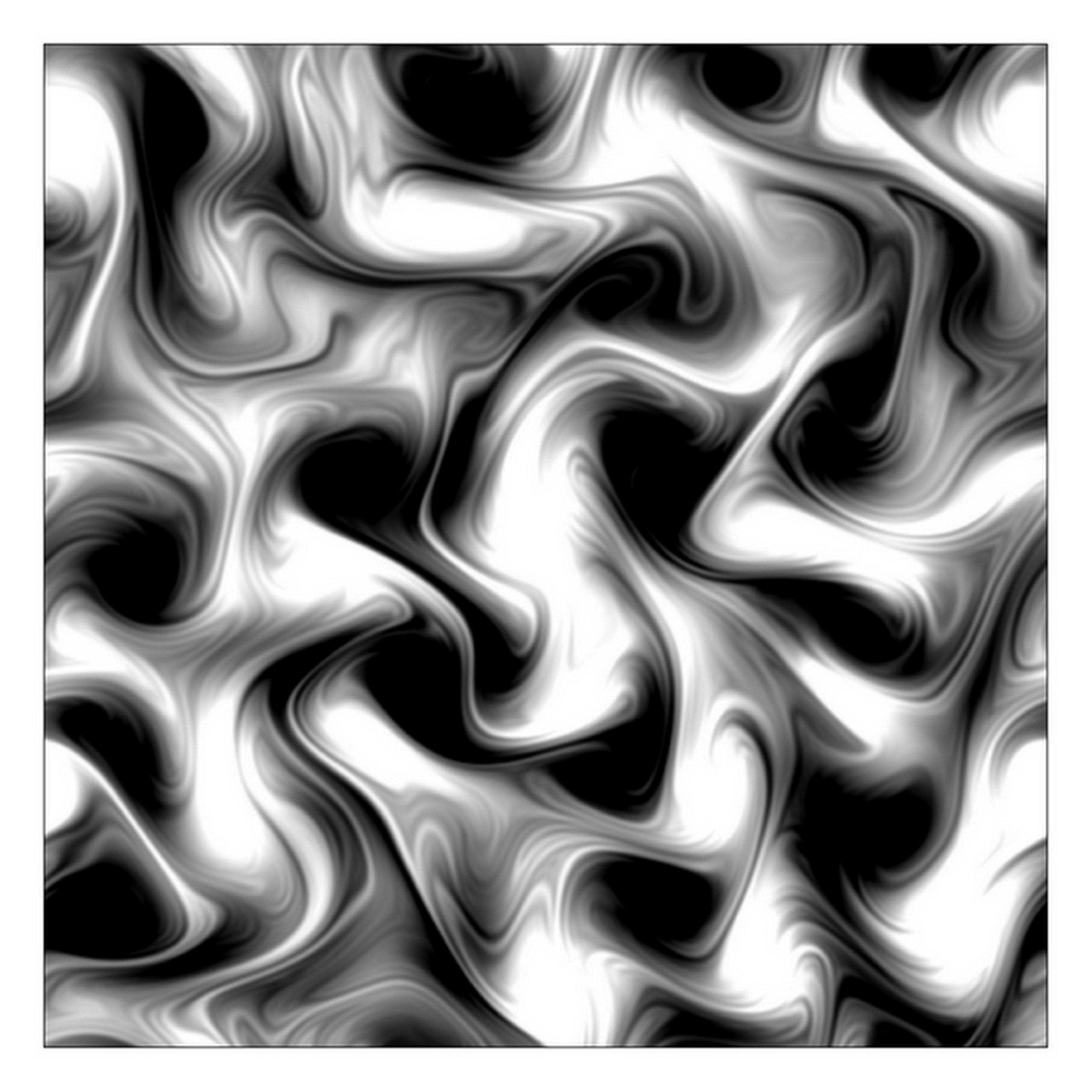}}
}
\caption{Instantaneous vorticity fields at time $t=100$ for varying the small scale dissipation coefficient using the forcing scale $k_f=5$ and the order of Laplacian $p=8$.}
\label{fig:field_nu_k5}
\end{figure*}

The scaling exponents of turbulence statistics that deviate from the values predicted by the theory have been observed in the literature \cite{saffman1971spectrum,mcwilliams1984emergence,legras1988high,benzi1990power,maltrud1991energy,danilov2000quasi,ishihara2001energy,kevlahan2007scaling}. The universality of the statistics of two-dimensional turbulence in both wave space and physical space have also been questioned \cite{danilov2001nonuniversal}. In the following analysis, statistical properties of the forced two-dimensional turbulence are systematically investigated by numerical simulations for different physical parameters. Our primary goal here is to elucidate the effects of the various forcing and dissipation mechanisms on the turbulence statistics.

\subsection{Effects of small scale dissipation mechanism}
\label{sec:disp-s}

First, we analyze the effects of small scales dissipation mechanism on turbulence statistics. Figure~\ref{fig:stat_nu_k5} shows the statistics for different values of small scale dissipation coefficient $\nu$. The order of the Laplacian, the large scale dissipation coefficient, and all the coefficients for forcing scheme remain fixed ($p=8$, $\lambda=0.05$, $k_f=5$, $\sigma=2$, and $\rho=0.5$). As we can see from the figure, an increase in the dissipation coefficient results in a slightly smaller inertial range in the energy spectrum; however, the scalings of the spectra remain the same. This deviation can also be seen from the comparison plot of the vorticity structure function for smaller scales. The levels of energy of the systems in the stationary regime are close to each other. This comparison clearly shows that the small scale dissipation mechanism has slight effect on the statistics, due to the sharp effect of the high order Laplacian in the dissipation mechanism. The instantaneous vorticity fields are also plotted in Figure~\ref{fig:field_nu_k5}, showing that there are no significant effects on flow field structures as well.

\begin{figure*}[!t]
\centering
\mbox{
\subfigure[]{\includegraphics[width=0.33\textwidth]{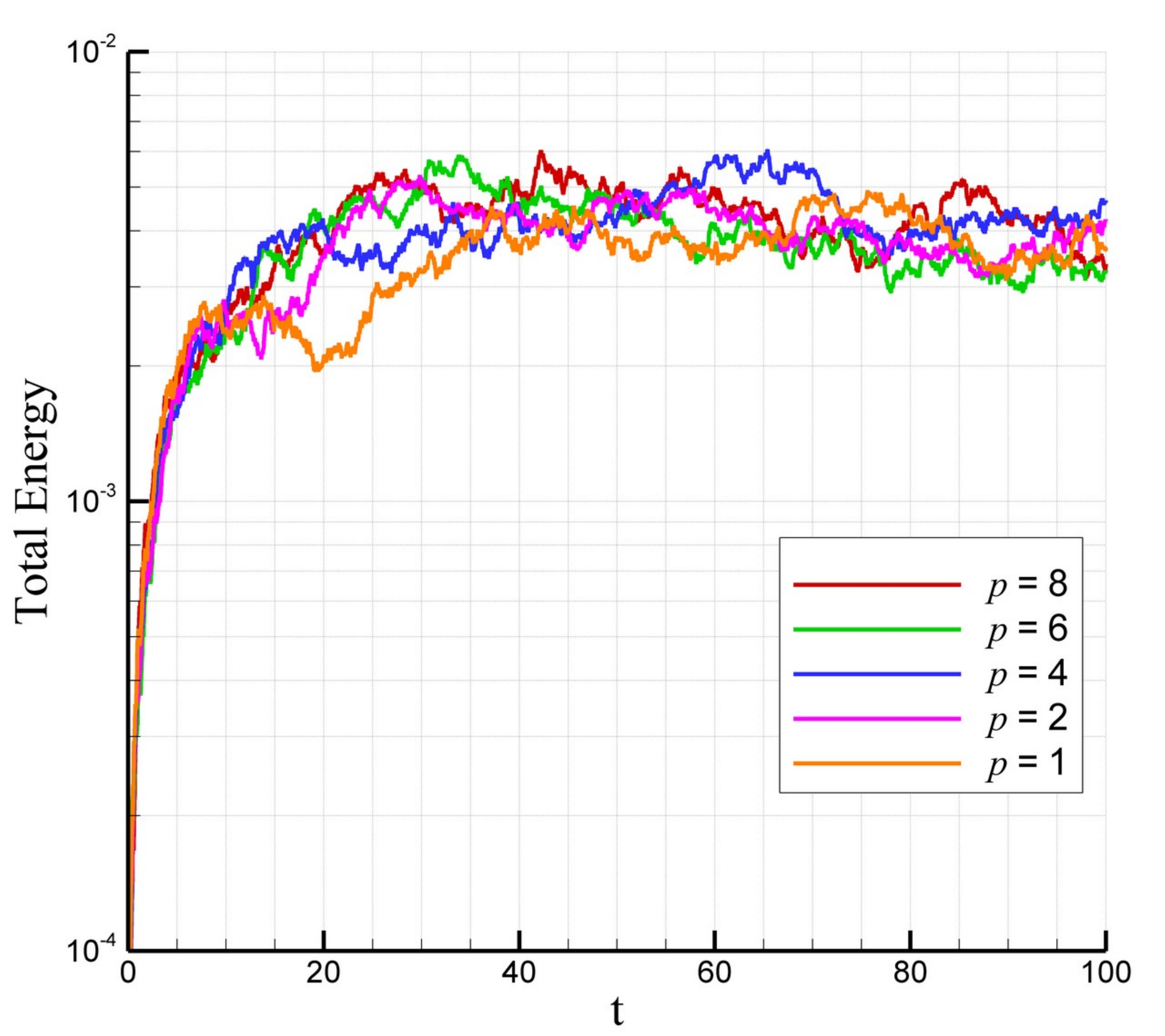}}
\subfigure[]{\includegraphics[width=0.33\textwidth]{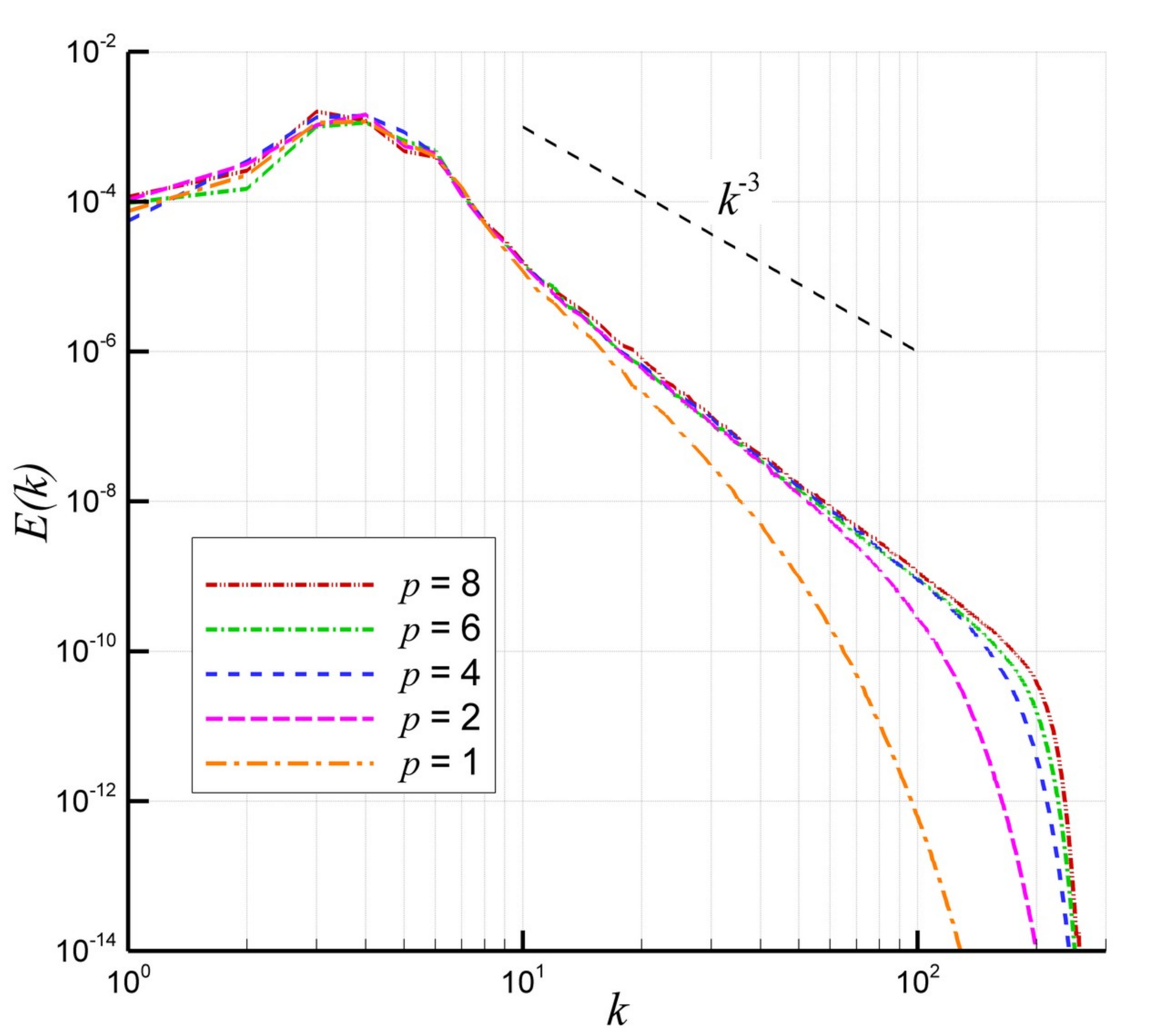}}
\subfigure[]{\includegraphics[width=0.33\textwidth]{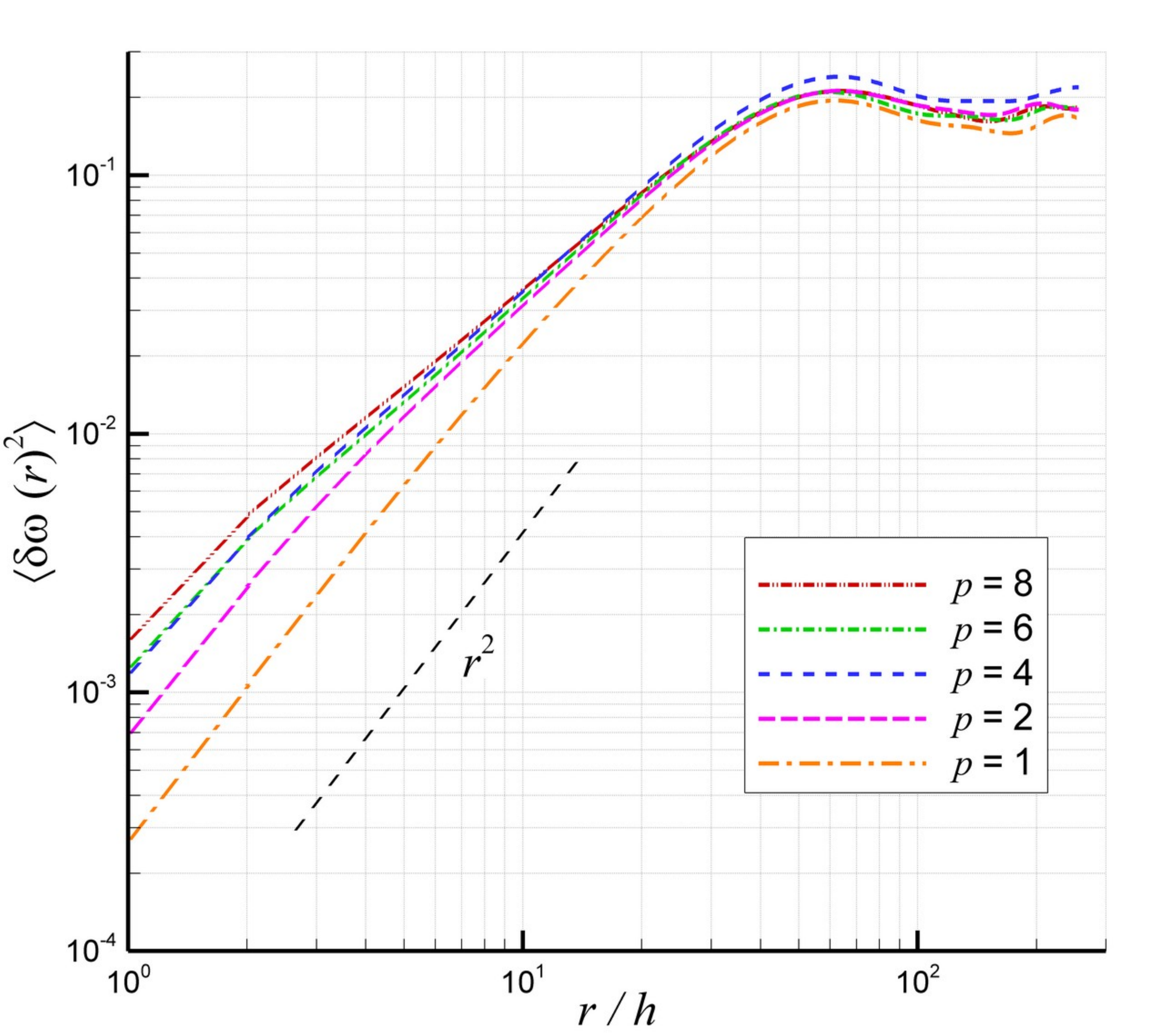}}
}
\caption{The effects of the order of hyperviscosity on the statistics ($\lambda=0.05$, $\nu=10$ and $k_f=5$); (a) time series of total energy, (b) angle averaged energy spectra, and (c) second-order vorticity structure functions.}
\label{fig:stat_nhyp_k5_nu10}
\end{figure*}

\begin{figure*}[!t]
\centering
\mbox{
\subfigure[$p=1$]{\includegraphics[width=0.245\textwidth]{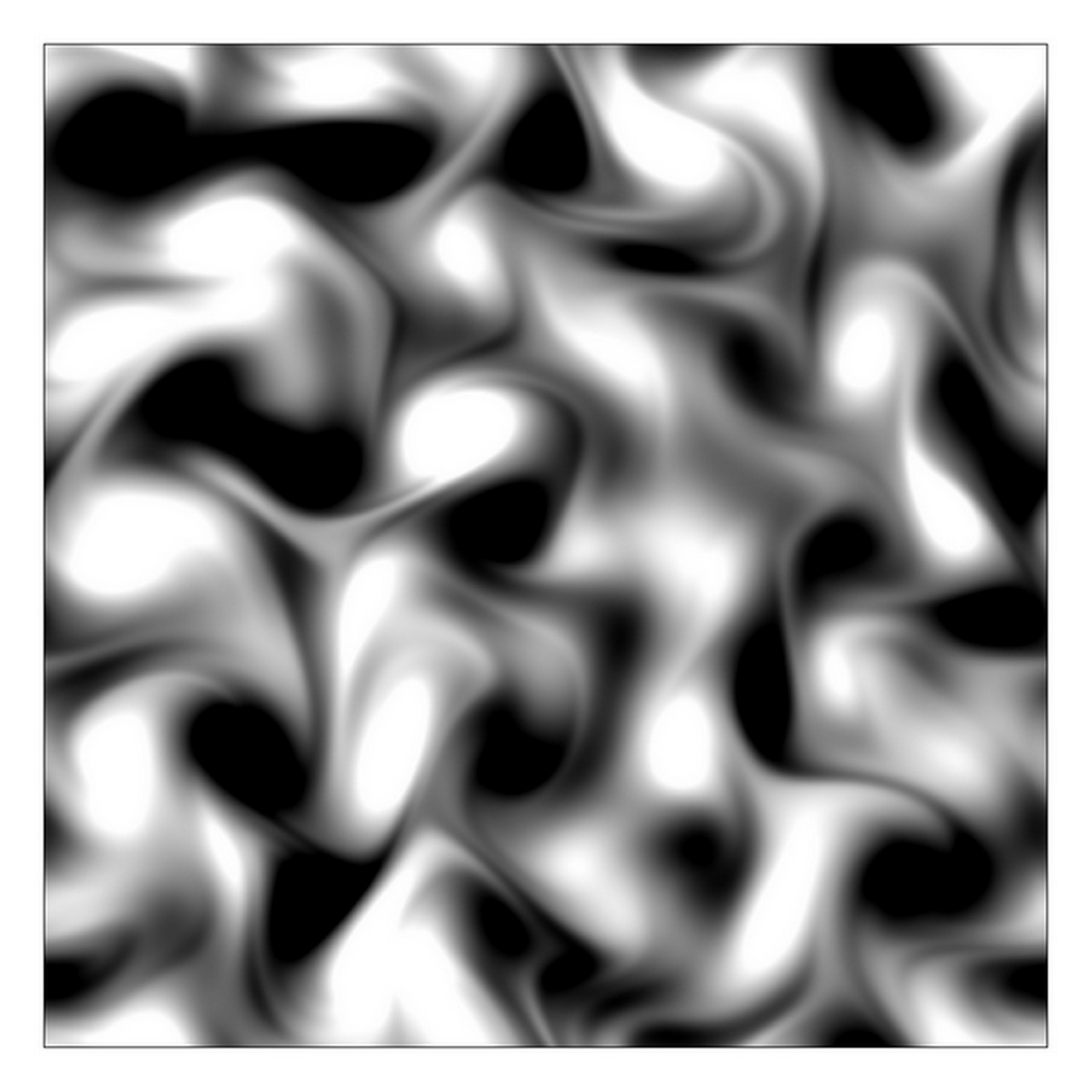}}
\subfigure[$p=2$]{\includegraphics[width=0.245\textwidth]{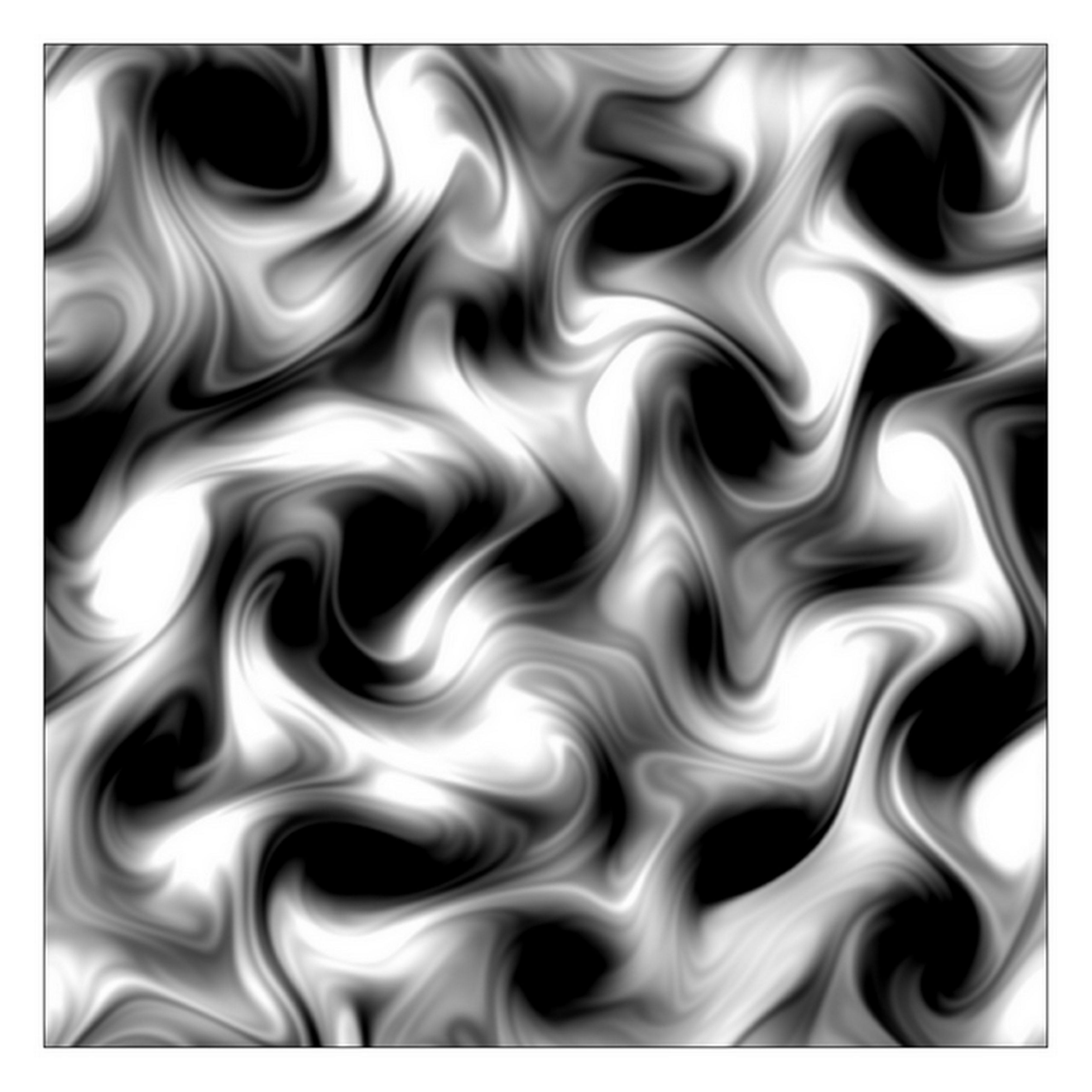}}
\subfigure[$p=4$]{\includegraphics[width=0.245\textwidth]{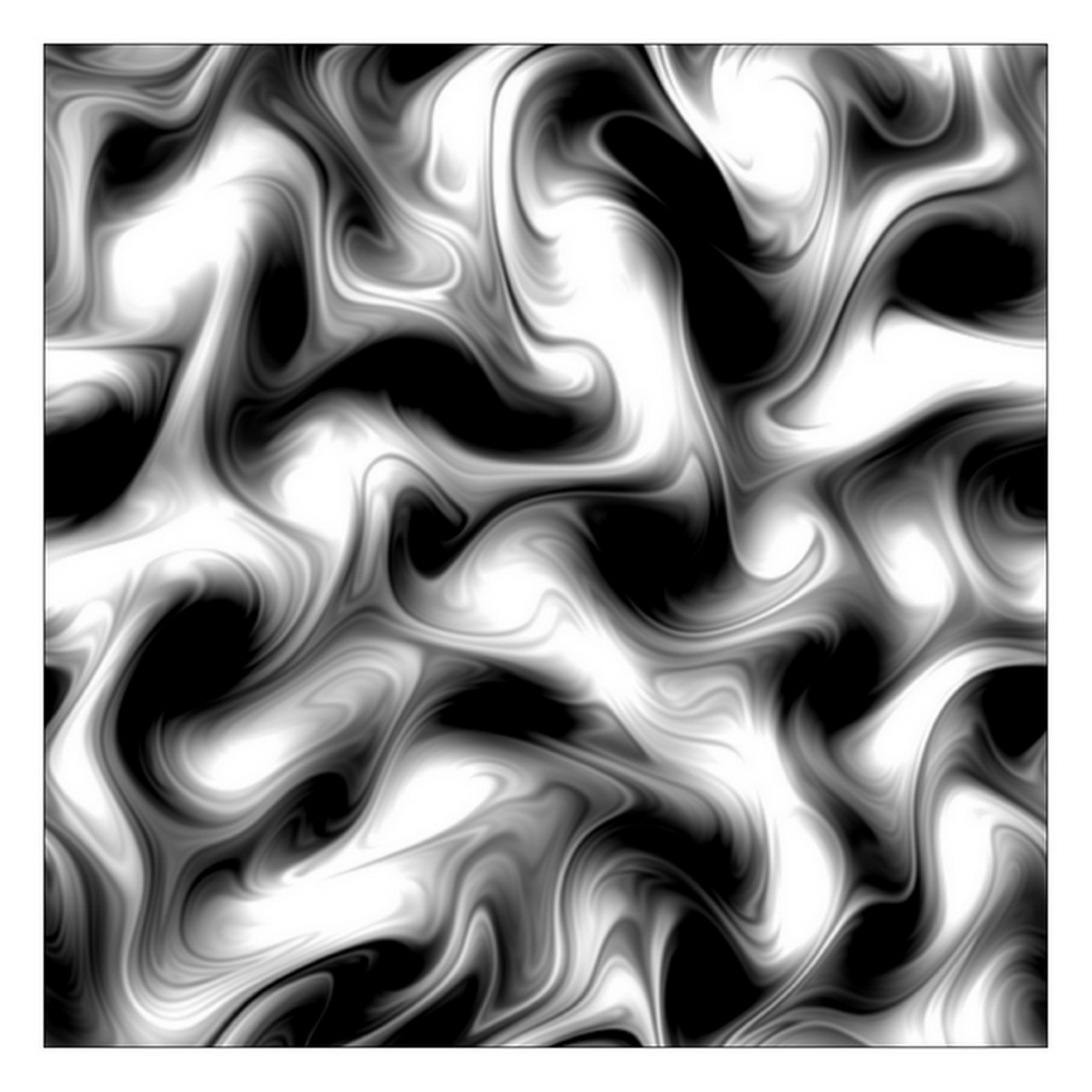}}
\subfigure[$p=6$]{\includegraphics[width=0.245\textwidth]{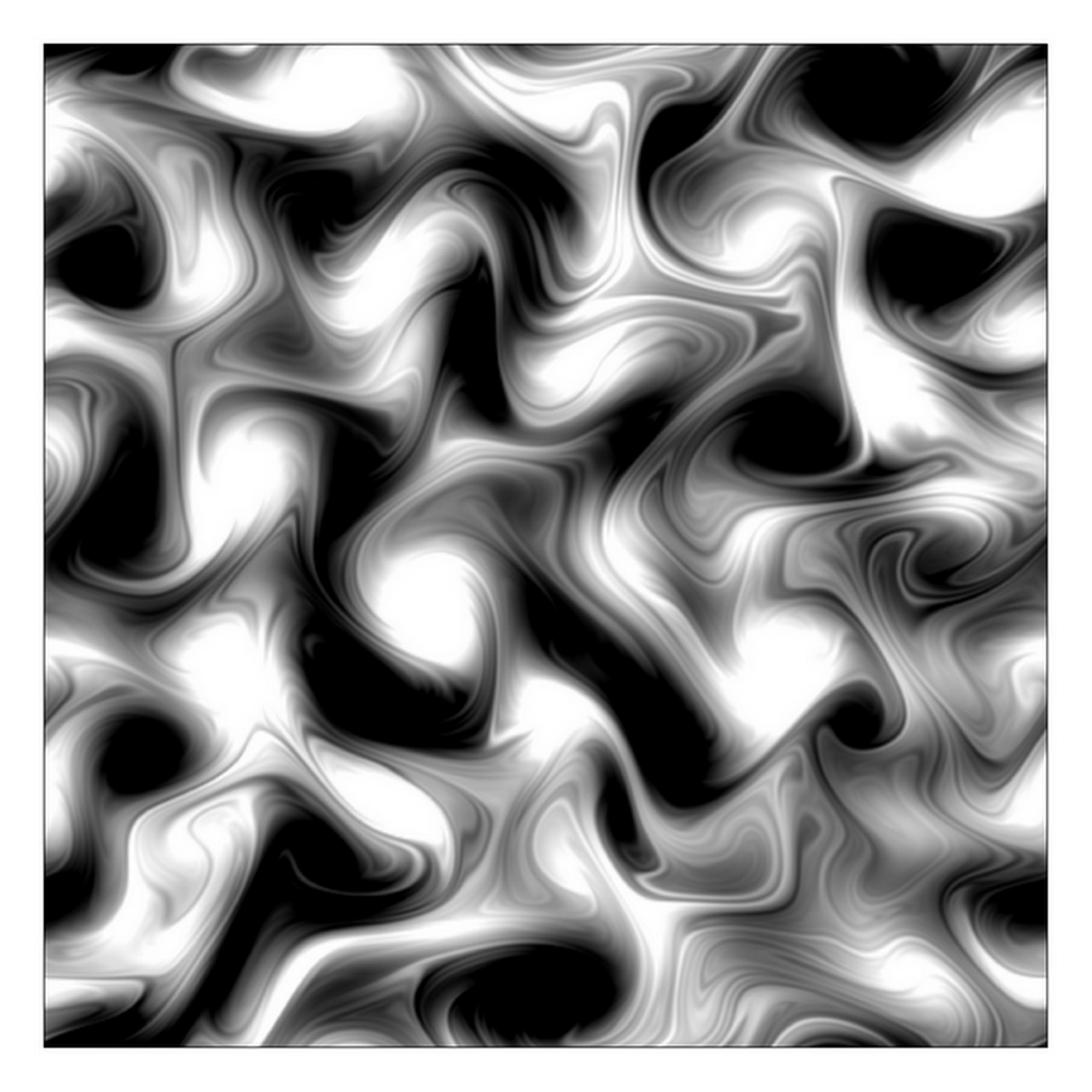}}
}
\caption{Instantaneous vorticity fields at time $t=100$ for various order of the hyperviscosity using the forcing scale of $k_f=5$ and the small scale dissipation coefficient of $\nu=10$.}
\label{fig:field_nhyp_k5_nu10}
\end{figure*}

Second, the effects of the order of the Laplacian in the small scale dissipation mechanism are investigated by using the same coefficients given above for other mechanisms ($\lambda=0.05$, $k_f=5$, $\sigma=2$, and $\rho=0.5$). The case for the order of Laplacian $p=1$ represents regular quadratic dissipation for Navier-Stokes equations, and all the other cases of $p>1$ represent the hyperviscous Navier-Stokes equations. As shown in Figure~\ref{fig:stat_nhyp_k5_nu10}, increasing the hyperviscosity order $p$ extends the inertial range and results in the small scale dissipation becomes effective in larger wave numbers. Its effect on the flow field is also illustrated in Figure~\ref{fig:field_nhyp_k5_nu10}, showing that the $p=1$ case has a less amount of filamentation compared to the hyperviscous cases. The comparison of vorticity structure functions in Figure~\ref{fig:stat_nhyp_k5_nu10} also demonstrates that the scaling for the case of $p=1$ is  $\langle \delta \omega (r)^{2}\rangle \sim r^{2}$ for smaller separation distances. It can also be seen that the vorticity structure functions flatten for large separation distances as predicted by the KBL theory of two-dimensional turbulence in the inviscid limit.

\begin{figure*}[!t]
\centering
\mbox{
\subfigure[]{\includegraphics[width=0.33\textwidth]{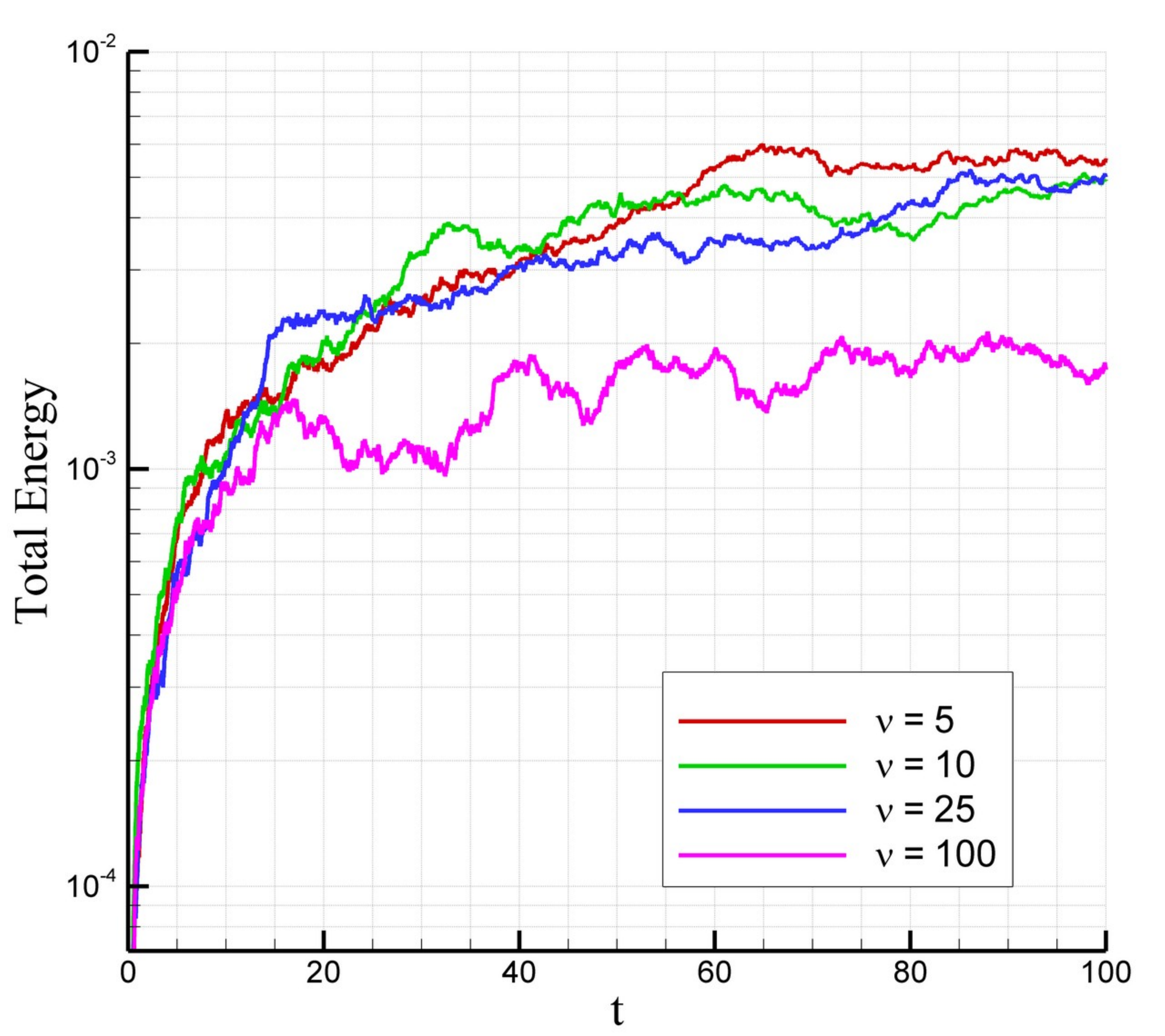}}
\subfigure[]{\includegraphics[width=0.33\textwidth]{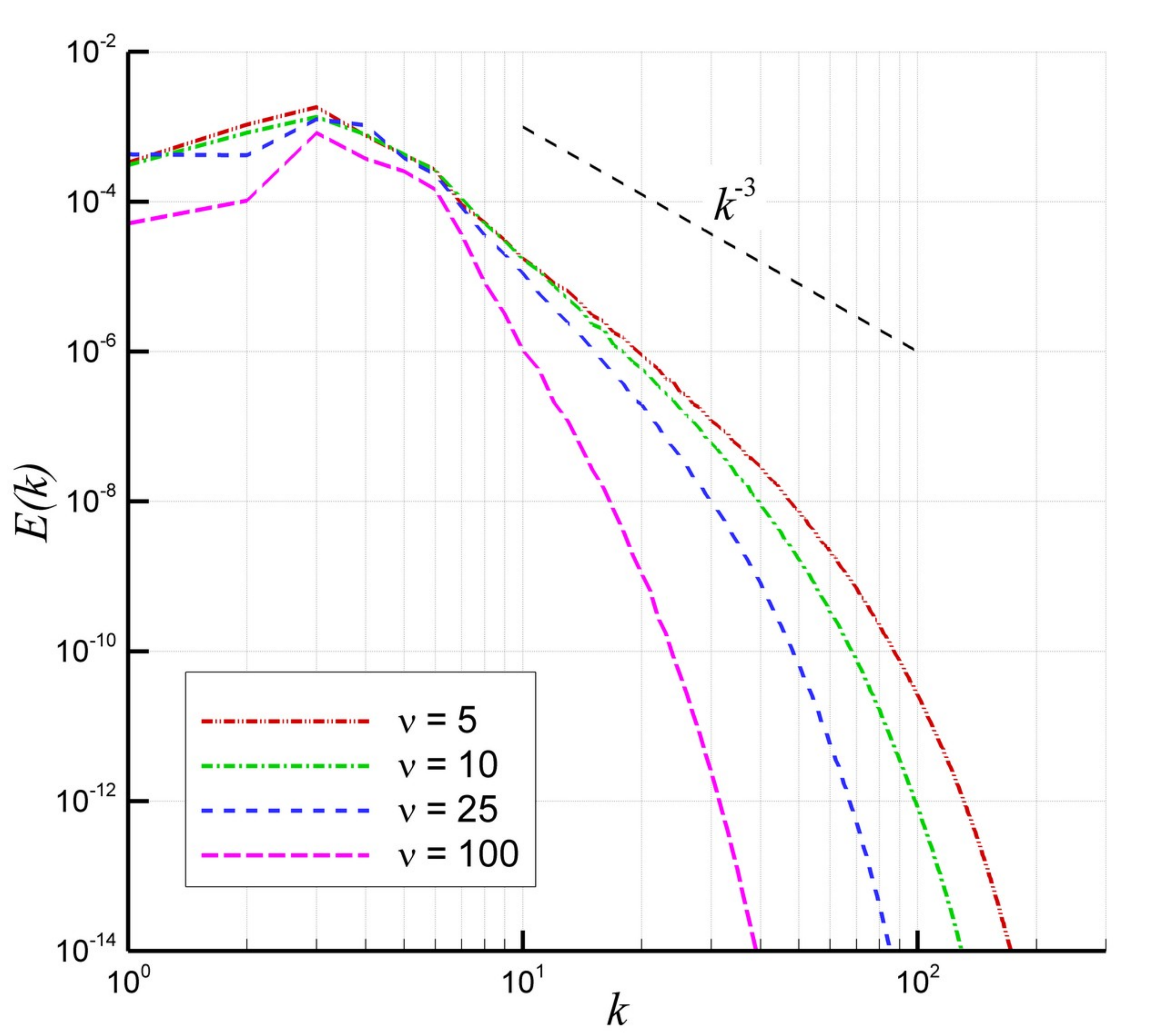}}
\subfigure[]{\includegraphics[width=0.33\textwidth]{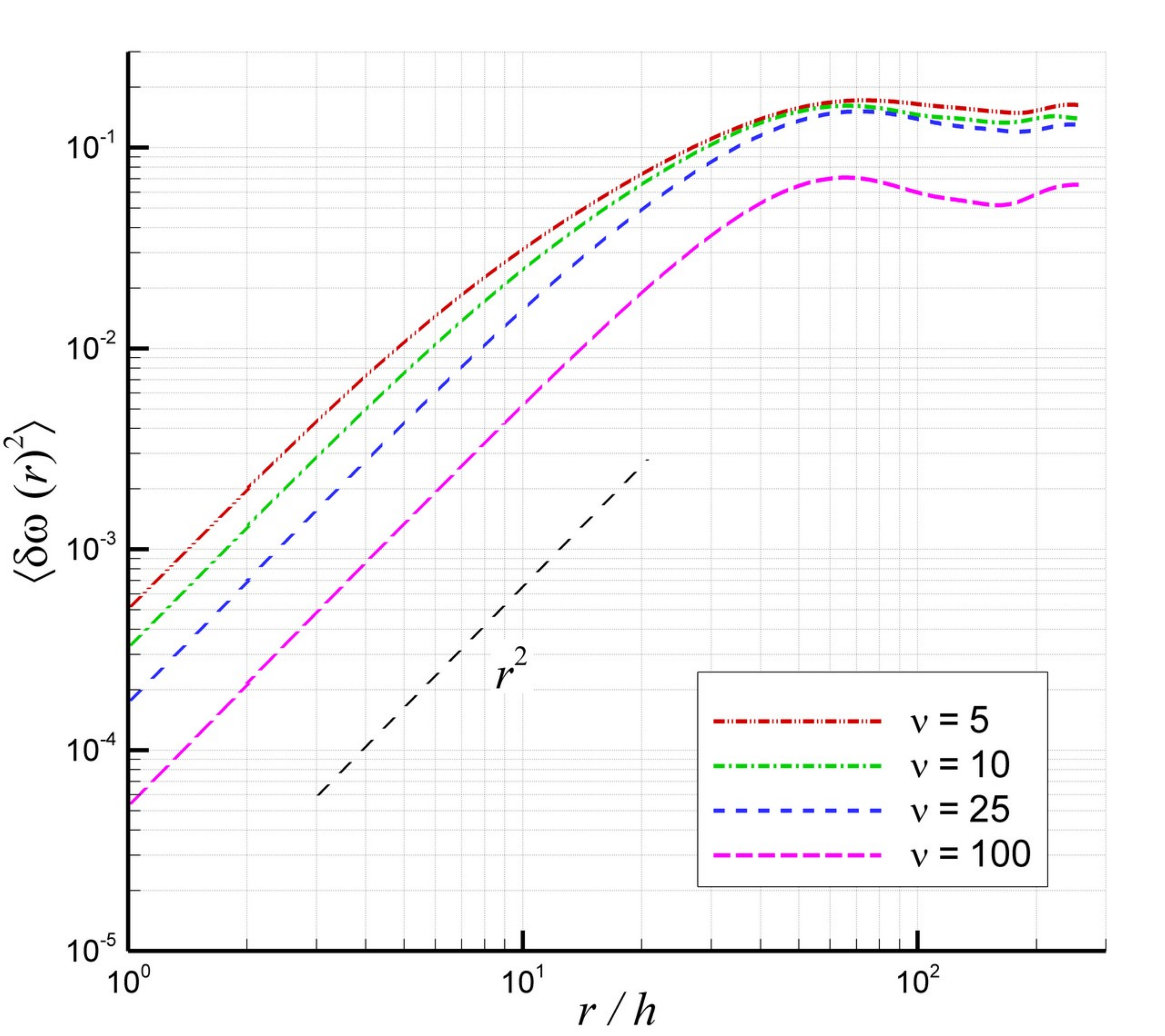}}
}
\caption{The effects of the small scale dissipation coefficient on the statistics without hyperviscosity ($\lambda=0.01$, $p=1$, $\rho=0$, $\sigma=2$, $f_0=0.1$ and $k_f=5$); (a) time series of total energy, (b) angle averaged energy spectra, and (c) second-order vorticity structure functions.}
\label{fig:stat_Re_lamda01}
\end{figure*}

\begin{figure*}[!t]
\centering
\mbox{
\subfigure[$\nu=5$ ($Re=12080$)]{\includegraphics[width=0.245\textwidth]{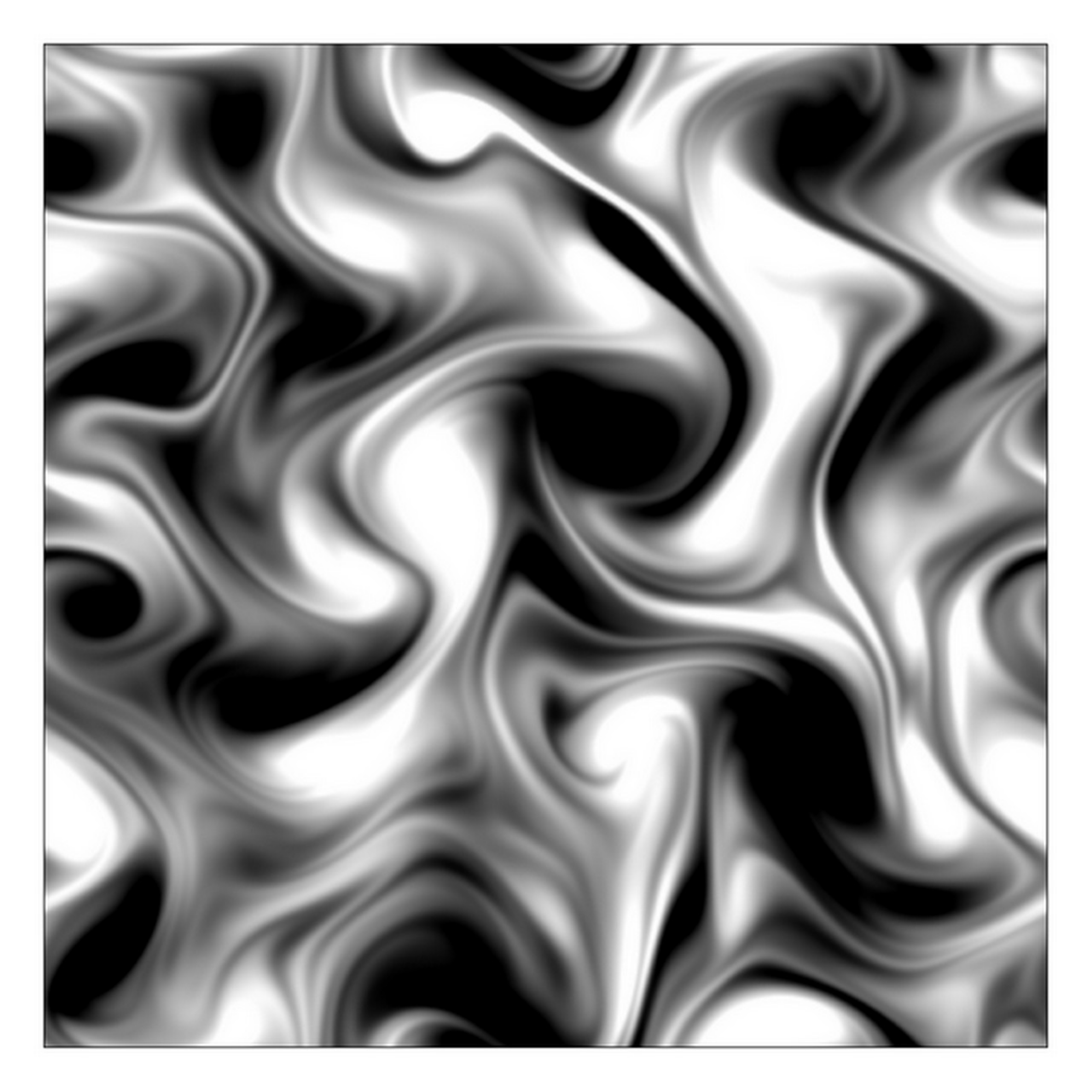}}
\subfigure[$\nu=10$ ($Re=6040$)]{\includegraphics[width=0.245\textwidth]{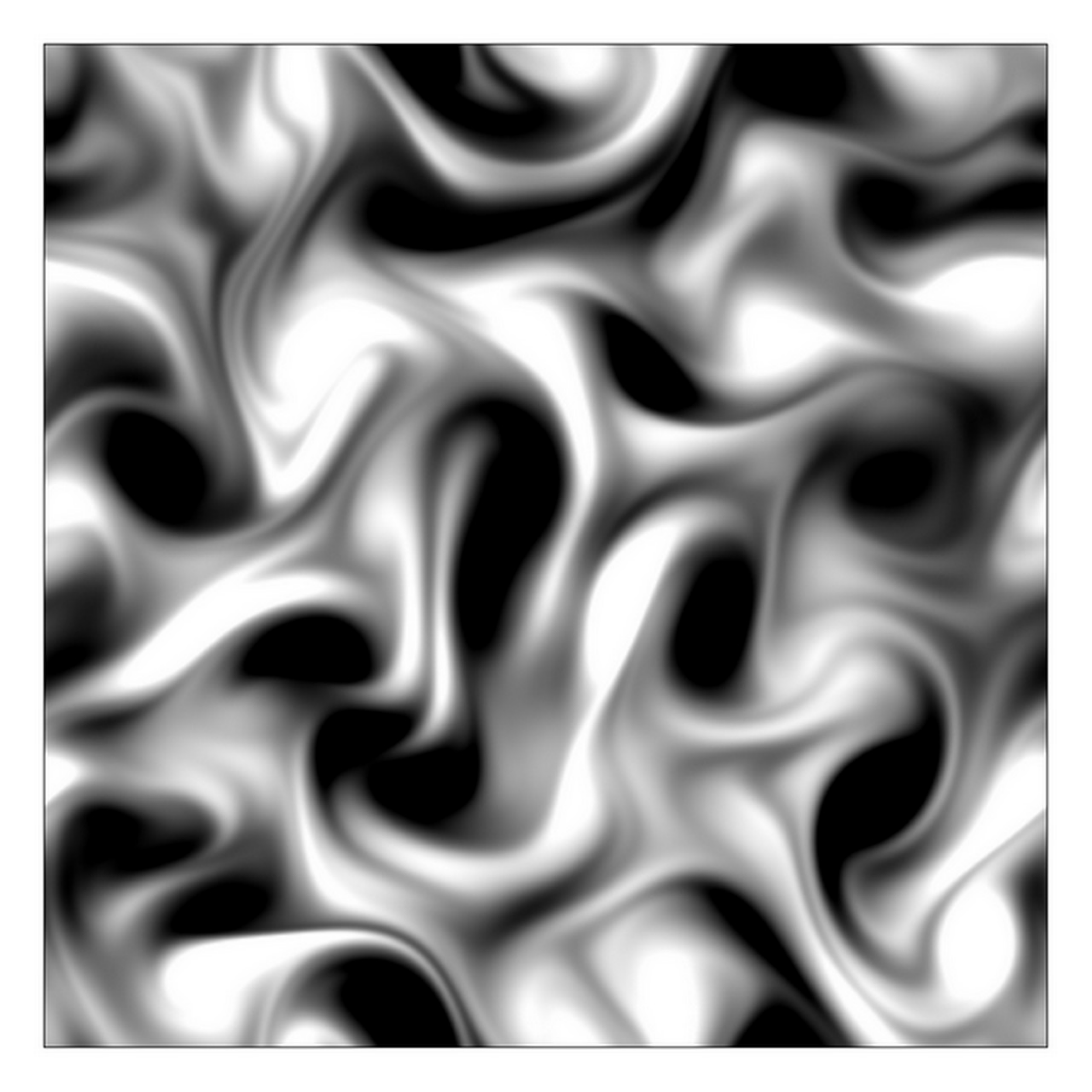}}
\subfigure[$\nu=25$ ($Re=2416$)]{\includegraphics[width=0.245\textwidth]{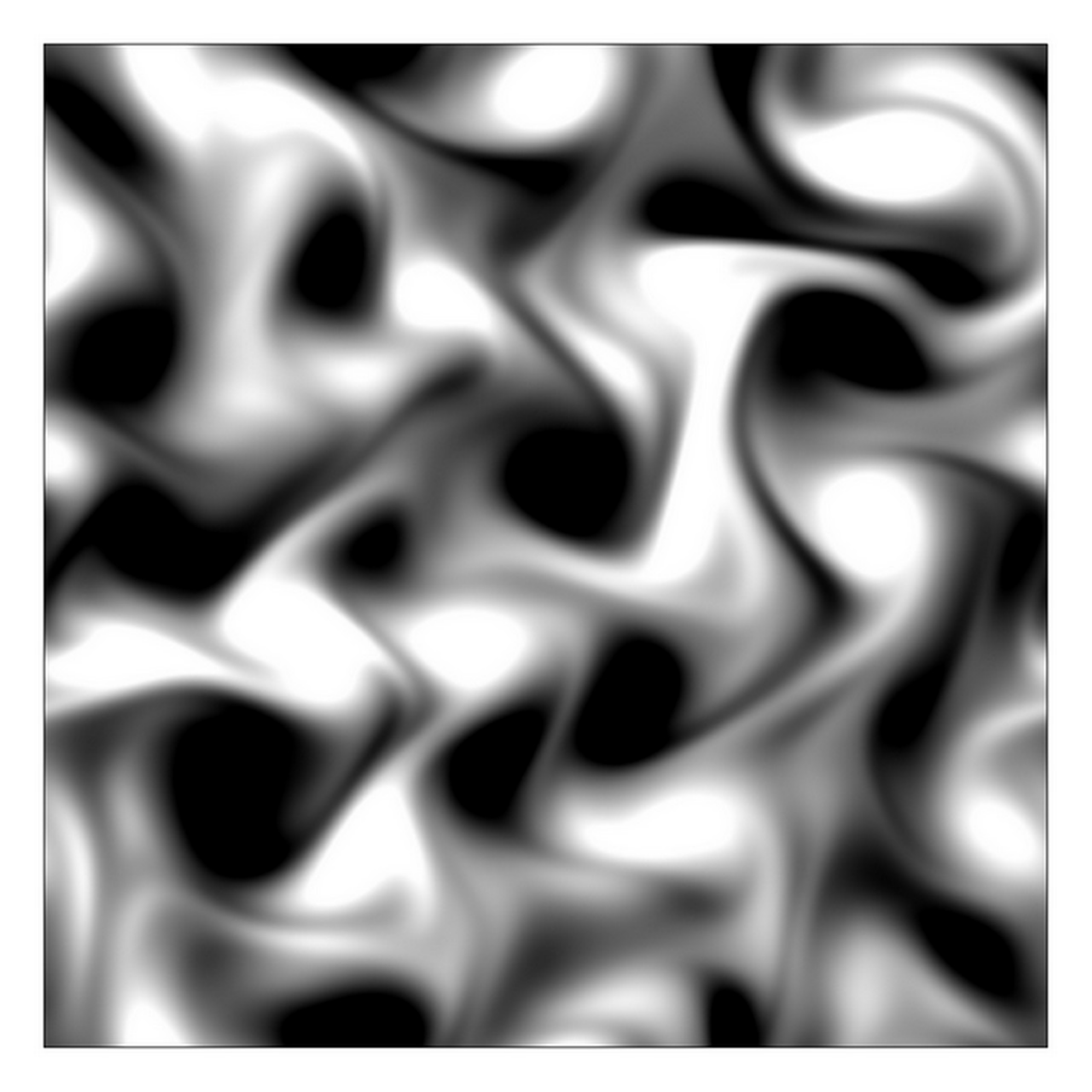}}
\subfigure[$\nu=100$ ($Re=604$)]{\includegraphics[width=0.245\textwidth]{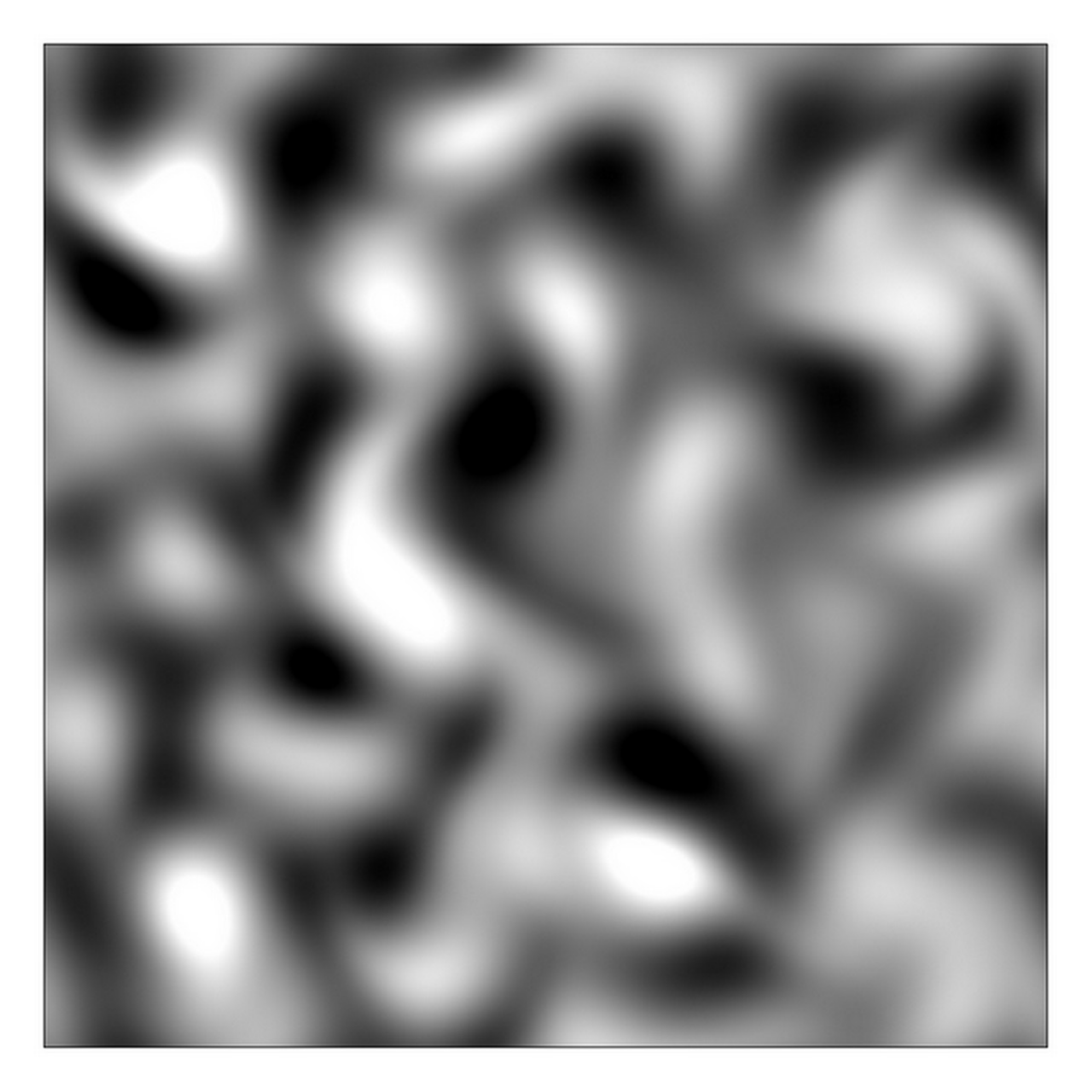}}
}
\caption{Instantaneous vorticity fields at time $t=100$ for varying Reynolds number using the forcing scale $k_f=5$ and the order of Laplacian $p=1$.}
\label{fig:field_Re_lamda01}
\end{figure*}

Next, we investigate the Reynolds number dependence of the statistics of stationary turbulence using the the order of Laplacian $p=1$ that represents the classical quadratic dissipation mechanism in Navier-Stokes equation without using hyperviscosity. As demonstrated in Figure~\ref{fig:stat_Re_lamda01}, the angle averaged energy spectrum asymptotically reaches the $k^{-3}$ scaling in the inertial range as $\nu$ decreases (i.e., $Re$ increases). We find that the Reynolds number dependency is more stringent if we look at the turbulence statistics in wave space using the angle averaged energy spectrum. The structure functions are proportional to $r^{2}$ for the smaller separations and flatten for higher separations. The corresponding instantaneous vorticity fields at time $t = 100$ are compared in Figure~\ref{fig:field_Re_lamda01} for the same set of Reynolds numbers. As we can see from Figure~\ref{fig:field_Re_lamda01}, the amount of filamentation increases for higher Reynolds numbers. Due to the smaller convection in lower Reynolds numbers, the interaction between two vortices is not as strong as that of the computations with higher Reynolds numbers.

\subsection{Effects of large scale dissipation mechanism}
\label{sec:disp-l}

\begin{figure*}[!t]
\centering
\mbox{
\subfigure[]{\includegraphics[width=0.33\textwidth]{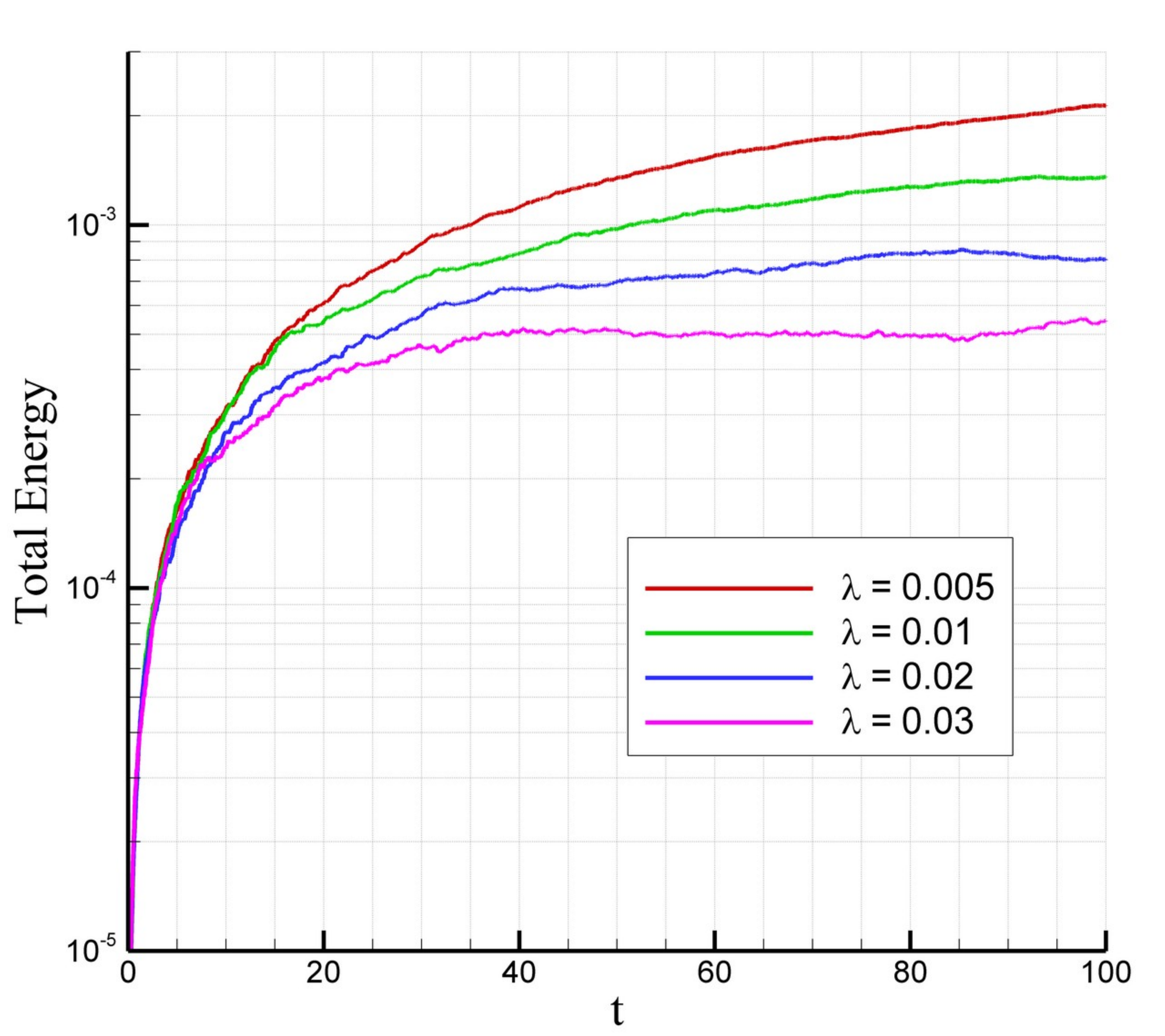}}
\subfigure[]{\includegraphics[width=0.33\textwidth]{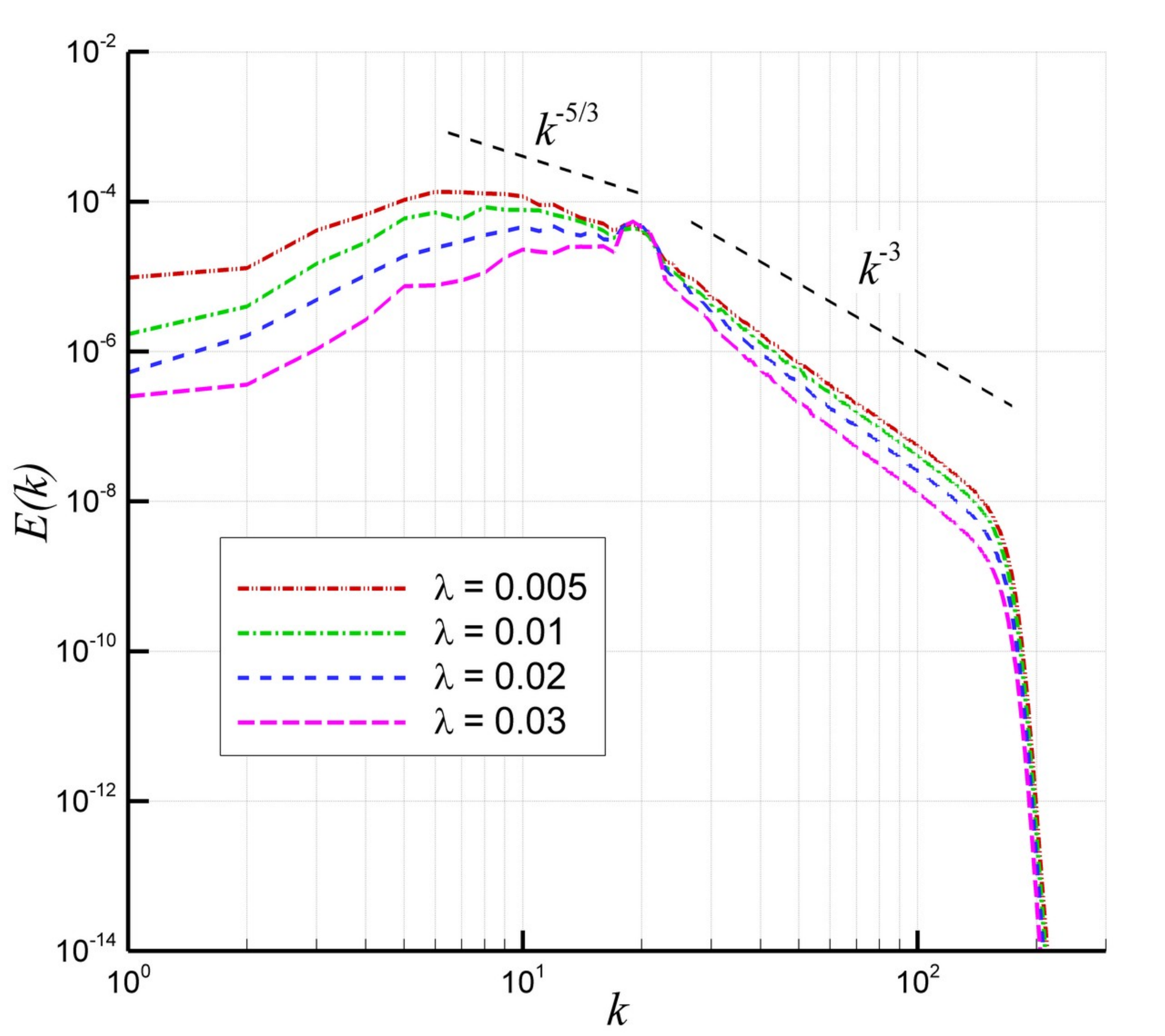}}
\subfigure[]{\includegraphics[width=0.33\textwidth]{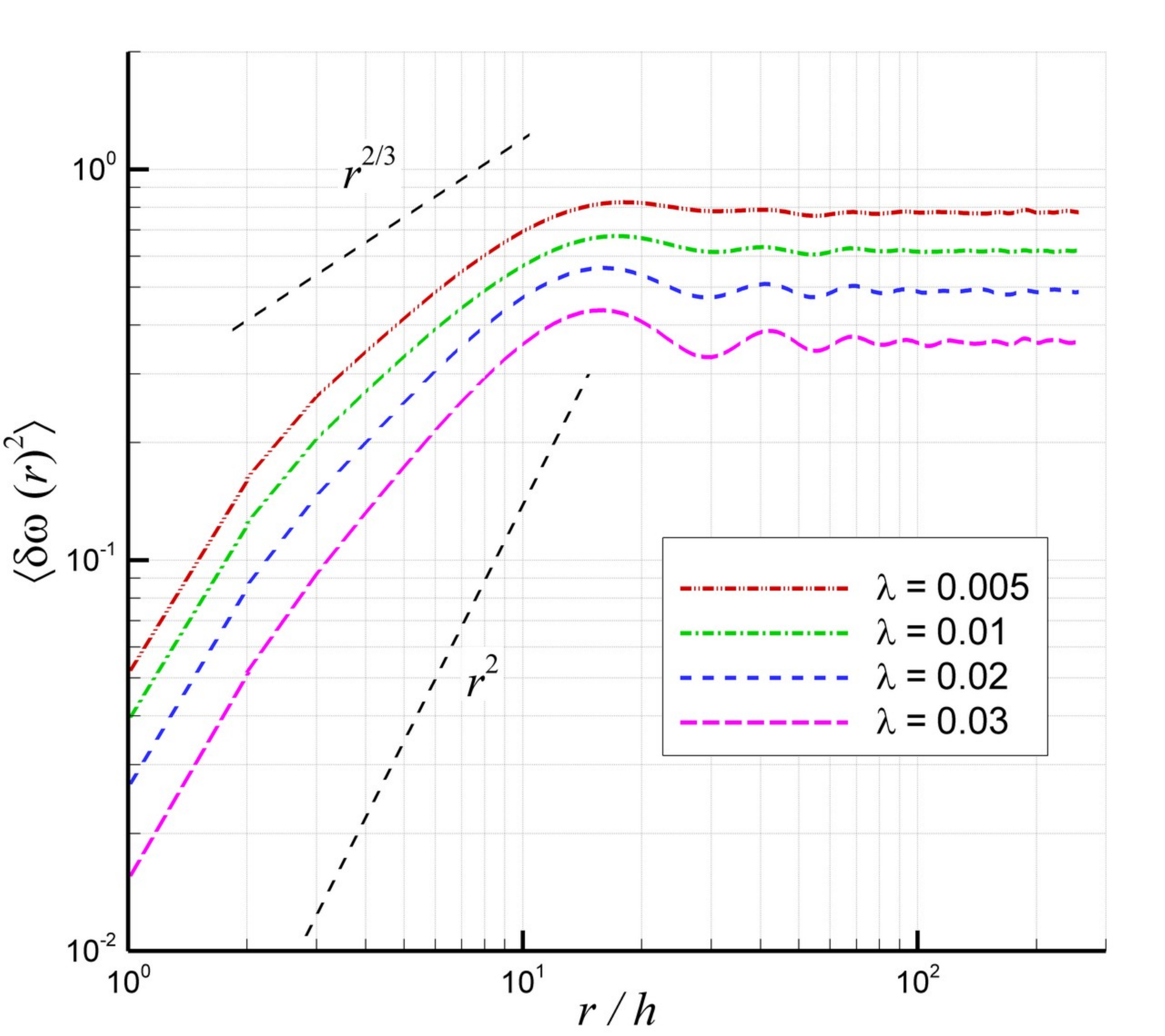}}
}
\caption{The effects of the large scale dissipation coefficient on the statistics ($k_f=20$, $f_0=0.1$, $\sigma=2$, $\rho=0.0$, $\nu=1000$ and $p=8$); (a) time series of total energy, (b) angle averaged energy spectra, and (c) second-order vorticity structure functions.}
\label{fig:stat_drag_k20}
\end{figure*}

\begin{figure*}[!t]
\centering
\mbox{
\subfigure[$\lambda=0.005$]{\includegraphics[width=0.245\textwidth]{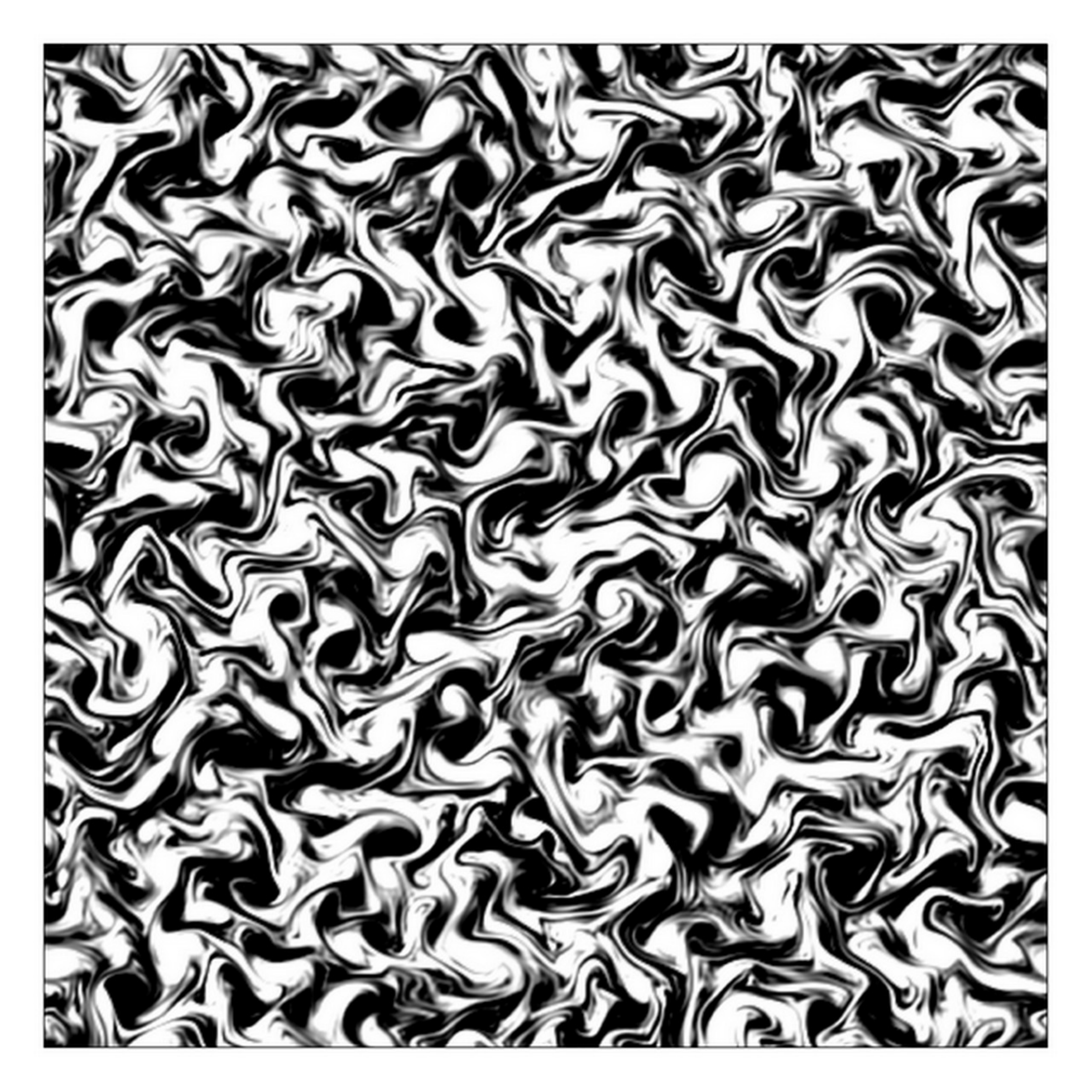}}
\subfigure[$\lambda=0.01$]{\includegraphics[width=0.245\textwidth]{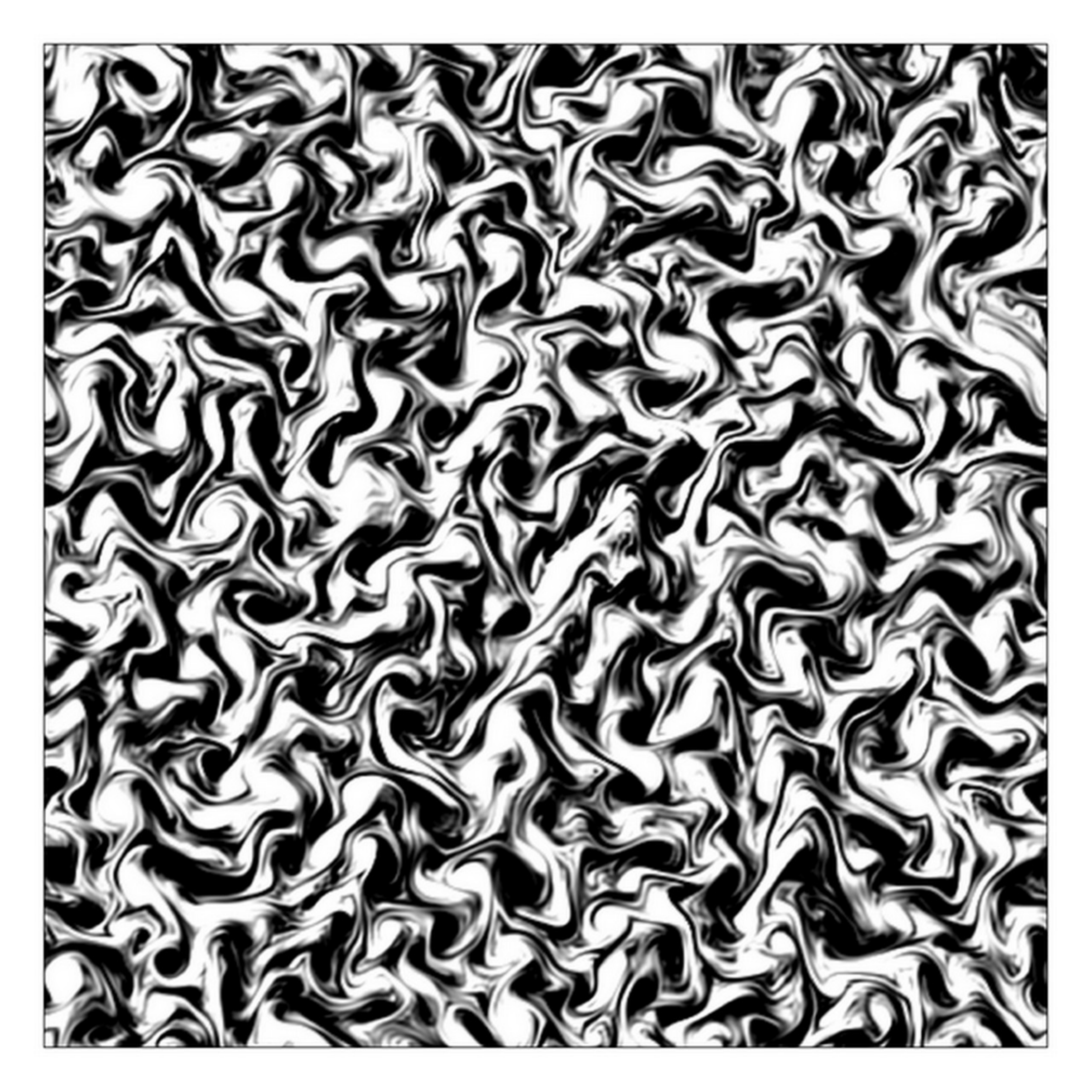}}
\subfigure[$\lambda=0.02$]{\includegraphics[width=0.245\textwidth]{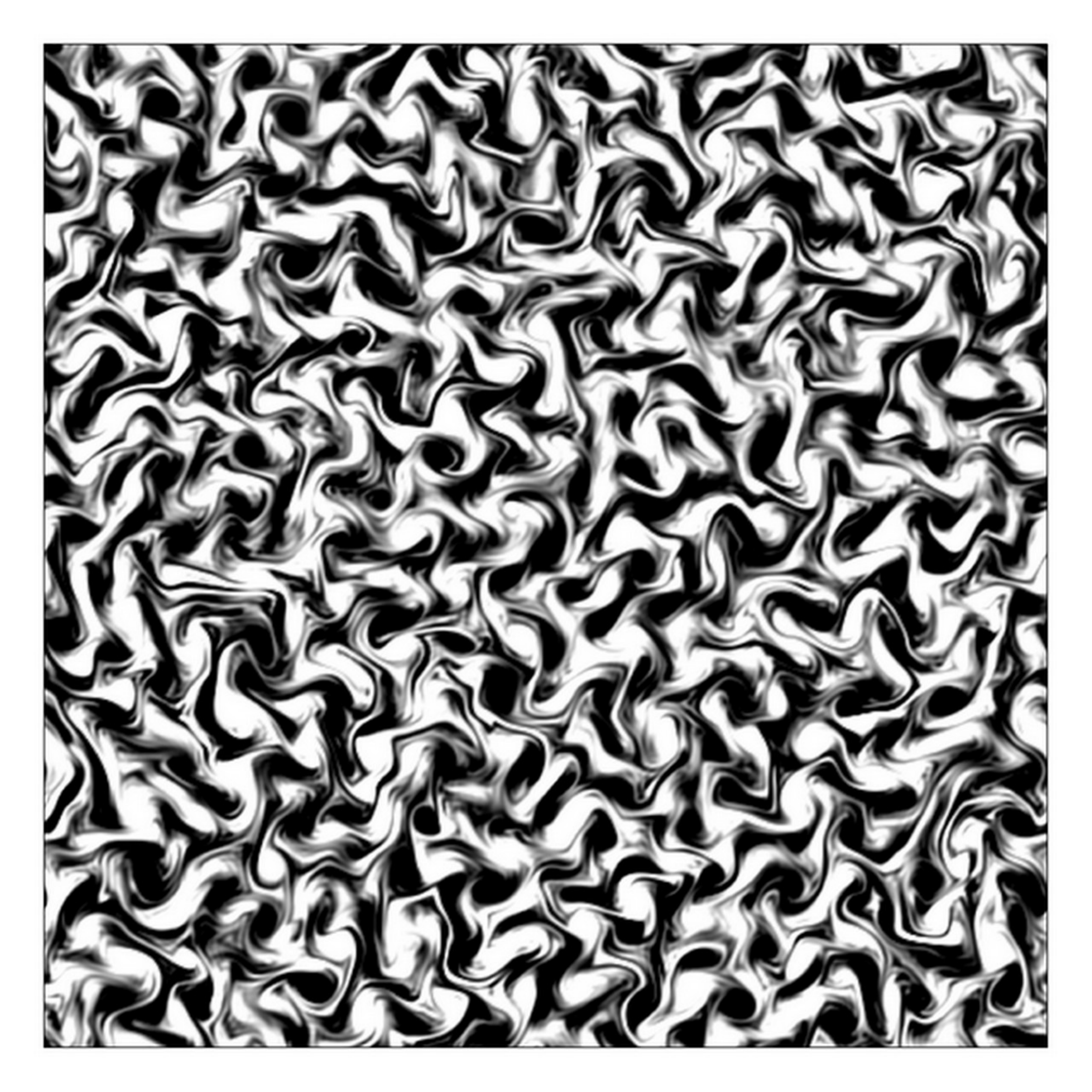}}
\subfigure[$\lambda=0.03$]{\includegraphics[width=0.245\textwidth]{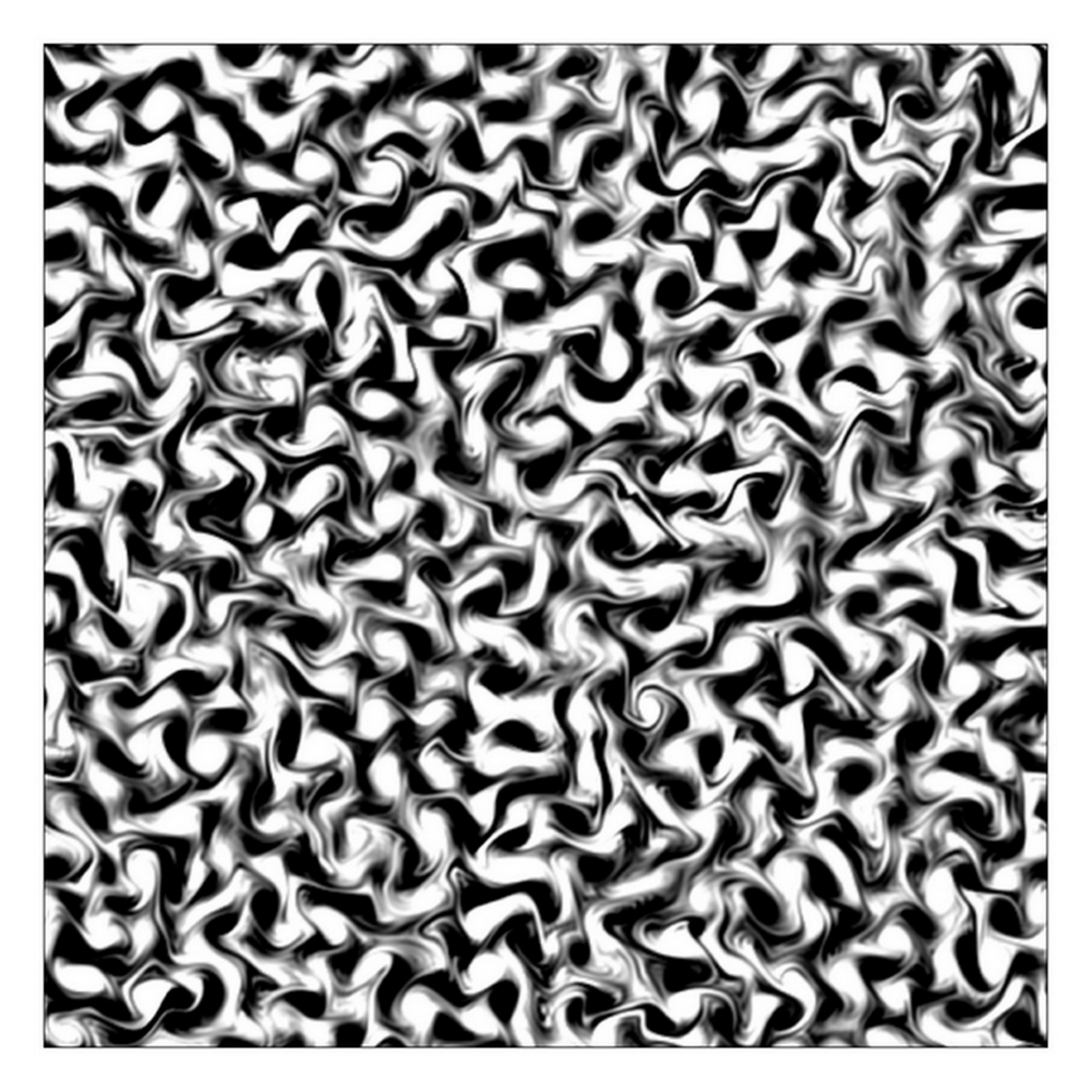}}
}
\caption{Instantaneous vorticity fields at time $t=100$ for varying large scale friction coefficient $\lambda$ using the forcing scale is $k_f=20$ and the small scale dissipation coefficients are $\nu=1000$ and $p=8$.}
\label{fig:field_drag_k20}
\end{figure*}

In order to investigate the effects of large scale dissipation mechanism in the dynamics of two-dimensional homogeneous turbulence, we perform a similar analysis by varying the large scale friction coefficient, while the other parameters associated with small scale friction and forcing mechanisms are held constant in the model. Figure~\ref{fig:stat_drag_k20} shows the computed statistics for the effective forcing scale $k_f=20$ applied in all cases. The statistics on the evolution of the total energy clearly demonstrate that the main mechanism that determines the time scale for stationary regime is the large-scale friction mechanism. Due to the inverse energy cascading in two-dimensional turbulence, a statistically steady state is established by damping mechanism in low-wavenumbers. Increasing the large scale damping coefficient $\lambda$ provides an earlier statistically steady state with a less amount of energy. It is interesting to note that the structure of the large scale spectrum also depends considerably on the amount of the large scale friction coefficient in which Kolmogorov scaling appears for smaller value of the $\lambda$. It can also be seen from Figure~\ref{fig:stat_drag_k20} that energy spectra in forward enstrophy cascade range appear steeper than $k^{-3}$ with increasing large scale damping coefficients. We found that the vorticity structure functions are independent of the separation length for larger $r$, as indicated by KBL theory in the inviscid limit, and change gradually from $r^{2}$ to $r^{2/3}$ for smaller separations. Thus, we conclude that the scaling exponents of the structure function are considerably influenced by the large scale damping mechanism for the small separation $r$. The instantaneous vorticity fields compared in Figure~\ref{fig:field_drag_k20} show another interesting observation that the flow pattern has more vorticity filaments for smaller large scale friction coefficients, whereas the amount of vortical structures seems equivalent in each case due to the use of the same forcing scale.

\begin{figure*}[!t]
\centering
\mbox{
\subfigure[]{\includegraphics[width=0.33\textwidth]{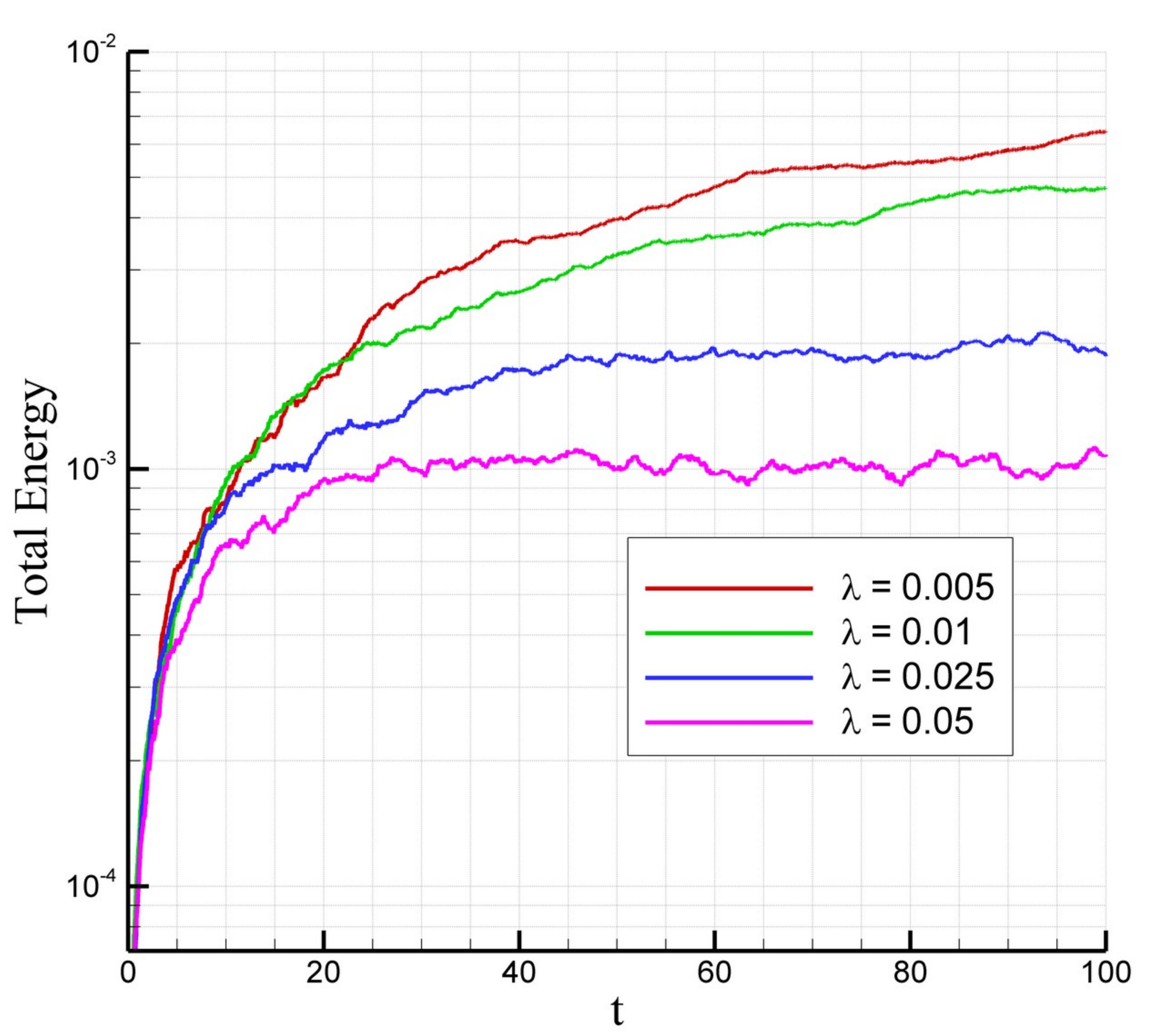}}
\subfigure[]{\includegraphics[width=0.33\textwidth]{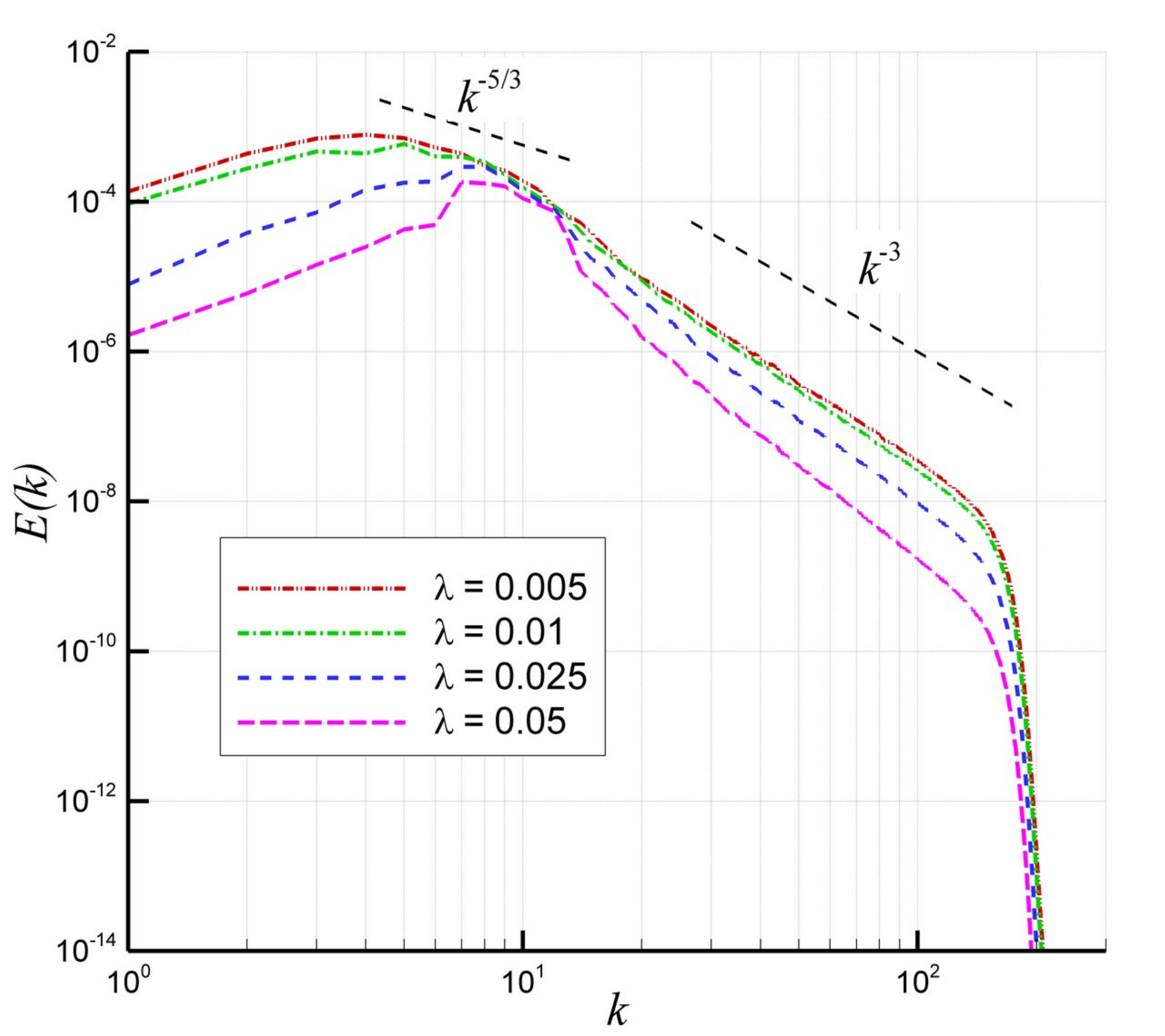}}
\subfigure[]{\includegraphics[width=0.33\textwidth]{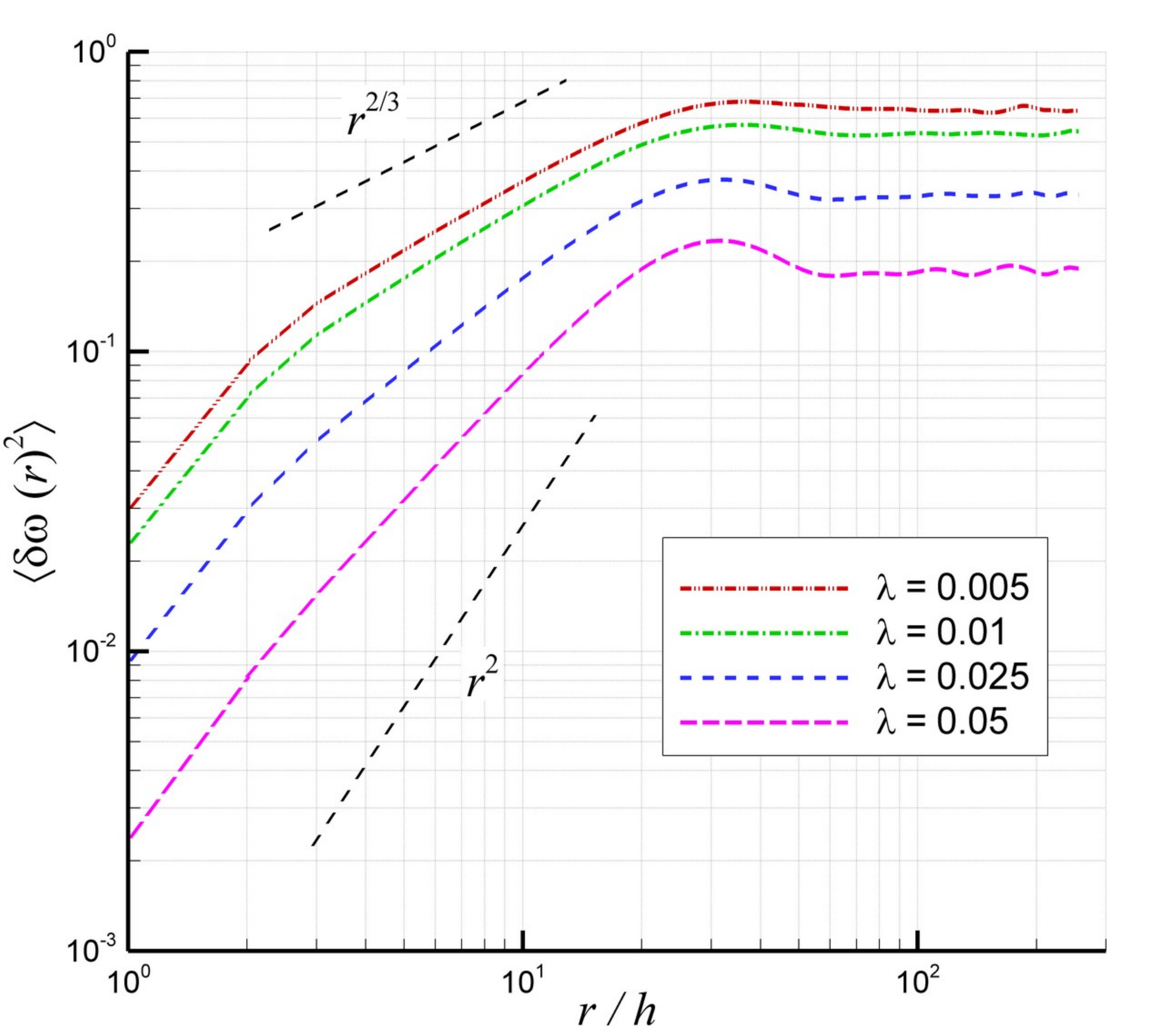}}
}
\caption{The effects of the large scale dissipation coefficient on the statistics ($k_f=10$, $f_0=0.1$, $\sigma=3$, $\rho=0.0$, $\nu=1000$ and $p=8$); (a) time series of total energy, (b) angle averaged energy spectra, and (c) second-order vorticity structure functions.}
\label{fig:stat_drag_k10}
\end{figure*}

\begin{figure*}[!t]
\centering
\mbox{
\subfigure[$\lambda=0.005$]{\includegraphics[width=0.245\textwidth]{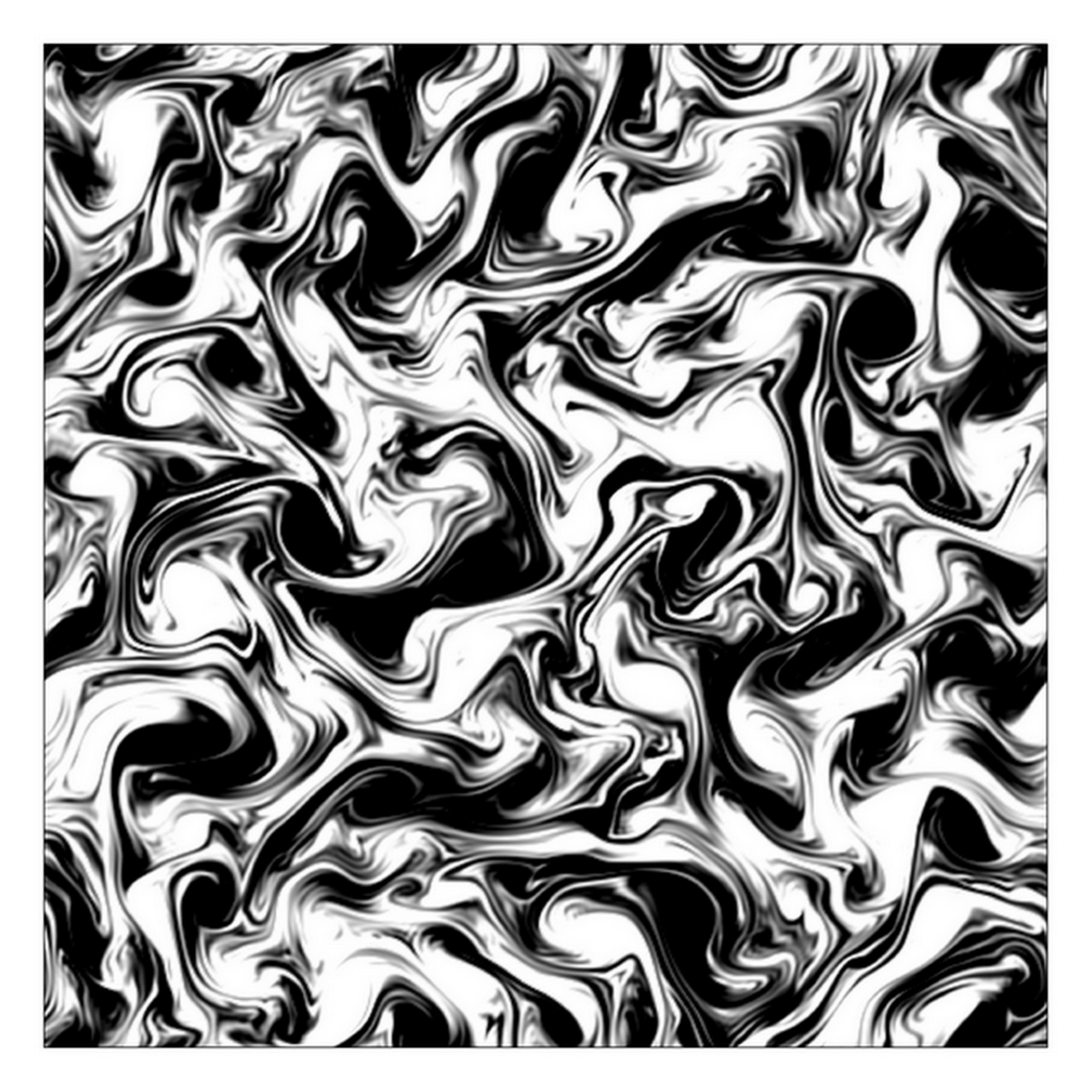}}
\subfigure[$\lambda=0.01$]{\includegraphics[width=0.245\textwidth]{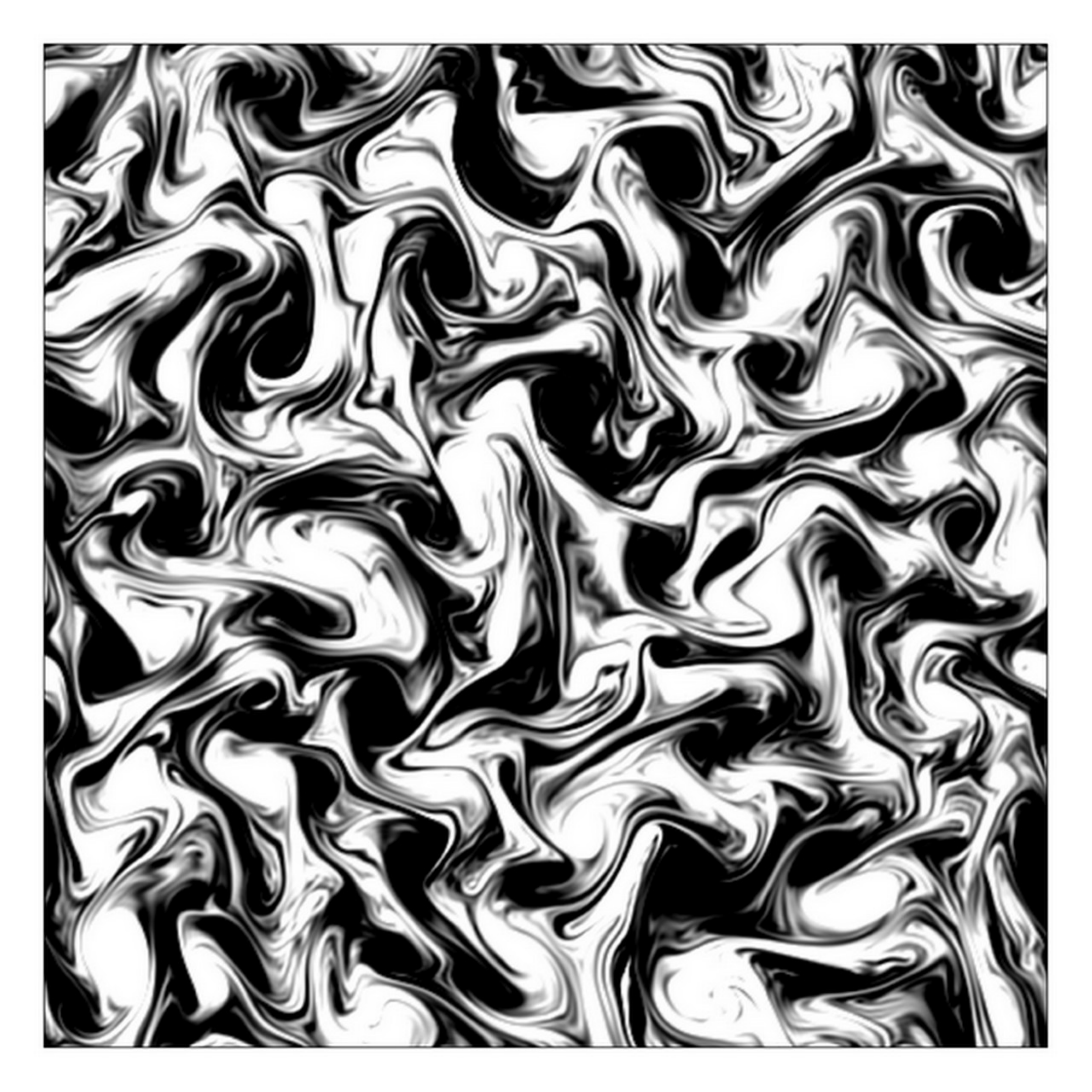}}
\subfigure[$\lambda=0.025$]{\includegraphics[width=0.245\textwidth]{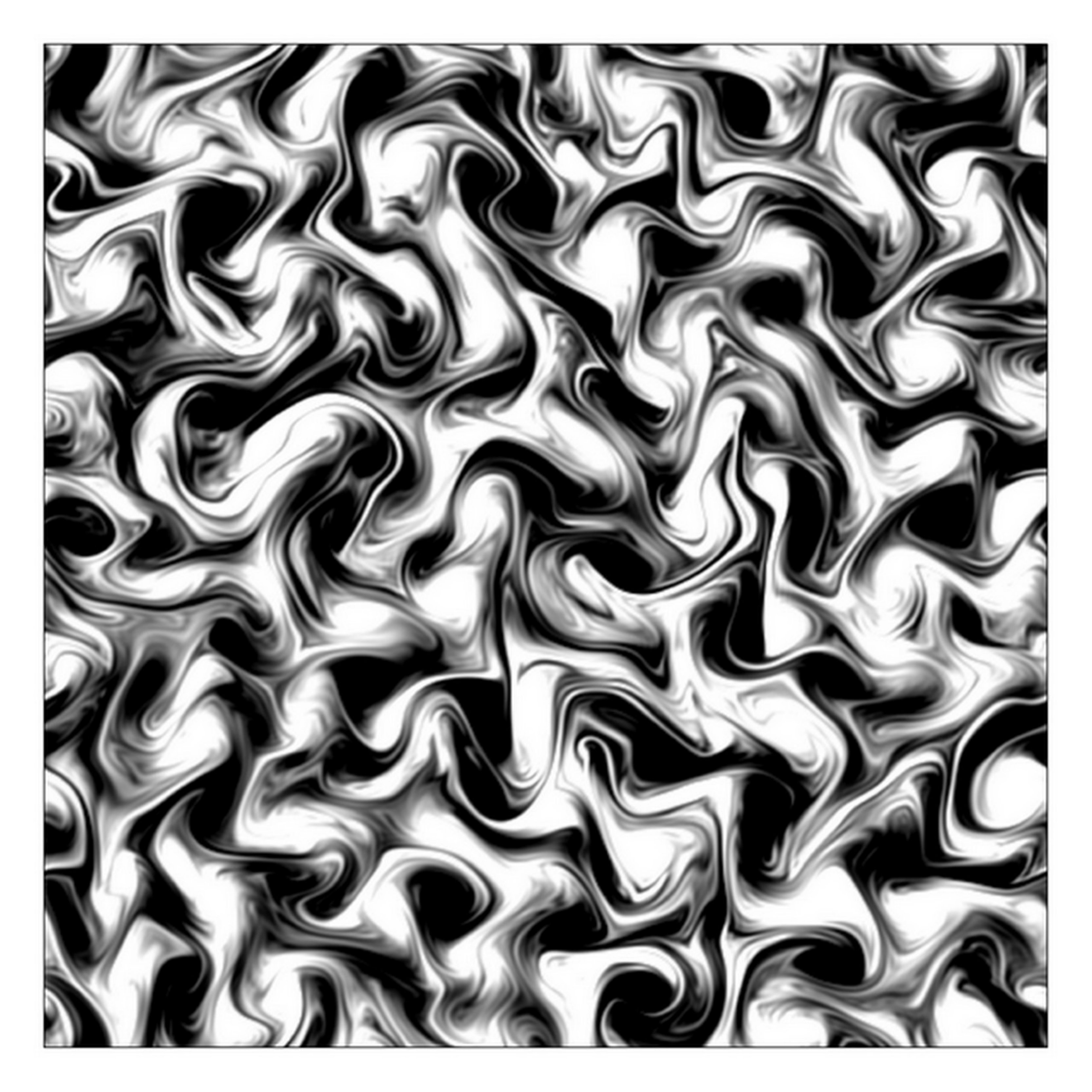}}
\subfigure[$\lambda=0.05$]{\includegraphics[width=0.245\textwidth]{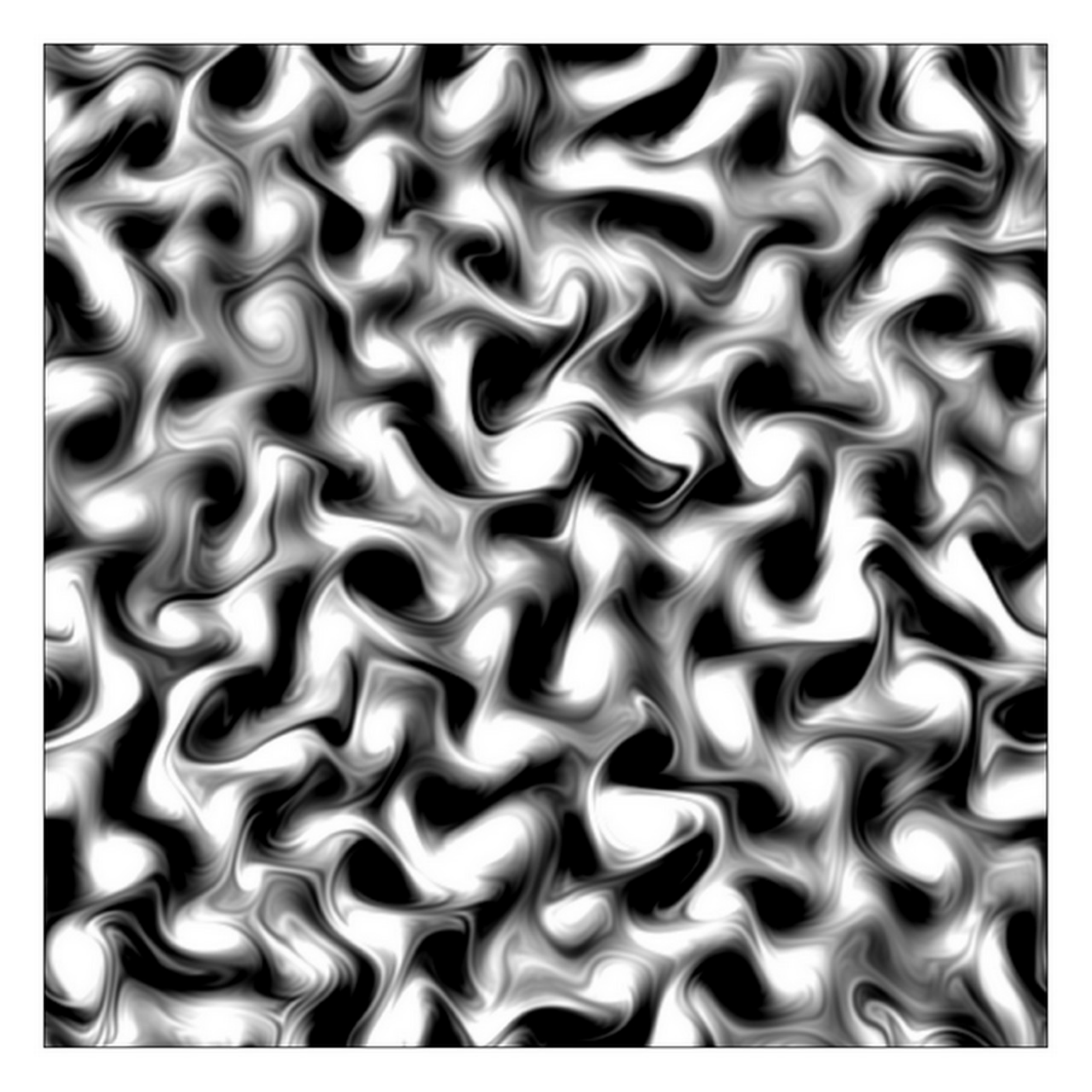}}
}
\caption{Instantaneous vorticity fields at time $t=100$ for varying the large scale friction coefficient $\lambda$ using the forcing scale $k_f=10$ and the small scale dissipation coefficients of $\nu=1000$ and $p=8$.}
\label{fig:field_drag_k10}
\end{figure*}

Figure~\ref{fig:stat_drag_k10} shows statistics for $k_f=10$. Similar to the previous comparisons with $k_f=20$, the dual cascading phenomenon, Kraichnan scaling for the scales smaller than the energy injection scale, and Kolmogorov scaling for the scales greater than the forcing scale, appears in statistics for energy spectra and structure functions. The lines $k^{-5/3}$ and $k^{-3}$ are shown in the plot for energy spectra and lines $r^{2/3}$ and $r^{2}$ are included in the plot for structure functions for comparison. The corresponding flow patterns are also shown in Figure~\ref{fig:field_drag_k10}, demonstrating a clear comparison for the effect of the large scale friction coefficient into the flow structure. For larger friction coefficient, flow tends to be concentrated in the centers of corresponding vortices, showing less interactions among them. The grid-like pattern of vortices emerges, corresponding to the accumulation of energy near forcing scales; this pattern is also evident in energy spectra.

\begin{figure*}[!t]
\centering
\mbox{
\subfigure[]{\includegraphics[width=0.33\textwidth]{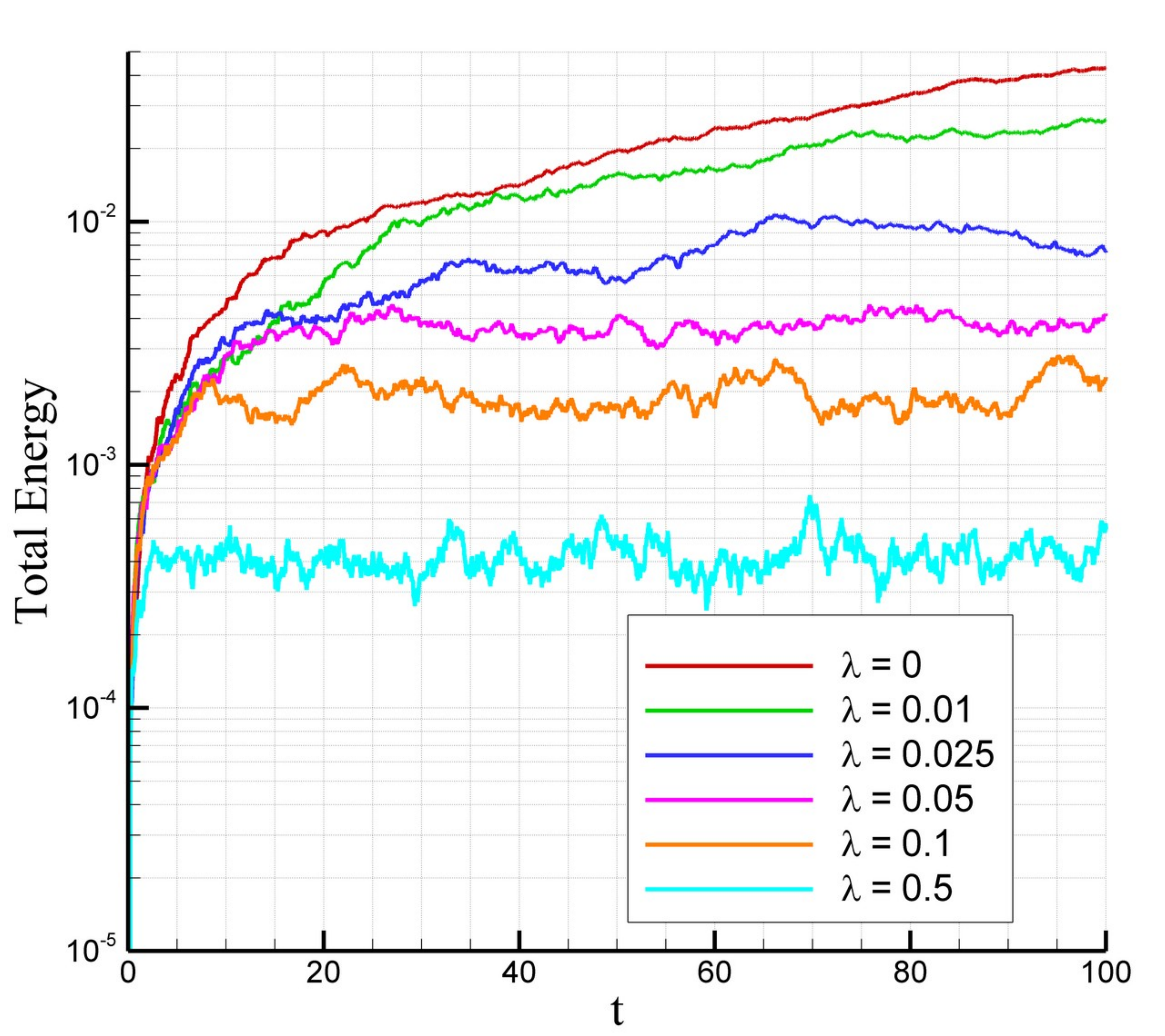}}
\subfigure[]{\includegraphics[width=0.33\textwidth]{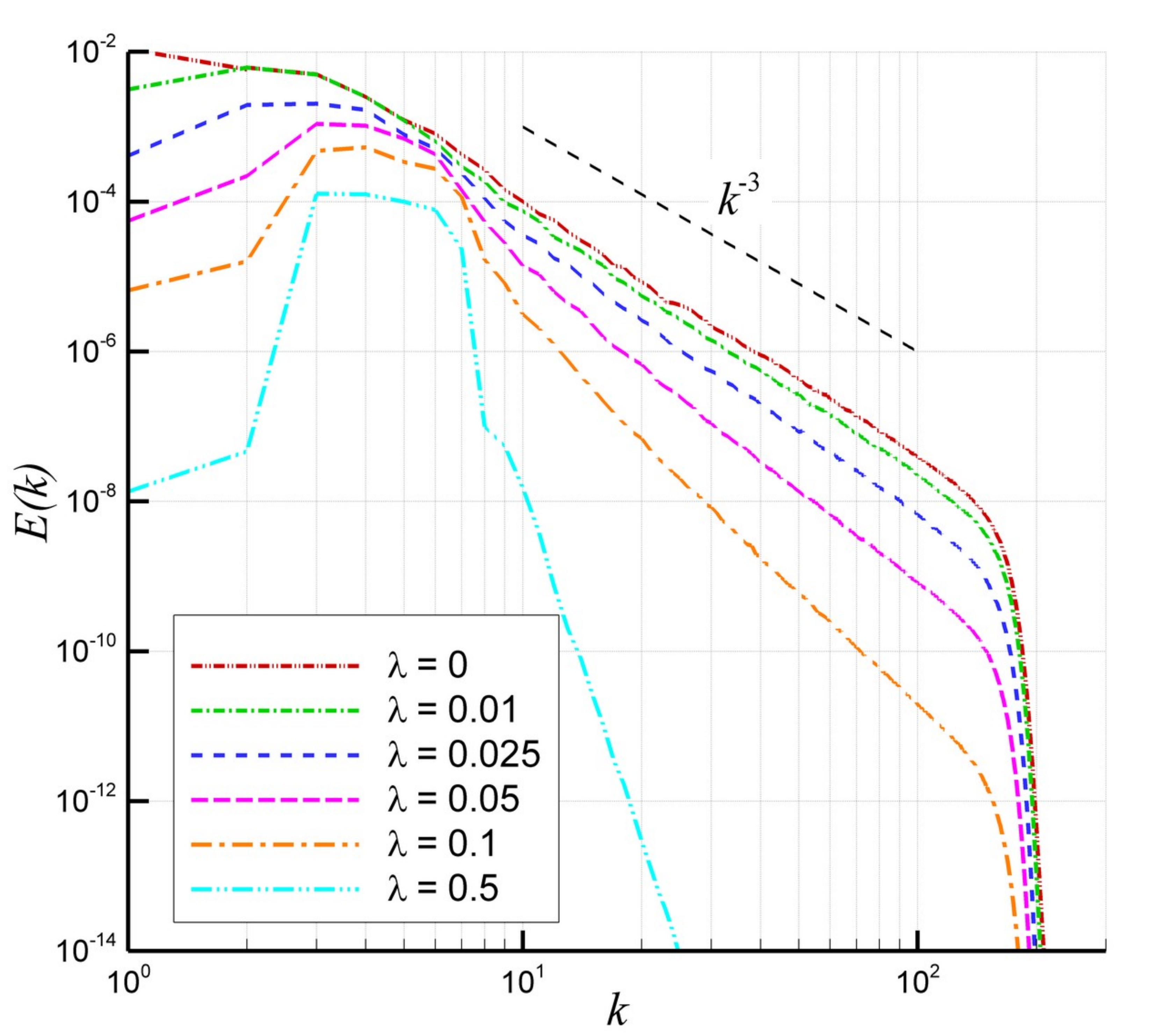}}
\subfigure[]{\includegraphics[width=0.33\textwidth]{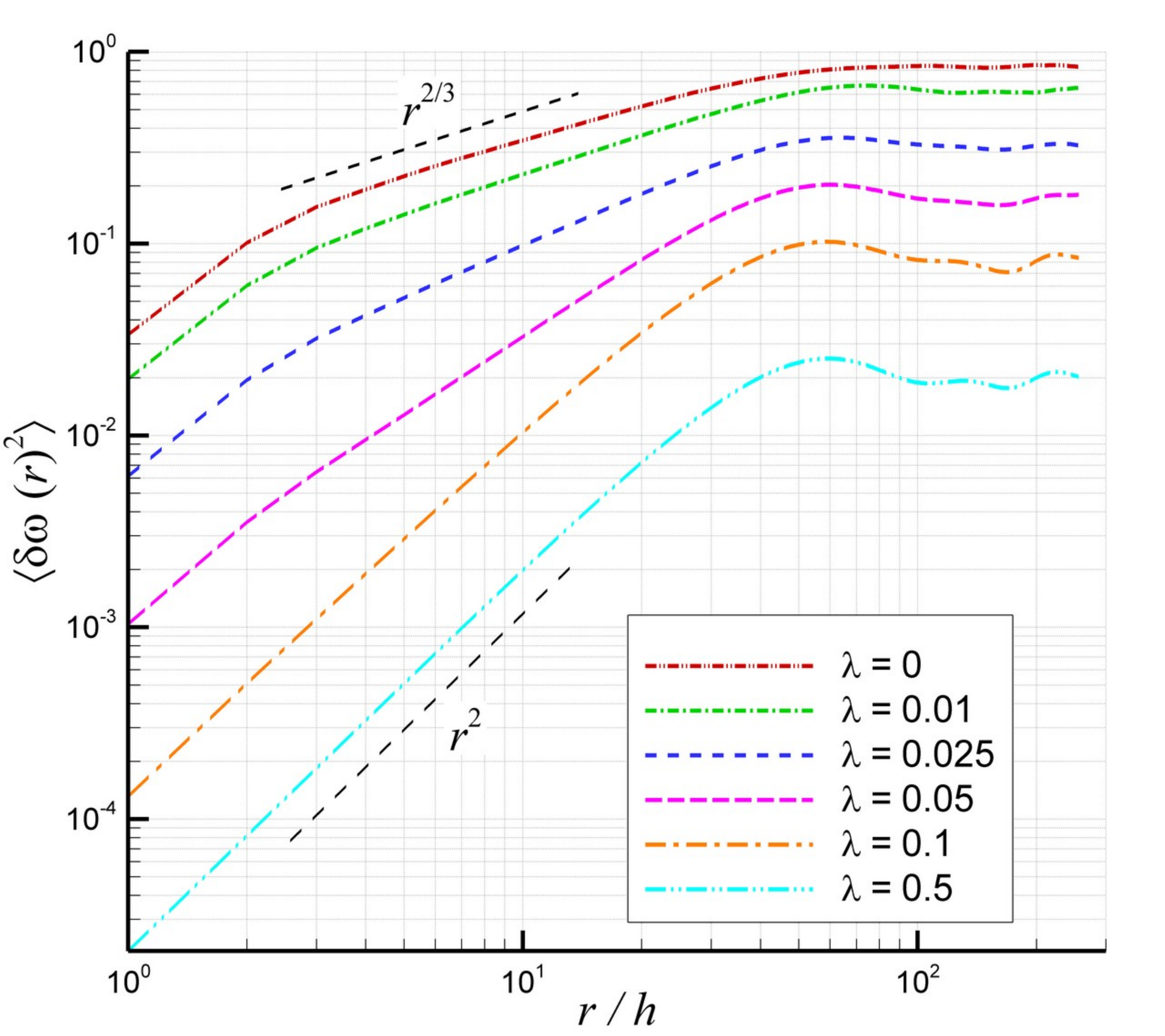}}
}
\caption{The effects of the large scale dissipation coefficient on the statistics ($k_f=5$, $f_0=0.1$, $\sigma=2$, $\rho=0.5$, $\nu=1000$ and $p=8$); (a) time series of total energy, (b) angle averaged energy spectra, and (c) second-order vorticity structure functions.}
\label{fig:stat_drag_k5}
\end{figure*}

\begin{figure*}[!t]
\centering
\mbox{
\subfigure[$\lambda=0.0$]{\includegraphics[width=0.245\textwidth]{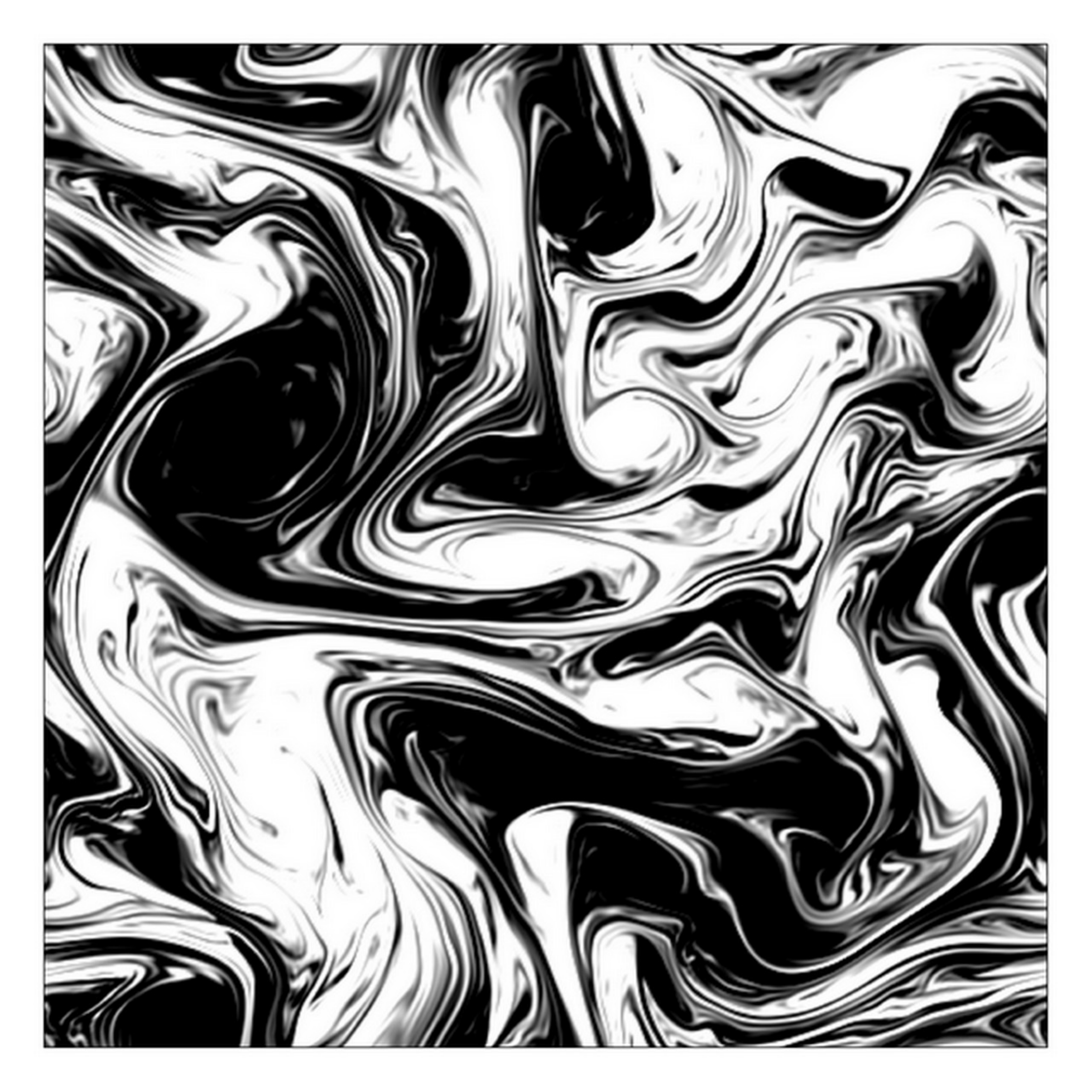}}
\subfigure[$\lambda=0.01$]{\includegraphics[width=0.245\textwidth]{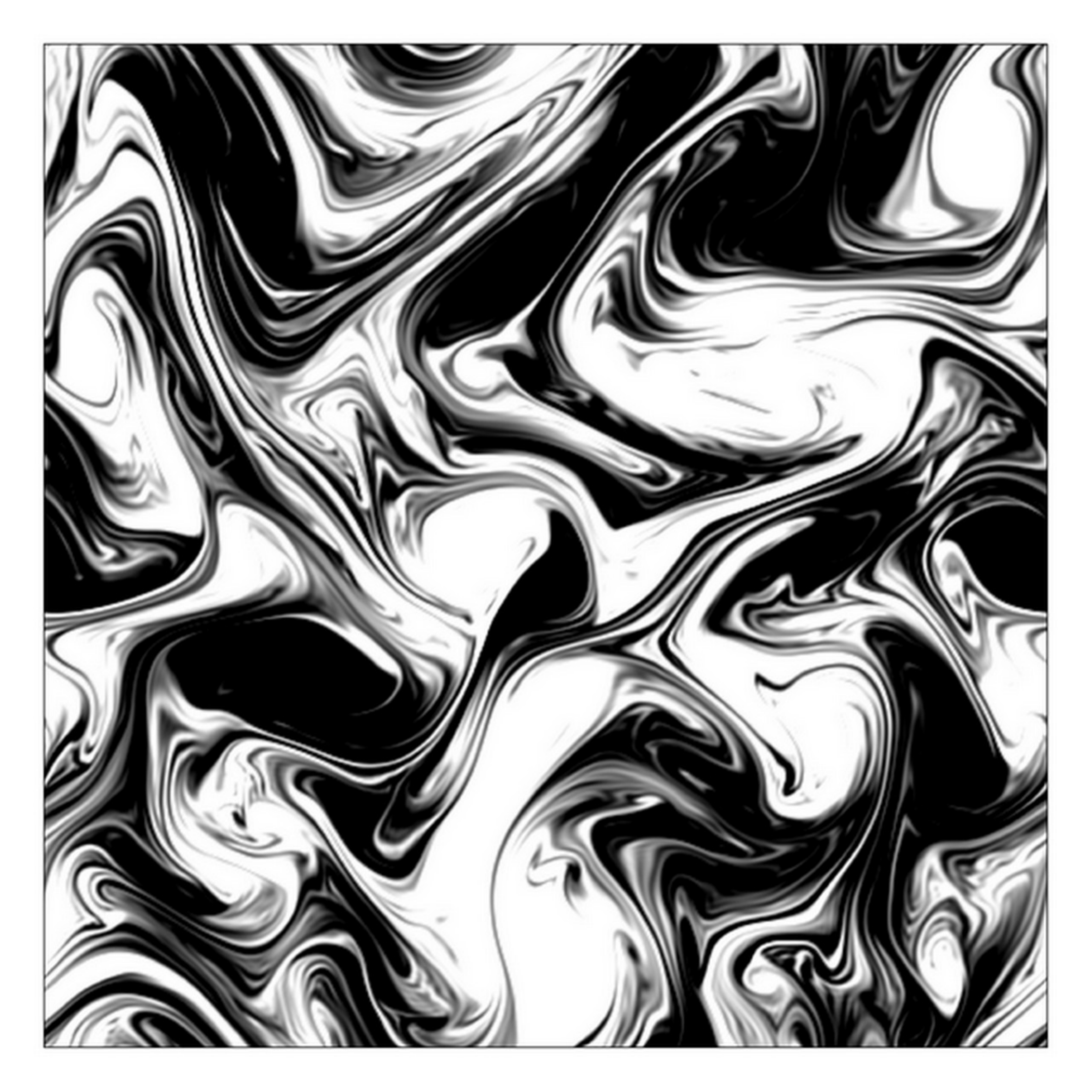}}
\subfigure[$\lambda=0.025$]{\includegraphics[width=0.245\textwidth]{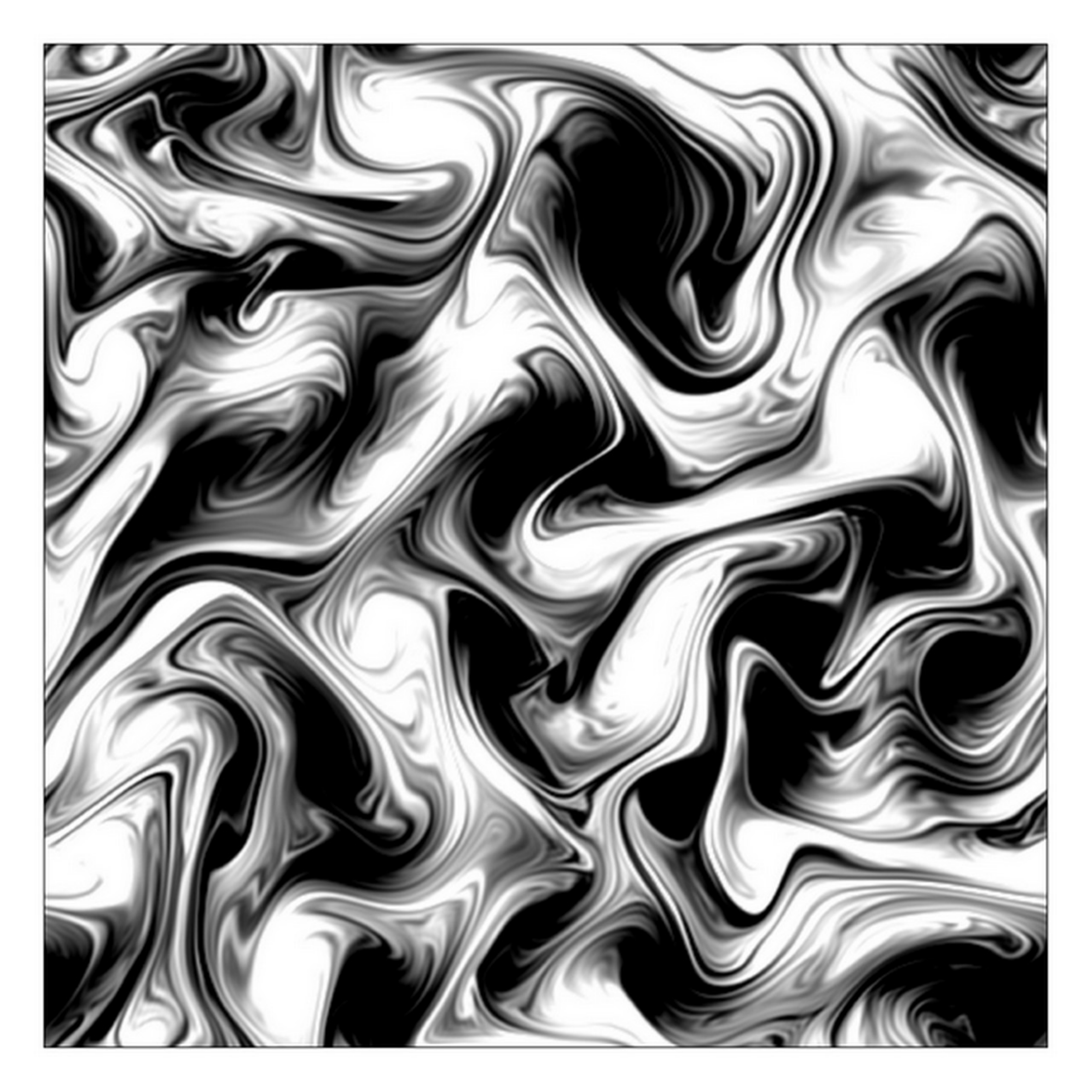}}
\subfigure[$\lambda=0.1$]{\includegraphics[width=0.245\textwidth]{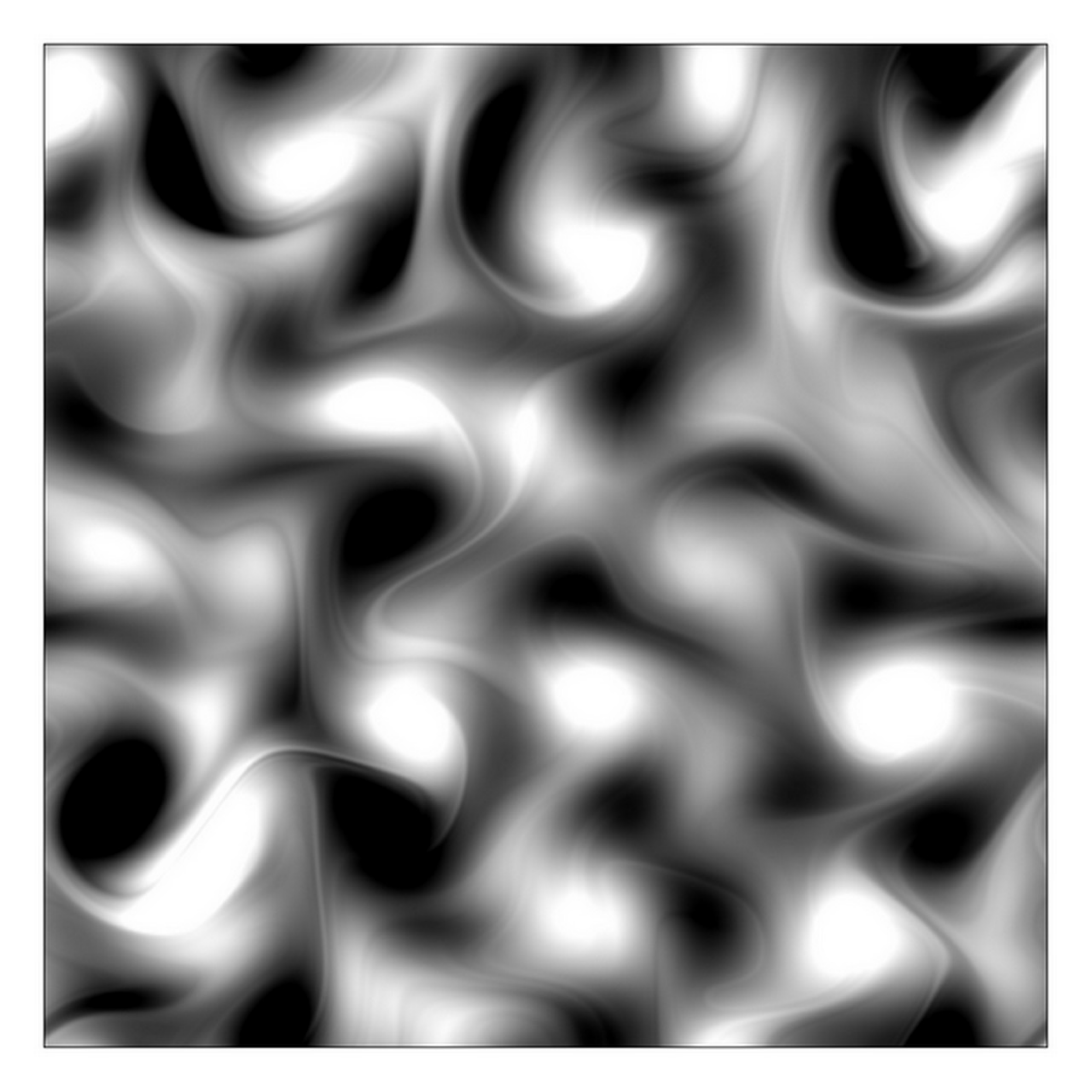}}
}
\caption{Instantaneous vorticity fields at time $t=100$ for varying the large scale friction coefficient $\lambda$ using the forcing scale $k_f=5$ and the small scale dissipation coefficients of $\nu=1000$ and $p=8$.}
\label{fig:field_drag_k5}
\end{figure*}

In order to further investigate the effects of large scale friction mechanism, we next perform a similar analysis for a larger energy injection scale $k_f=5$. The results are summarized in Figure~\ref{fig:stat_drag_k5}, showing evolution of total kinetic energy, mean energy spectra, and mean second-order vorticity structure functions for a series of runs with varying large scale friction coefficient. For $\lambda=0$, the flow approaches two-dimensional homogenous turbulence with a forward cascade of enstrophy and an inverse cascade of energy. After an initial period of nonlinear adjustment, the total energy grows linearly in time, $dE/dt\approx t$. Some part of the energy input injected by forcing mechanism at forcing scale $k_f=5$ is transferred to scales smaller than the forcing scale, and dissipated by viscosity. In this finite system for $\lambda=0$, however, there is no statistically steady state due to inverse energy cascade in which energy is continuously transferring from the forcing scale to the larger scales. Figure~\ref{fig:stat_drag_k5} also shows that as the large scale friction coefficient is decreased, the growth of energy is decreased, until at a value $\lambda\approx0.025$, the growth is completely suppressed by the large scale damping, and stationary turbulence is obtained. Comparing energy spectra in Figure~\ref{fig:stat_drag_k5}, Kraichnan scaling is obtained for $\lambda \rightarrow 0$,  though the spectra for larger $\lambda$ appear steeper than $k^{-3}$. It is also important to note that the tails of corresponding second-order vorticity structure functions scale as $r^{2}$ for larger $\lambda$. Therefore, we confirm that the scaling for the structure functions shows an asymptotical limit for flows having energy spectrum is steeper than $k^{-3}$. Based on these findings, we conjecture that looking for turbulence statistics only in terms of structure functions does not guarantee the proper scaling exponents in energy cascading predictions. Figure~\ref{fig:stat_drag_k5} also shows that the second-order vorticity structure function flattens for larger separation distances $r$, as predicted by the KBL theory in the inviscid limit. Figure~\ref{fig:field_drag_k5} illustrates the instantaneous vorticity field at time $t=100$ for a set of values for $\lambda$. Similar to our previous outcomes, we observe more layered flow patterns for smaller $\lambda$, while energy is concentrated in the vortical structures associated with the forcing scale for larger $\lambda$.

\begin{figure*}[!t]
\centering
\mbox{
\subfigure[$k_f=5$, $\lambda=0.0$]{\includegraphics[width=0.5\textwidth]{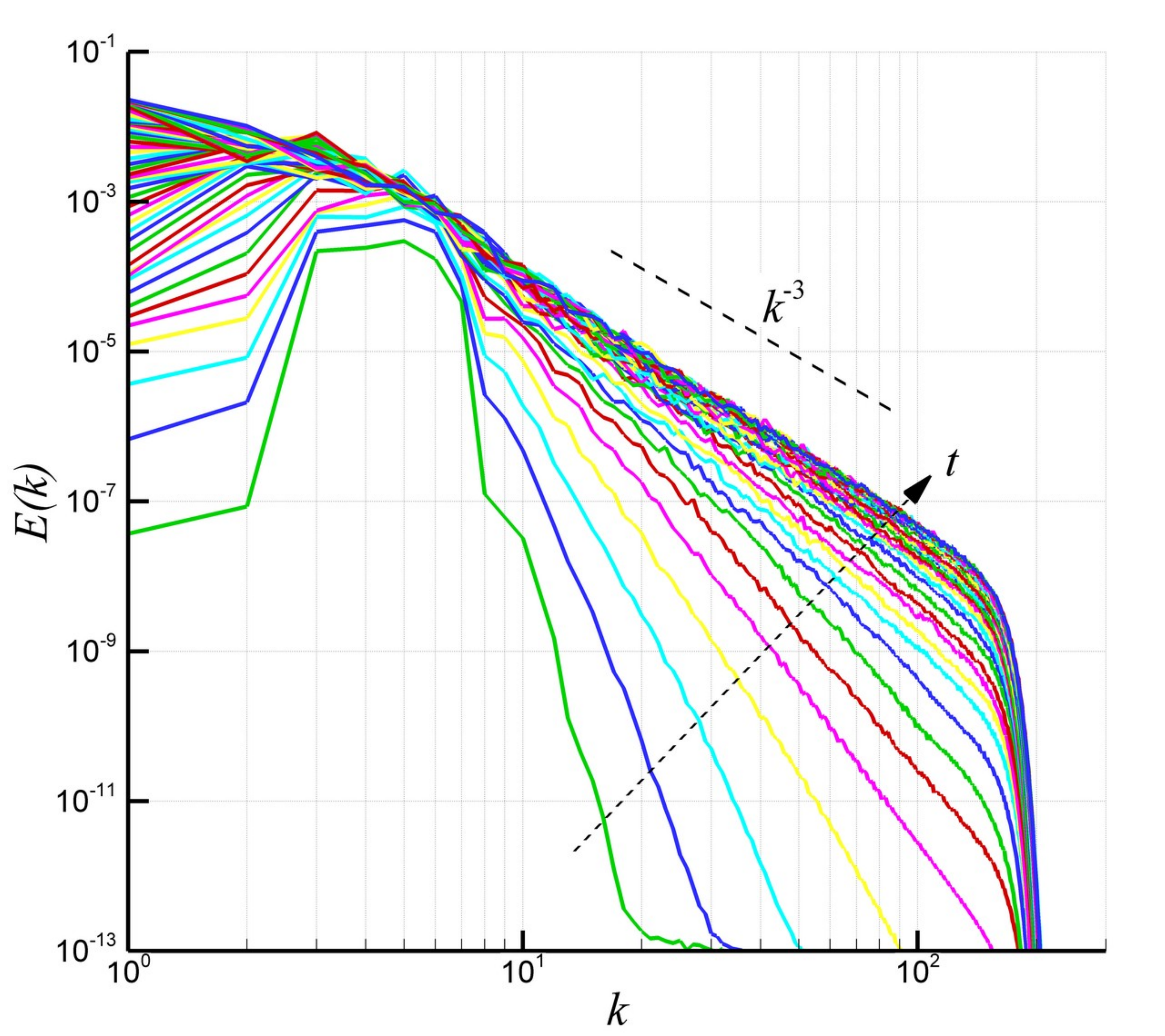}}
\subfigure[$k_f=5$, $\lambda=0.01$]{\includegraphics[width=0.5\textwidth]{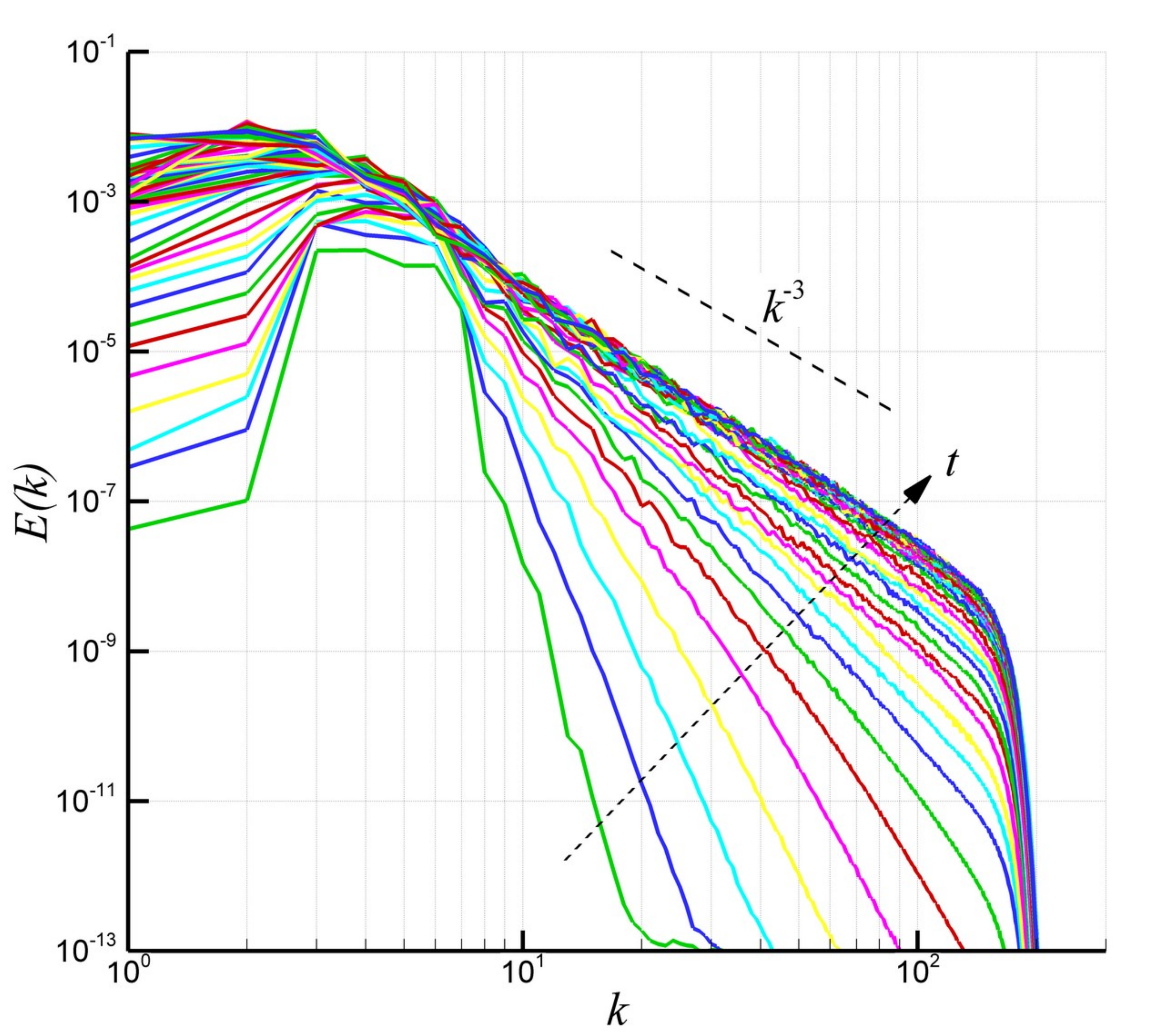}}
}\\
\mbox{
\subfigure[$k_f=15$, $\lambda=0.0$]{\includegraphics[width=0.5\textwidth]{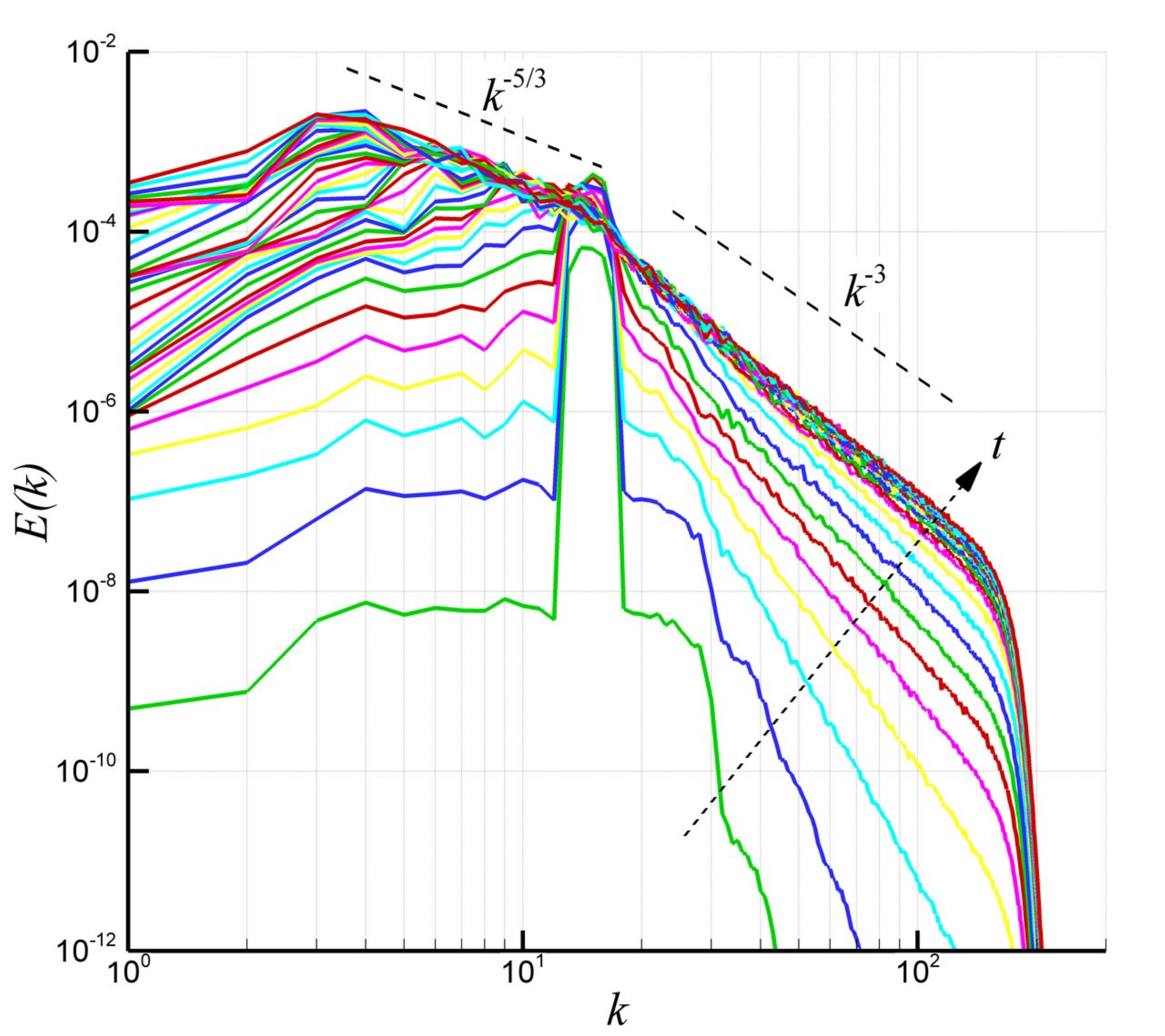}}
\subfigure[$k_f=15$, $\lambda=0.01$]{\includegraphics[width=0.5\textwidth]{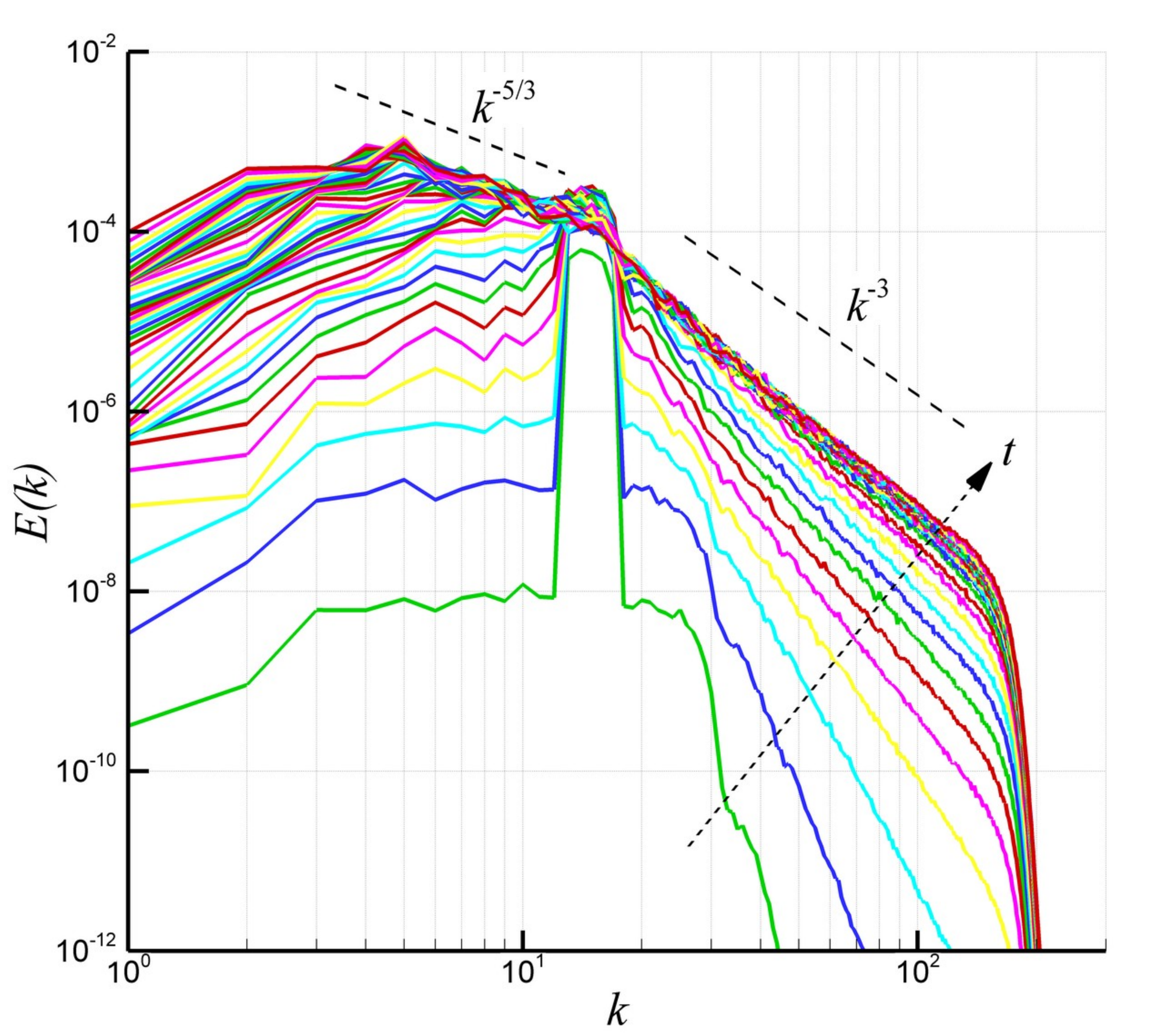}}
}
\caption{Evolution of the angle averaged energy spectrum ($f_0=0.1$, $\sigma=2$, $\rho=0.5$, $\nu=1000$ and $p=8$) for varying the large scale friction coefficient $\lambda$ and the effective forcing scale $k_f$.}
\label{fig:spectra-time}
\end{figure*}

Figure~\ref{fig:spectra-time} shows energy spectra at several times to illustrate the spectral energy density evolution for a series of cases with two values of $\lambda$ and $k_f$. The time interval between two adjacent lines is $\delta t=2$. The damping effect of the large scale friction mechanism in the development of the energy cascade is observable for the $k_f=5$ cases in the early time evolution. We also observe that a separation of time scales exists in the dynamics of the energy spectrum. This can be seen more clearly for the cases in which $k_f=15$. Once the forcing of the initially at-rest fluid begins, the entire spectrum immediately fills out with noise and a strong forcing peak appears centered around a wave number commensurate with the forcing length scale. Certain wave number ranges in a typical energy spectrum in a developing turbulent flow are populated immediately, while certain characteristics, the $k^{-5/3}$ scaling in the wave numbers smaller than the forcing scale, for example, evolve slowly over time. The spectrum fills out slowly until, after a long time, the final statistically steady state Kolmogorov spectrum appears if there exist even a small amount of large scale damping. This may be contrasted with the time evolution of a three-dimensional homogeneous isotropic turbulent flow, which develops its structure immediately, and with time transfers this structure to progressively smaller length scales.

\subsection{Effects of forcing mechanism}
\label{sec:forcing}

\begin{figure*}[!t]
\centering
\mbox{
\subfigure[]{\includegraphics[width=0.33\textwidth]{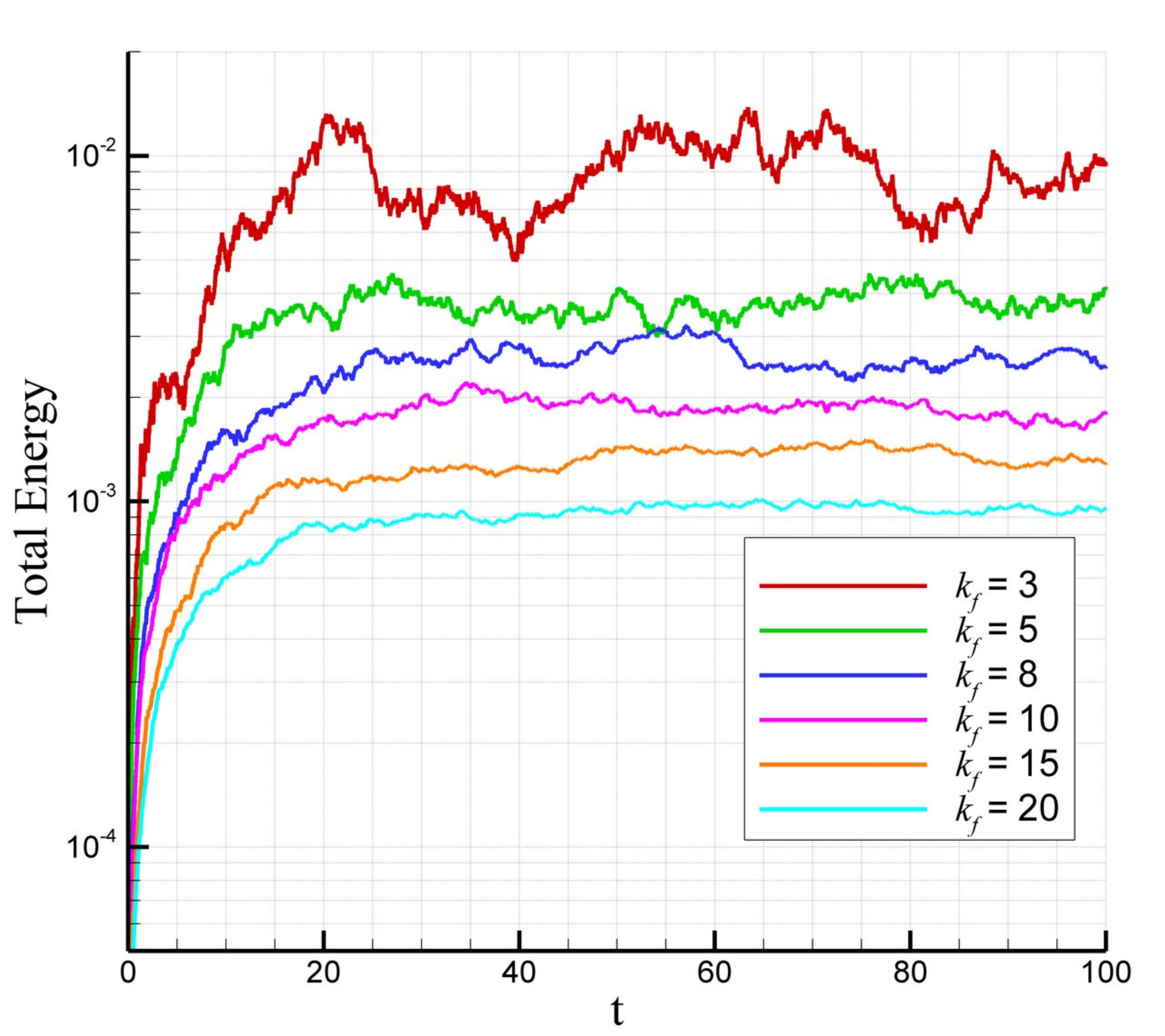}}
\subfigure[]{\includegraphics[width=0.33\textwidth]{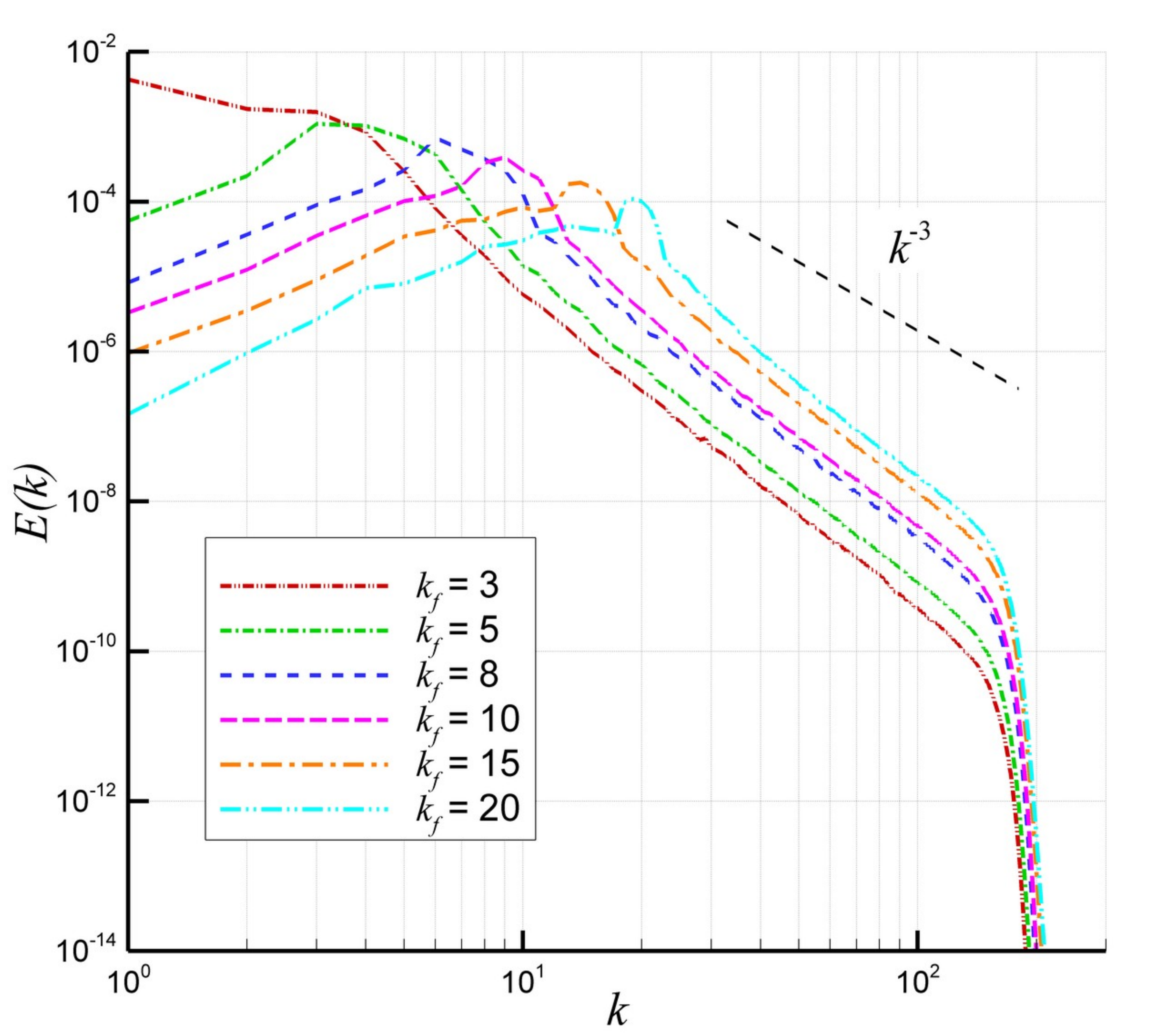}}
\subfigure[]{\includegraphics[width=0.33\textwidth]{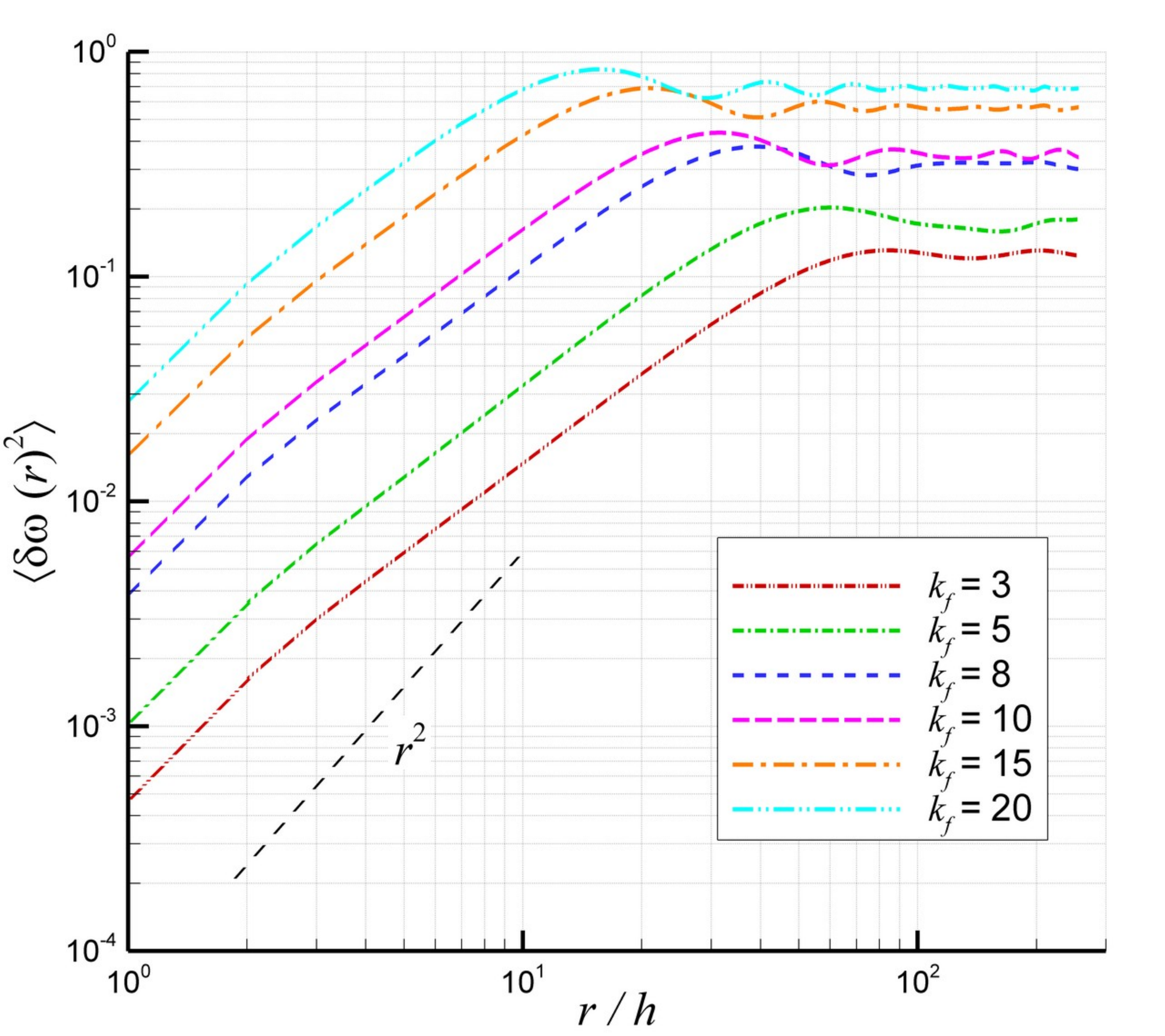}}
}
\caption{The effects of the forcing scale on the statistics ($\lambda=0.05$, $\sigma=2$, $\rho=0.5$, and $f_0=0.1$); (a) time series of total energy, (b) angle averaged energy spectra, and (c) second-order vorticity structure functions.}
\label{fig:stat_forcing_lamda05}
\end{figure*}

\begin{figure*}[!t]
\centering
\mbox{
\subfigure[$k_f=3$]{\includegraphics[width=0.245\textwidth]{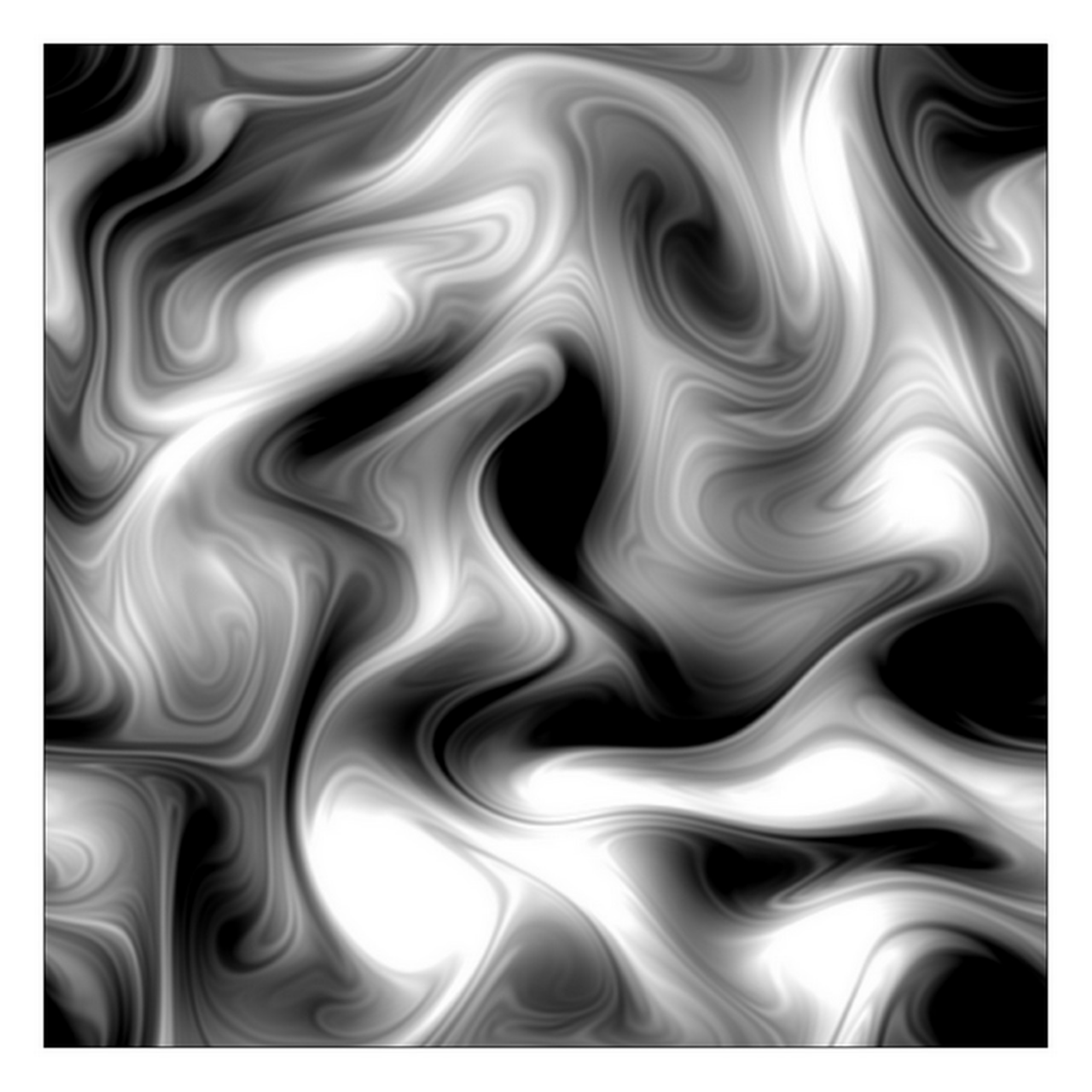}}
\subfigure[$k_f=5$]{\includegraphics[width=0.245\textwidth]{64.pdf}}
\subfigure[$k_f=10$]{\includegraphics[width=0.245\textwidth]{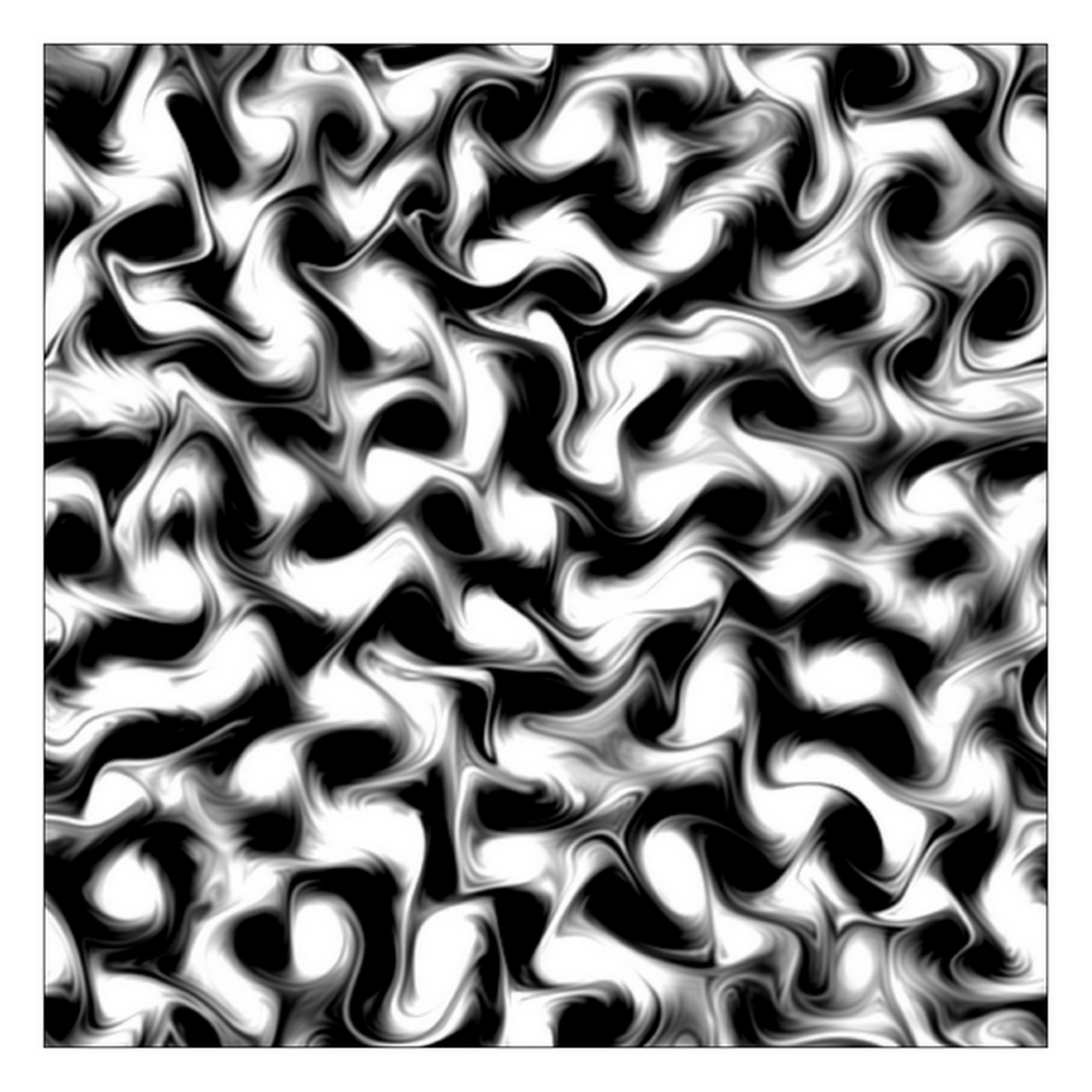}}
\subfigure[$k_f=15$]{\includegraphics[width=0.245\textwidth]{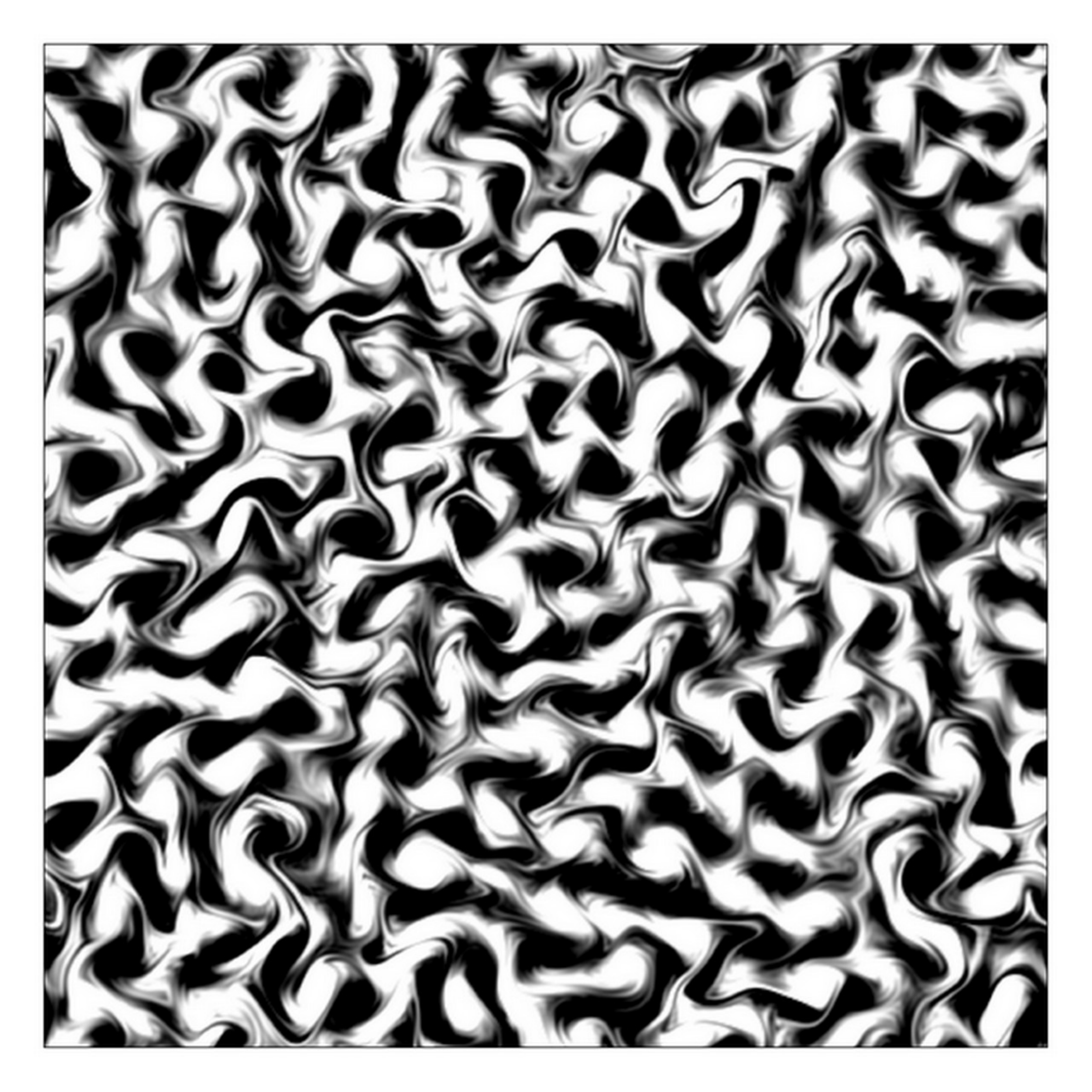}}
}
\caption{Instantaneous vorticity fields at time $t=100$ for varying the forcing scale $k_f$ using the large scale dissipation coefficient of $\lambda=0.05$.}
\label{fig:field_forcing_lamda05}
\end{figure*}

Here, we systematically analyze the effects of the parameters associated with Markovian forcing scheme. The underlying settings for the small scale dissipation mechanism are same for all the cases (i.e., $\nu=1000$ and $p=8$). The use of hyperviscosity maintains a constant flux of enstrophy in a wider interval $k_f > k > k_d$ by eliminating as much as possible the effects of viscosity at intermediate scales, thus extending the inertial ranges. First, we compute the statistics by varying the corresponding scale for Markovian forcing using a large scale friction coefficient of $\lambda=0.05$. Figure~\ref{fig:stat_forcing_lamda05} shows the statistics in terms of evolution of the total energy, mean energy spectrum, and mean second-order vorticity structure function for different forcing scales with the same memory correlation coefficient, forcing amplitude, and forcing bandwidth. The comparison of time series clearly demonstrates that all computations reach quasistationary regime at the same time but having a different level of energy. Increasing the forcing scale result in a decrease in quasistationary energy level. It is also shown that the variability increases by decreasing forcing scale. On the other hand, the scaling exponents in energy spectra and structure functions are invariant of the forcing scale. Instantaneous vorticity fields at time $t=100$ for varying the forcing scale $k_f$ are also illustrated in Figure~\ref{fig:field_forcing_lamda05} showing the similar pattern of mixing and filamentation.

\begin{figure*}[!t]
\centering
\mbox{
\subfigure[]{\includegraphics[width=0.33\textwidth]{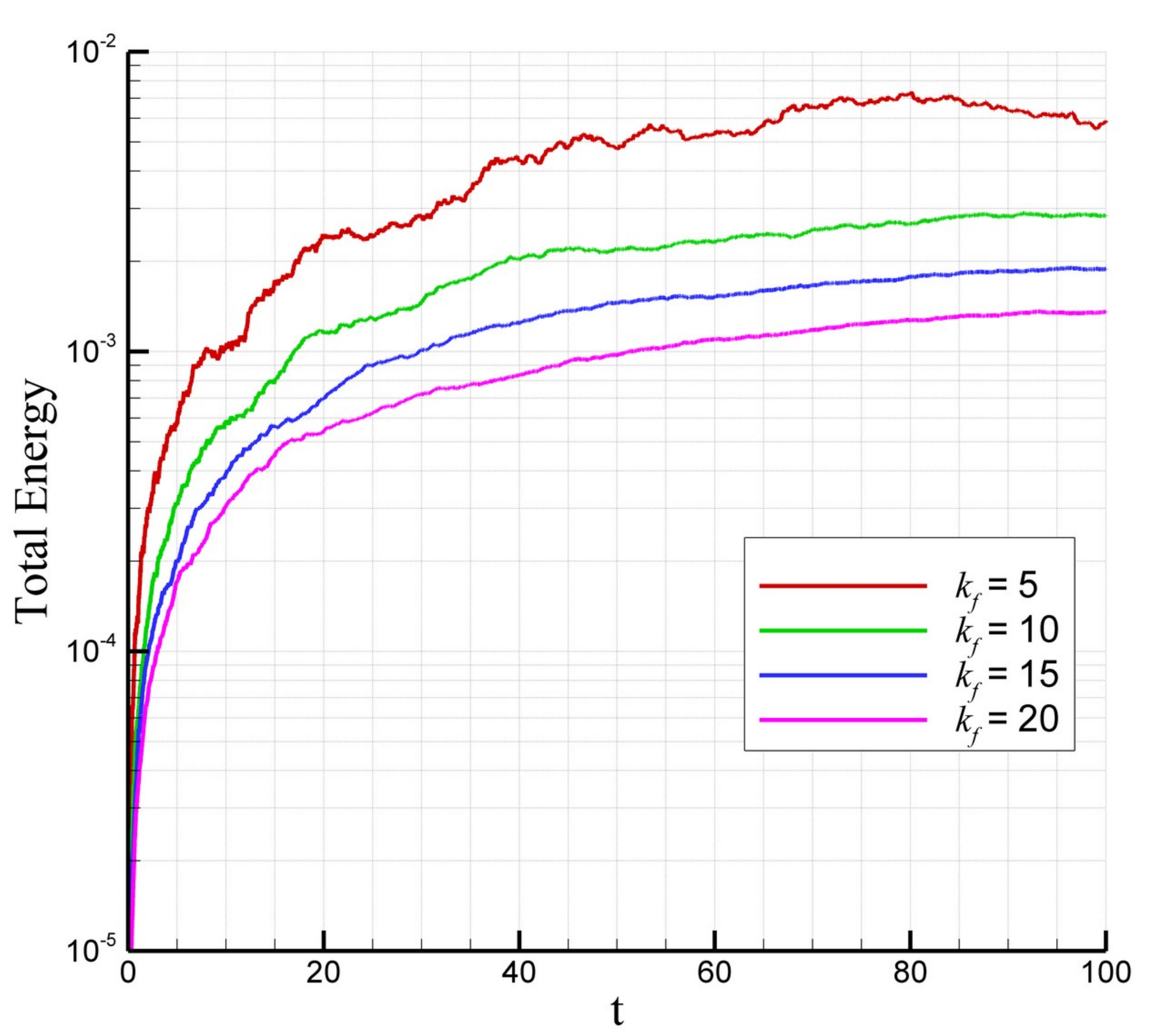}}
\subfigure[]{\includegraphics[width=0.33\textwidth]{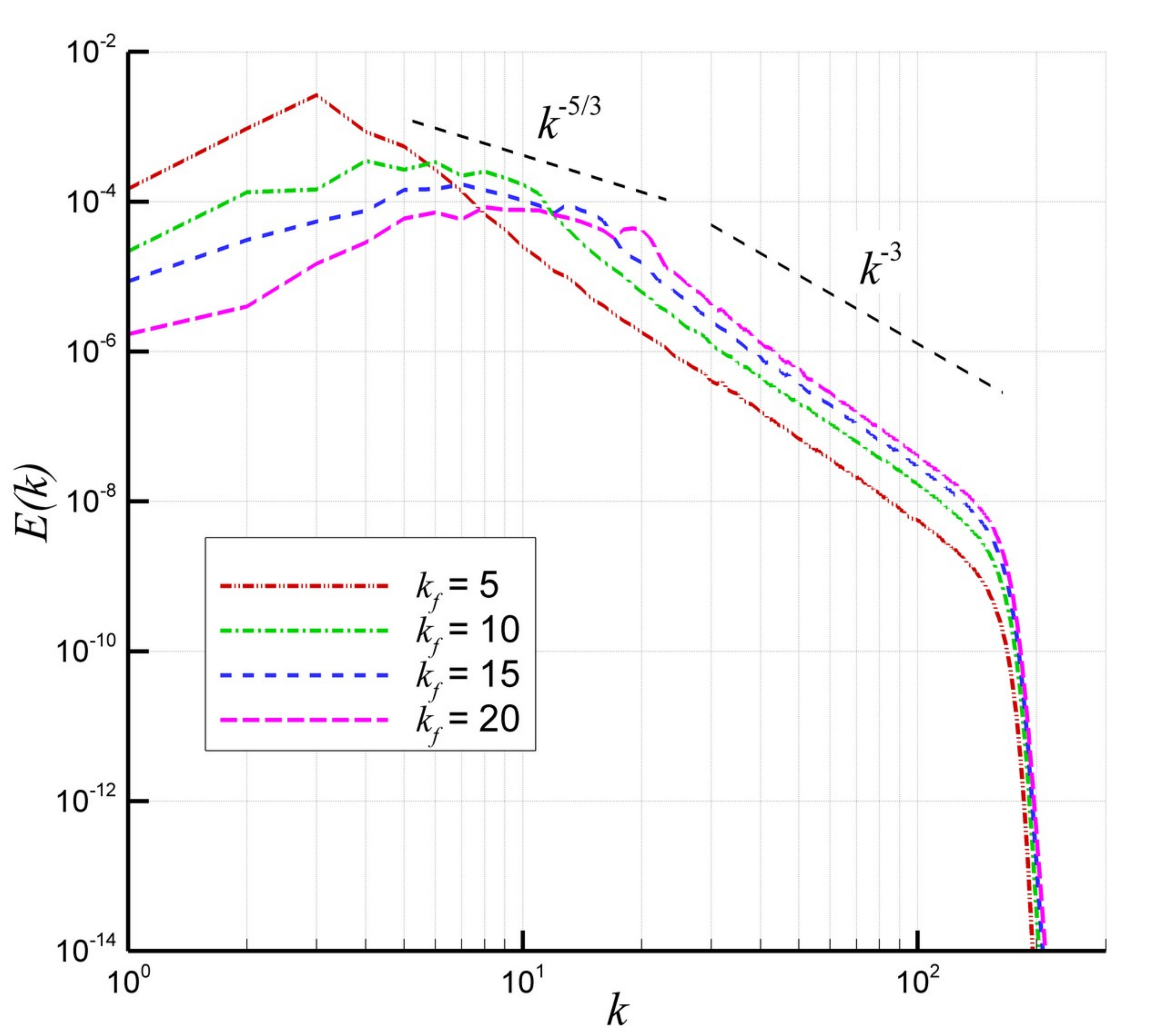}}
\subfigure[]{\includegraphics[width=0.33\textwidth]{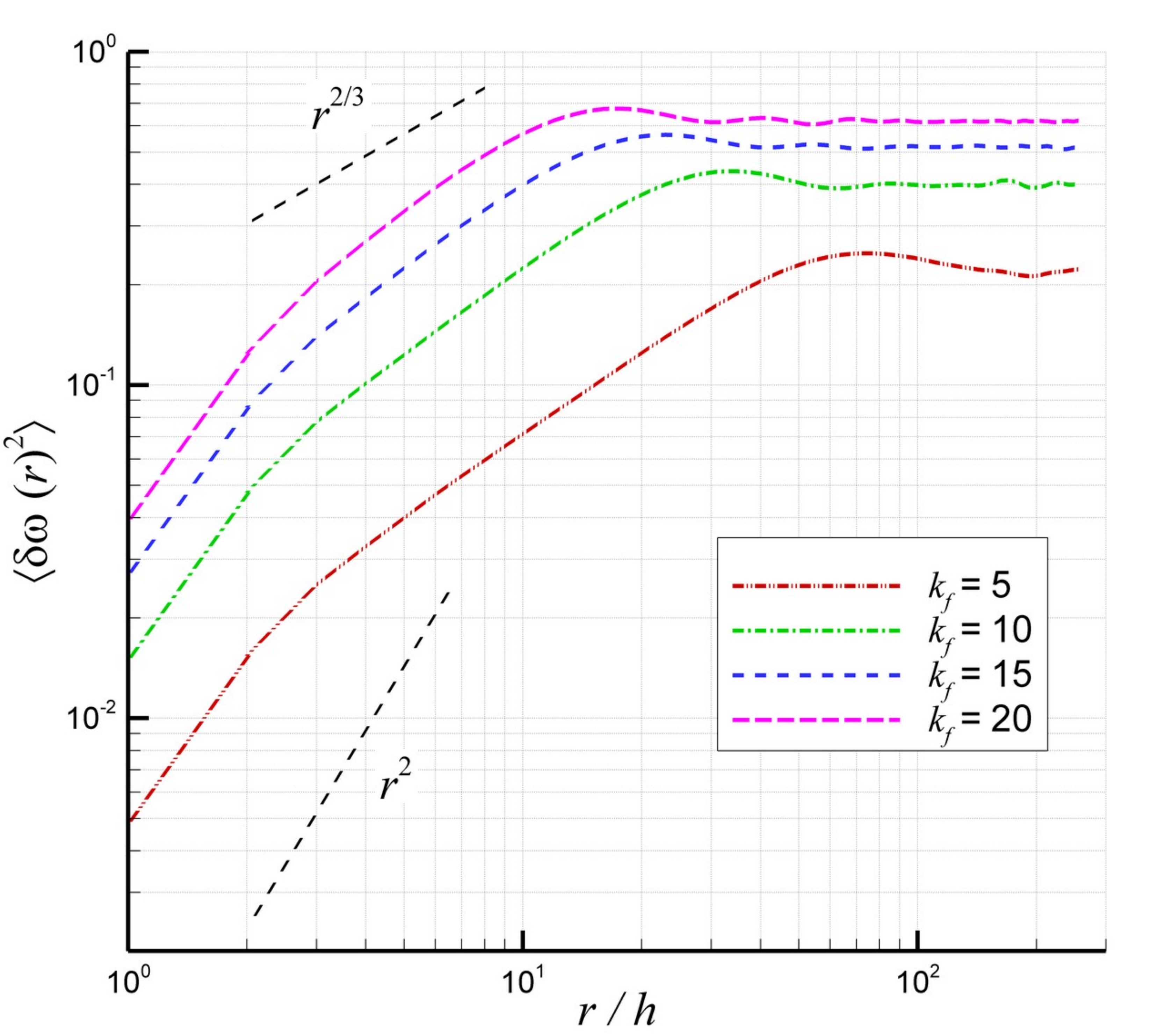}}
}
\caption{The effects of the forcing scale on the statistics ($\lambda=0.01$, $\sigma=2$, $\rho=0.0$, and $f_0=0.1$); (a) time series of total energy, (b) angle averaged energy spectra, and (c) second-order vorticity structure functions.}
\label{fig:stat_forcing_lamda01}
\end{figure*}

\begin{figure*}[!t]
\centering
\mbox{
\subfigure[$k_f=5$]{\includegraphics[width=0.245\textwidth]{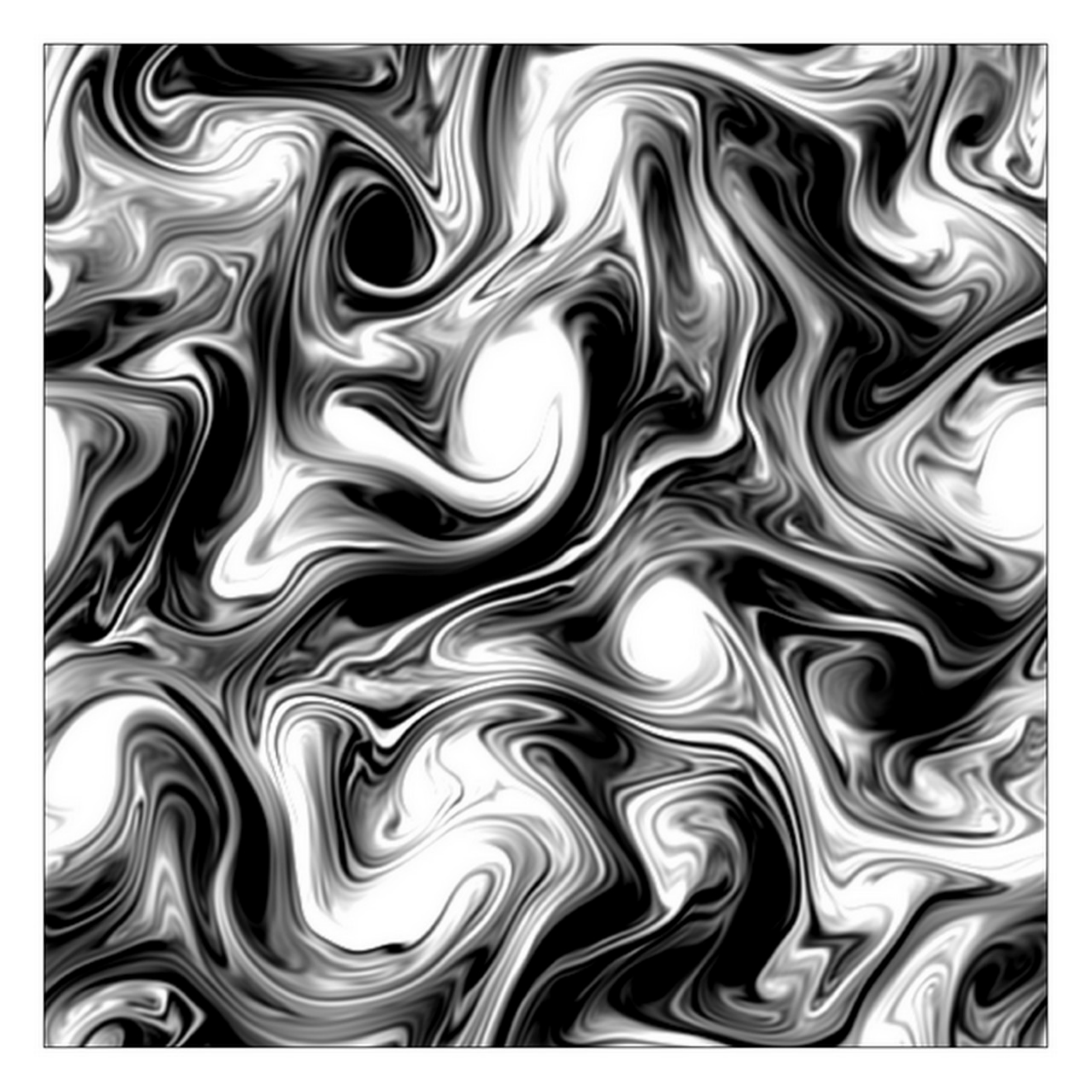}}
\subfigure[$k_f=10$]{\includegraphics[width=0.245\textwidth]{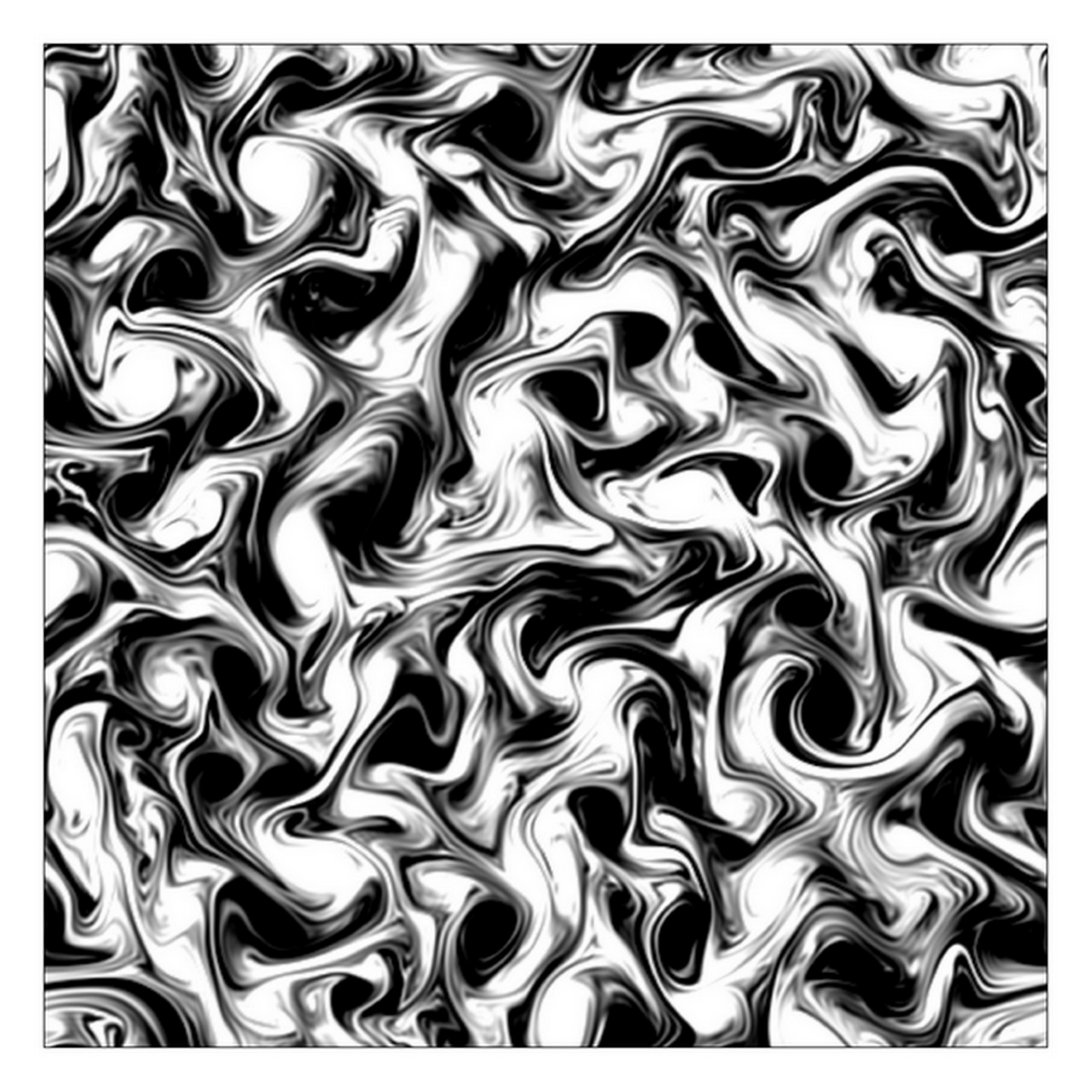}}
\subfigure[$k_f=15$]{\includegraphics[width=0.245\textwidth]{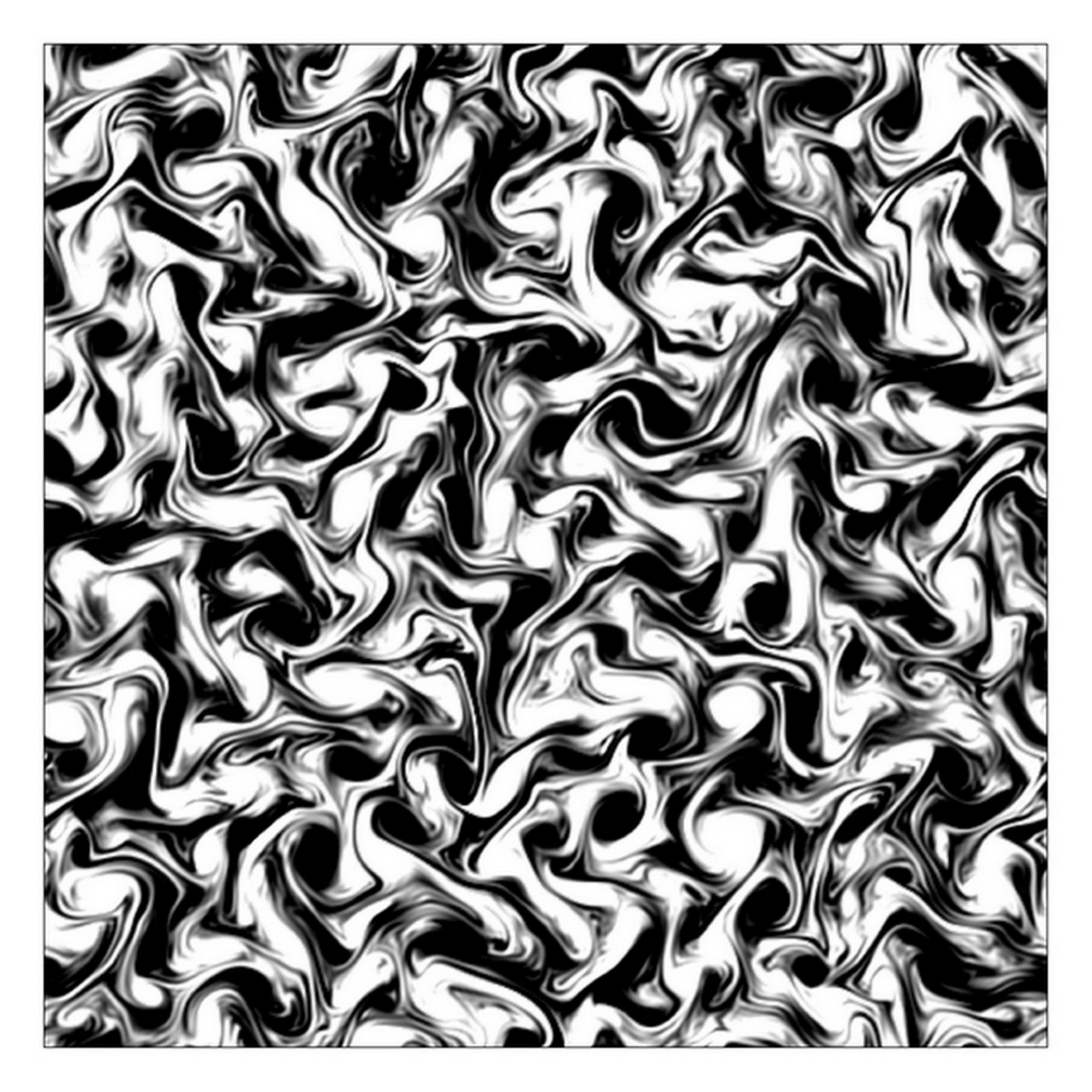}}
\subfigure[$k_f=20$]{\includegraphics[width=0.245\textwidth]{159.pdf}}
}
\caption{Instantaneous vorticity fields at time $t=100$ for varying the forcing scale $k_f$ using the large scale dissipation coefficient of $\lambda=0.01$.}
\label{fig:field_forcing_lamda01}
\end{figure*}

Similarly, we compute the statistics by varying the effective scale for Markovian forcing using a smaller large scale friction coefficient of $\lambda=0.01$. The statistics and flow field patterns are shown in Figure~\ref{fig:stat_forcing_lamda01} and Figure~\ref{fig:field_forcing_lamda01}, respectively. Due to the reduction in damping coefficient we observe more interaction between vortical structures for all the cases with varying $k_f$. In the inertial range, especially for $k_f=5$,  it is also interesting to see that energy spectrum scales as $k^{-3}$, while structure functions scales as $r^{2/3}$. The main reason for energy spectrum appearing steeper than $k^{-3}$ for increasing $k_f$ is that the relative importance of the large scale damping coefficient in larger $k_f$ cases is greater. This can also be seen from the energy levels shown in Figure~\ref{fig:stat_forcing_lamda01}.

\begin{figure*}[!t]
\centering
\mbox{
\subfigure[]{\includegraphics[width=0.33\textwidth]{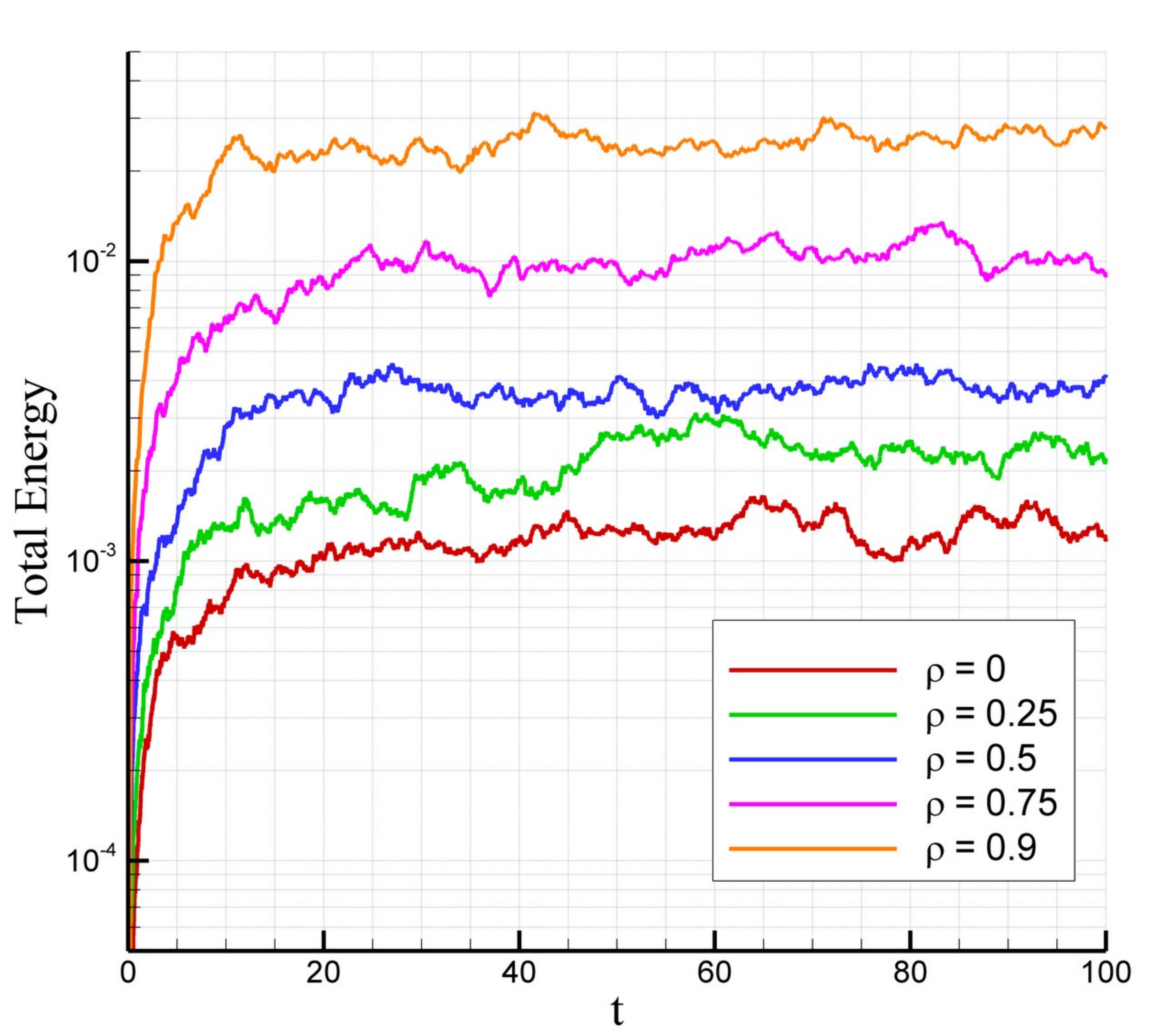}}
\subfigure[]{\includegraphics[width=0.33\textwidth]{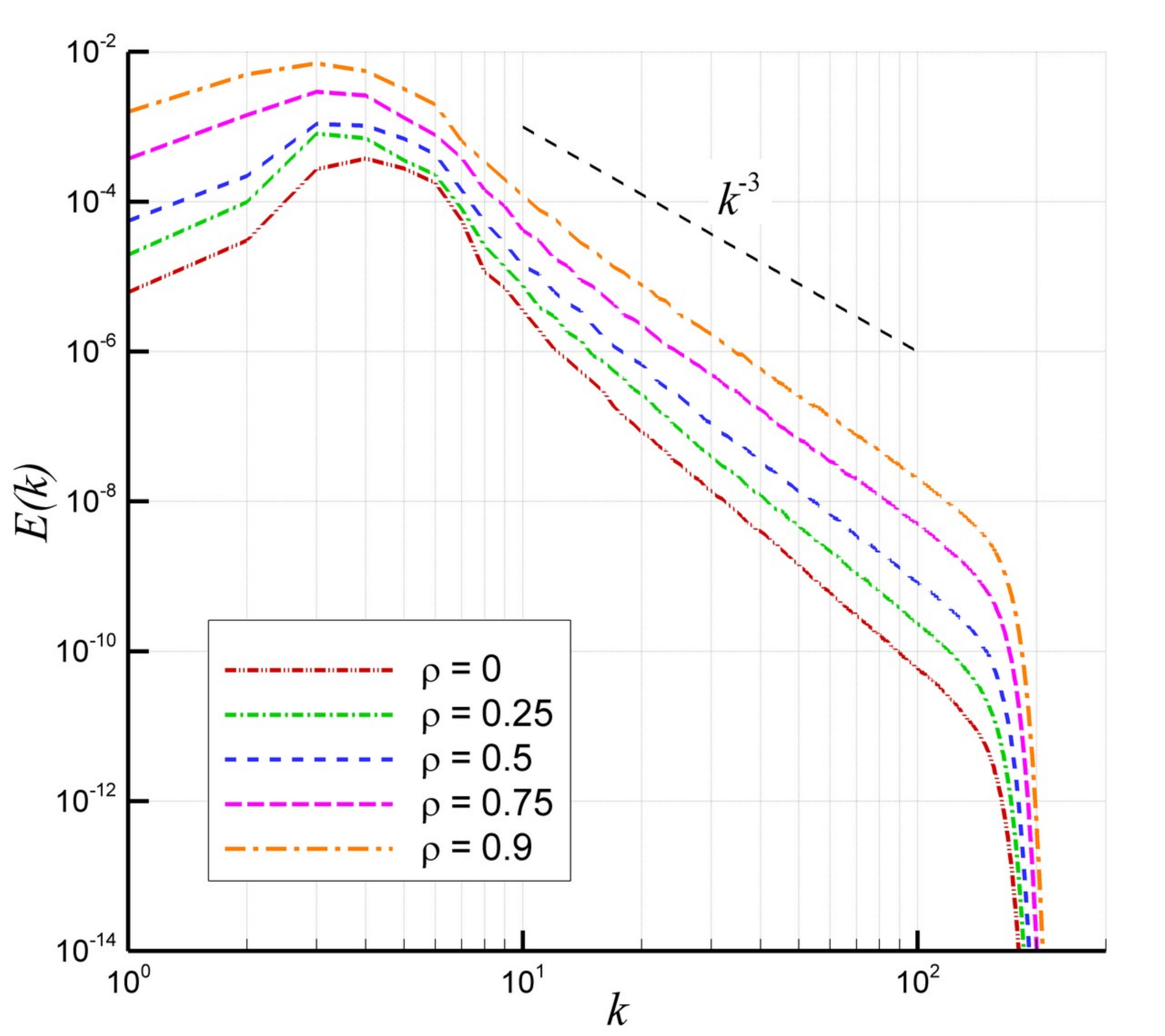}}
\subfigure[]{\includegraphics[width=0.33\textwidth]{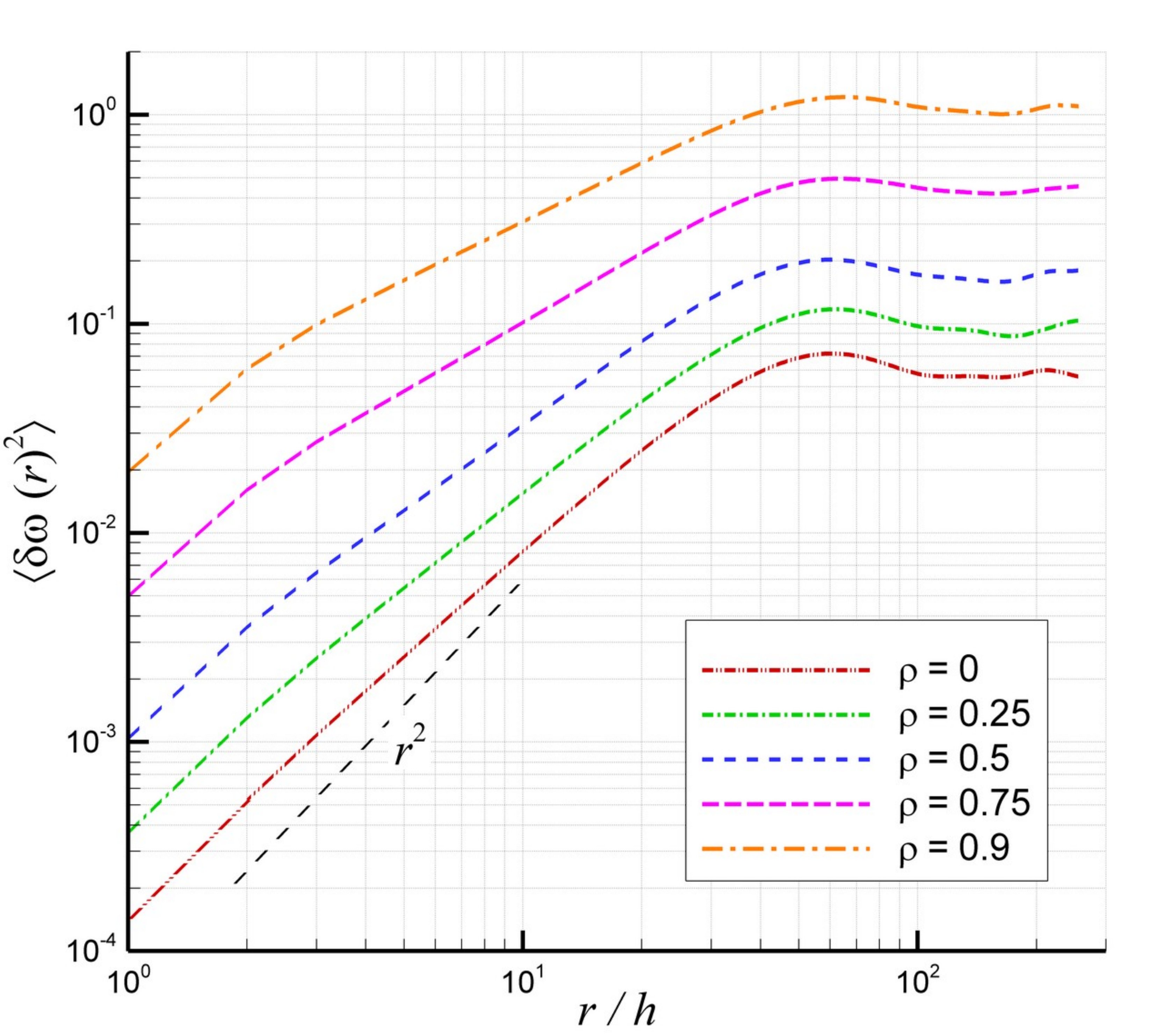}}
}
\caption{The effects of the memory coefficient on the statistics ($\lambda=0.05$, $k_f=5$, $\sigma=2$, $\nu=1000$ and $p=8$); (a) time series of total energy, (b) angle averaged energy spectra, and (c) second-order vorticity structure functions.}
\label{fig:stat_r_k5}
\end{figure*}

Next, we analyze the effects of the memory coefficient $\rho$ in the Markovian forcing scheme. As shown in Figure~\ref{fig:stat_r_k5}, its main effect on the statistics is the translation of the level of energy. There is no significant effect into the scaling exponents of the energy spectra and structure functions. The vorticity fields are also illustrated in Figure~\ref{fig:stat_r_k5} showing that the amplitude of vorticity increases with increasing memory coefficient $\rho$. In this study, we focus on the random forcing mechanism controlled by this coefficient that measures the stochastic process in the system for which the forcing is purely random (i.e., $\rho=0$). It will be interesting to further investigate the effects of deterministic forcing in the context of generation of coherent vortices. However, we particulary use a random forcing mechanism to concentrate on the statistical deviations from the theoretical scalings which are solely due to viscous effects and eliminate possible situations that coherent vortices might play a role.

\begin{figure*}[!t]
\centering
\mbox{
\subfigure[$\rho=0.0$]{\includegraphics[width=0.245\textwidth]{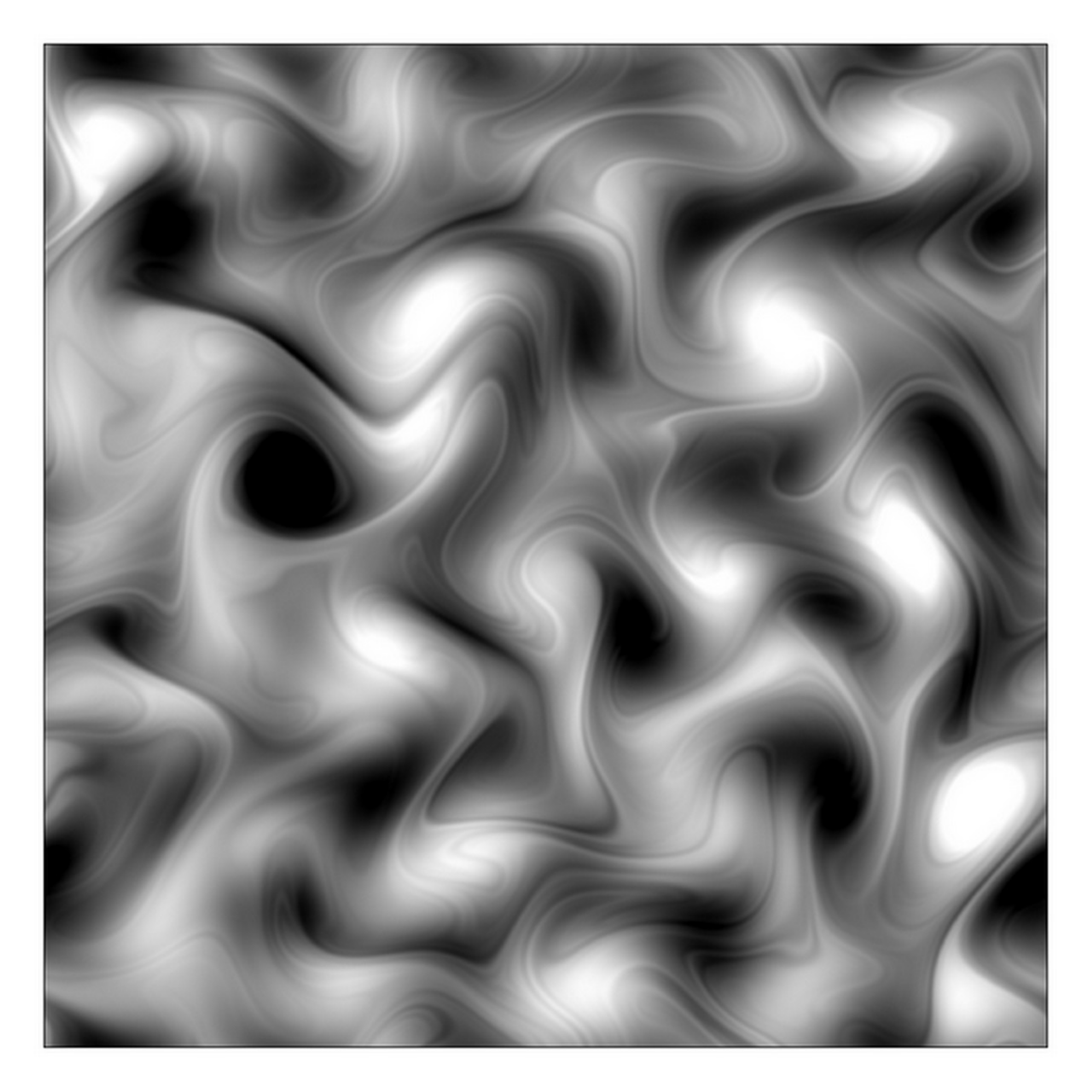}}
\subfigure[$\rho=0.25$]{\includegraphics[width=0.245\textwidth]{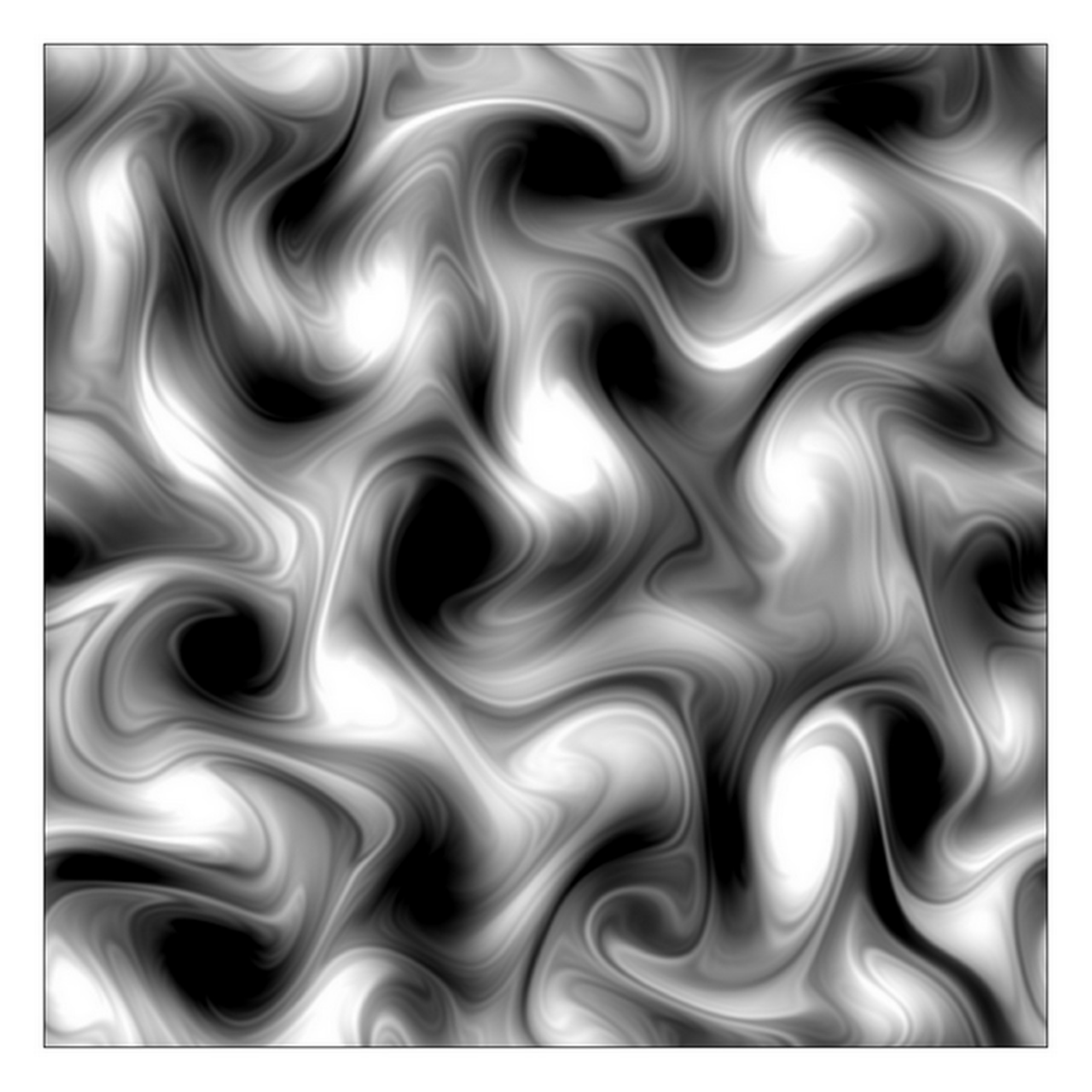}}
\subfigure[$\rho=0.5$]{\includegraphics[width=0.245\textwidth]{64.pdf}}
\subfigure[$\rho=0.9$]{\includegraphics[width=0.245\textwidth]{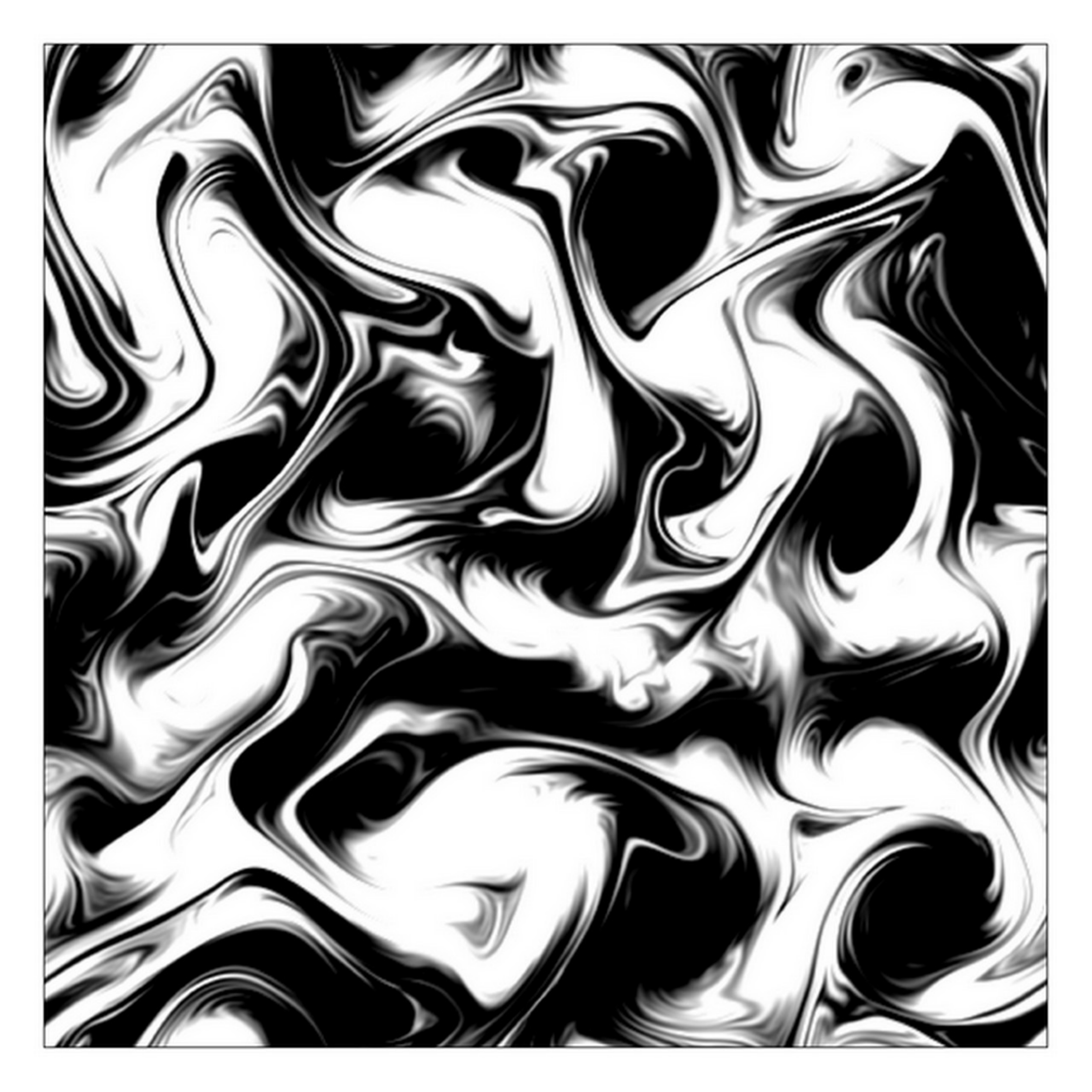}}
}
\caption{Instantaneous vorticity fields at time $t=100$ for varying the memory coefficient using the large scale dissipation coefficient of $\lambda=0.05$ and the small scale dissipation coefficients of $\nu=1000$ and $p=8$.}
\label{fig:field_r_k5}
\end{figure*}

\begin{figure*}[!t]
\centering
\mbox{
\subfigure[]{\includegraphics[width=0.33\textwidth]{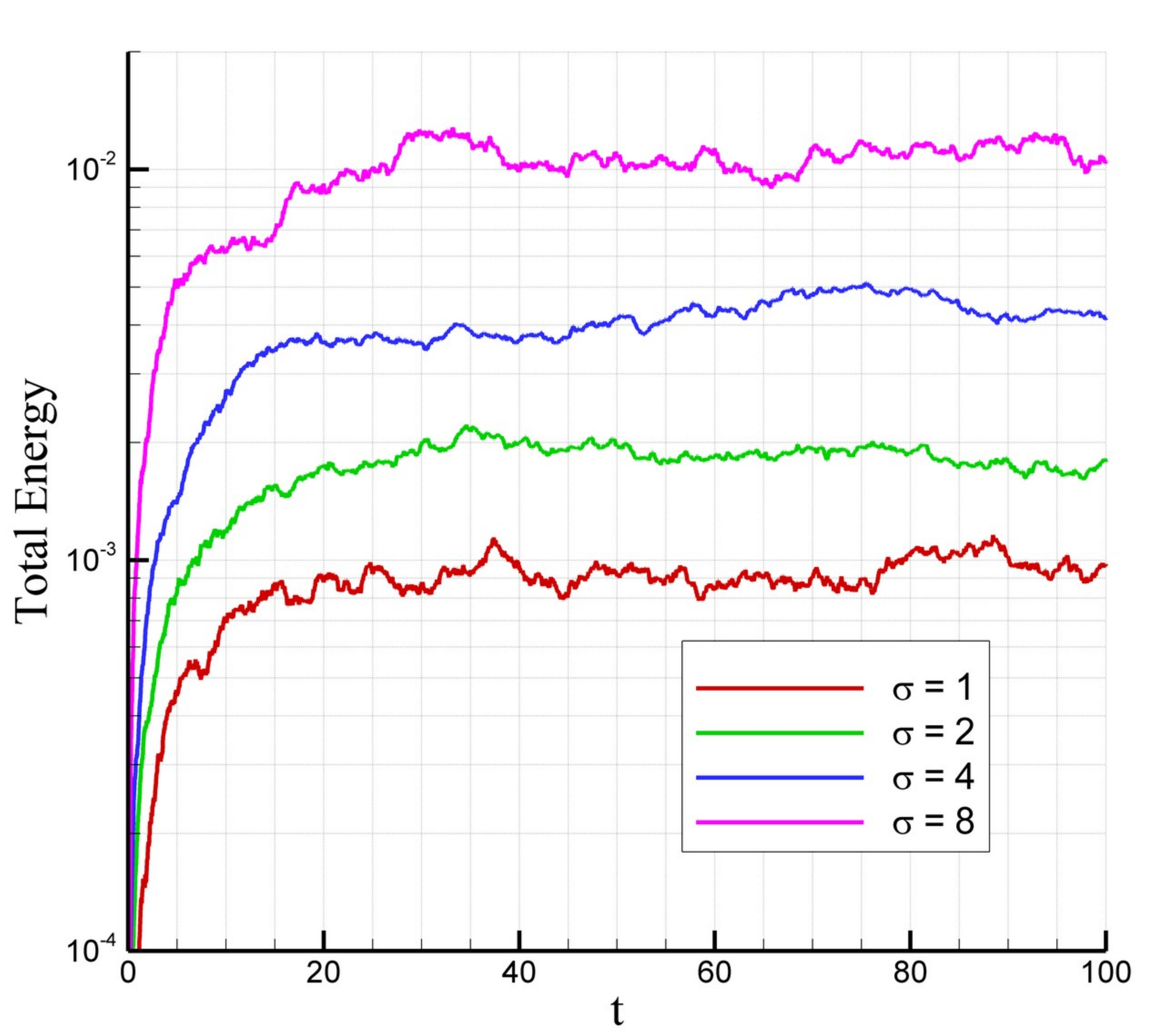}}
\subfigure[]{\includegraphics[width=0.33\textwidth]{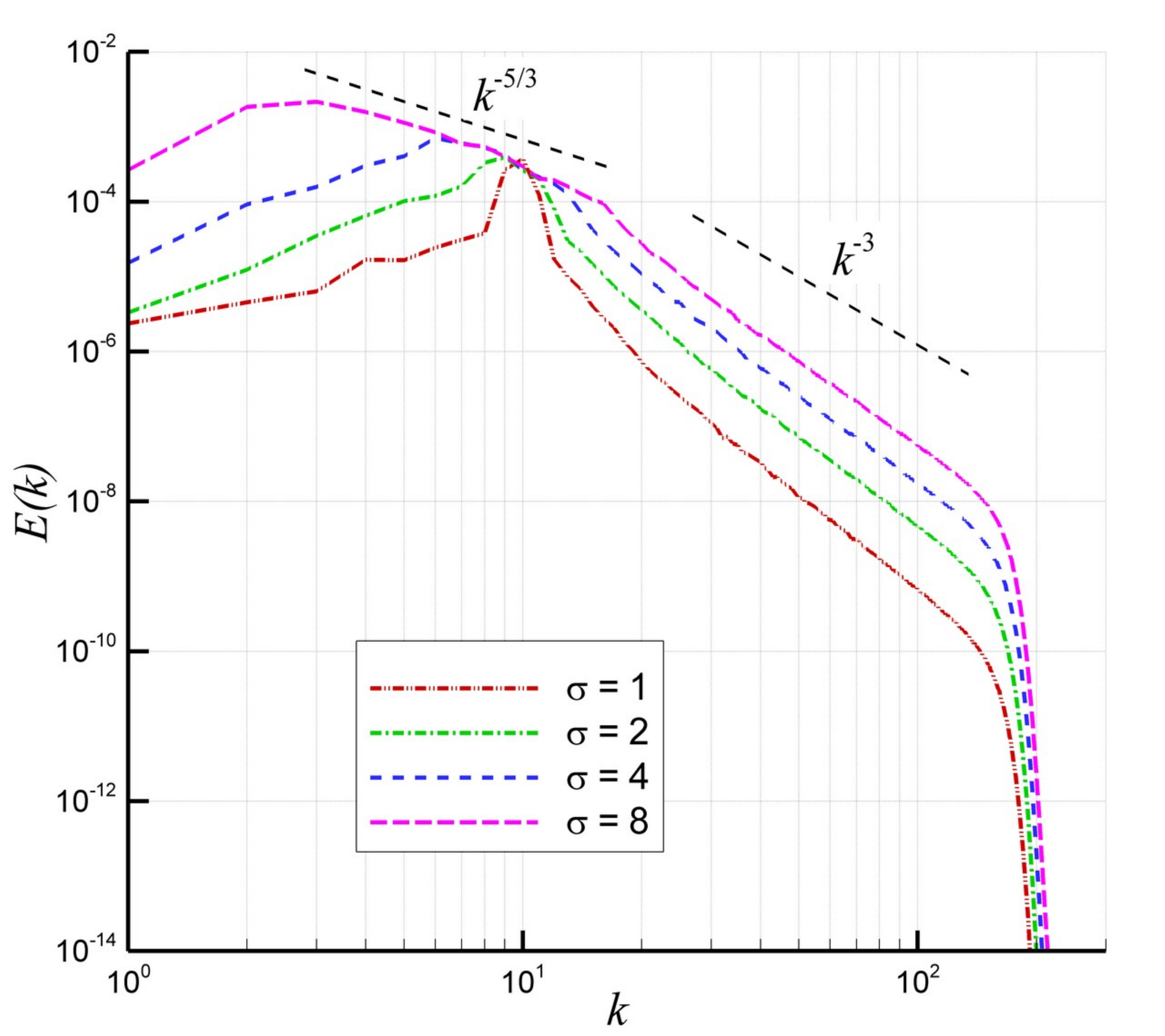}}
\subfigure[]{\includegraphics[width=0.33\textwidth]{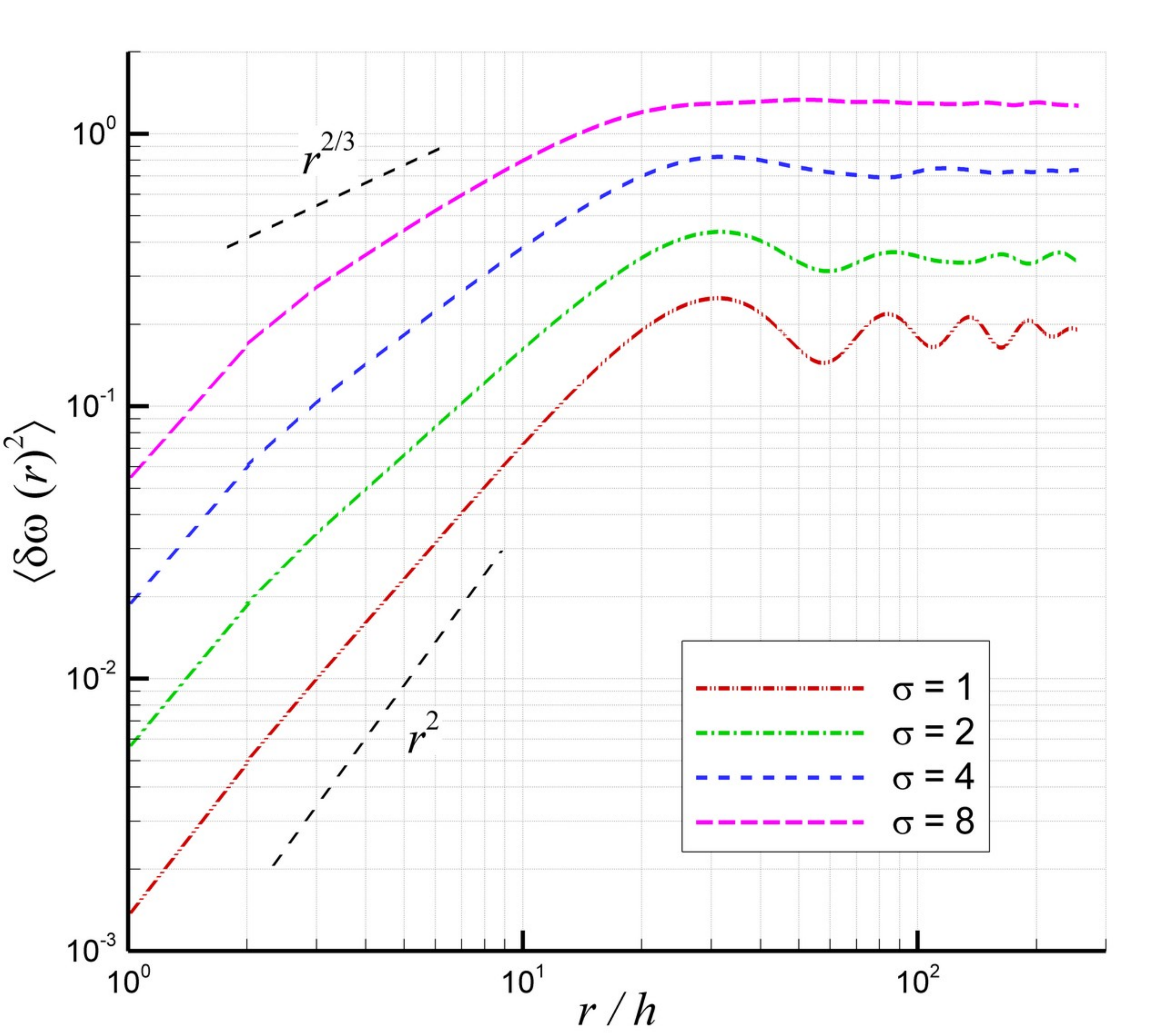}}
}
\caption{The effects of the forcing bandwidth on the statistics ($\lambda=0.05$, $k_f=10$, $\rho=0.5$, $\nu=1000$ and $p=8$); (a) time series of total energy, (b) angle averaged energy spectra, and (c) second-order vorticity structure functions.}
\label{fig:stat_sigma_k10}
\end{figure*}

\begin{figure*}[!t]
\centering
\mbox{
\subfigure[$\sigma=1$]{\includegraphics[width=0.245\textwidth]{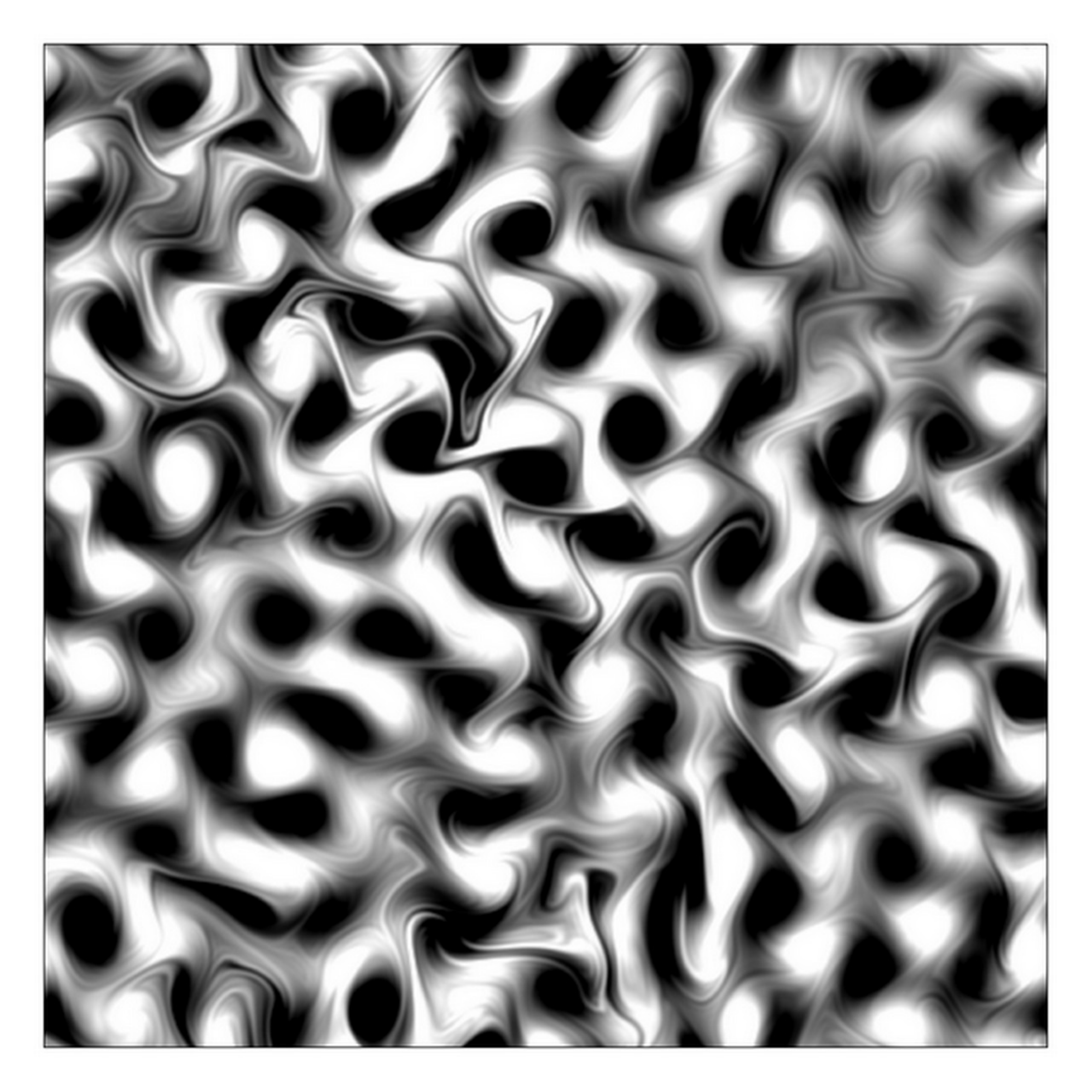}}
\subfigure[$\sigma=2$]{\includegraphics[width=0.245\textwidth]{78.pdf}}
\subfigure[$\sigma=4$]{\includegraphics[width=0.245\textwidth]{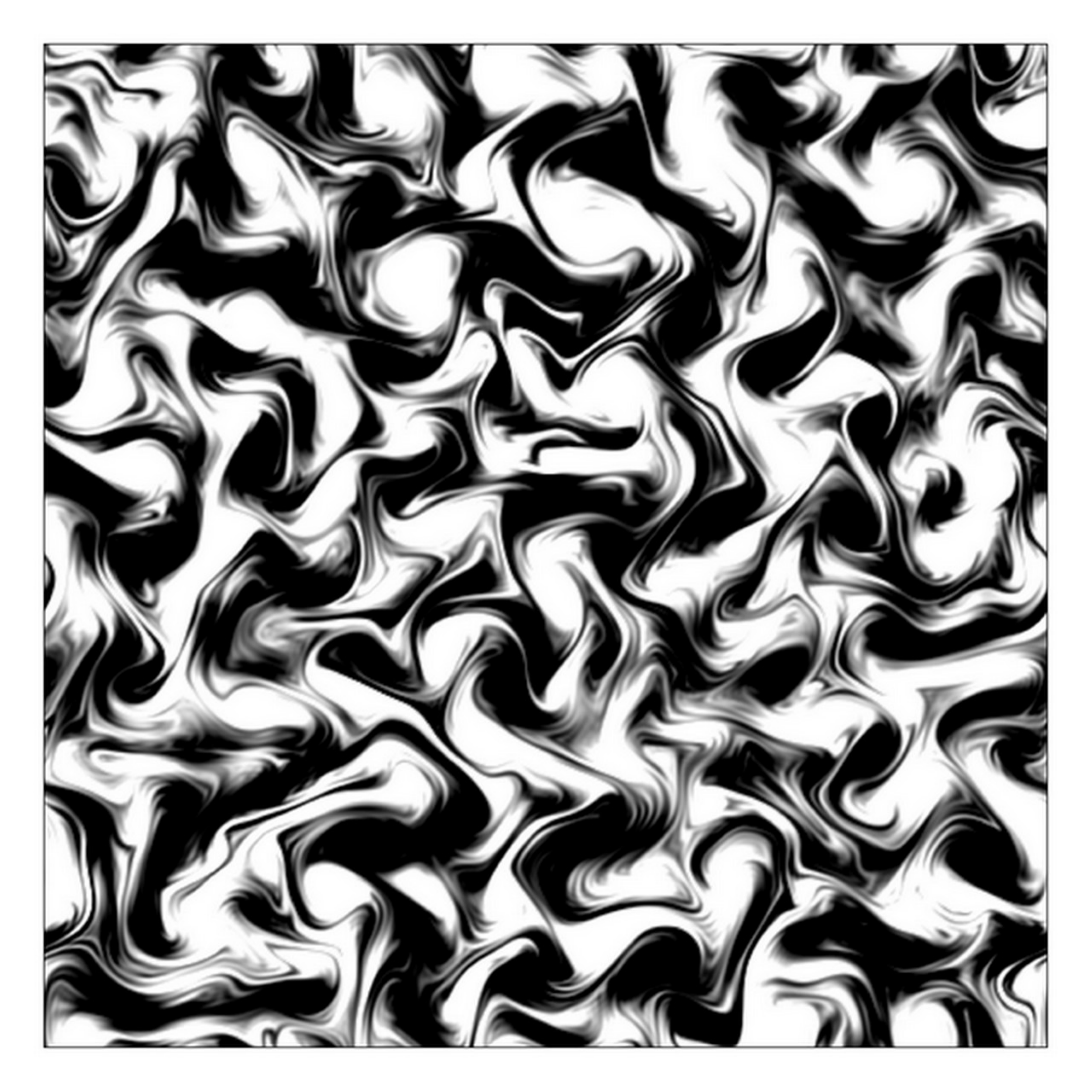}}
\subfigure[$\sigma=8$]{\includegraphics[width=0.245\textwidth]{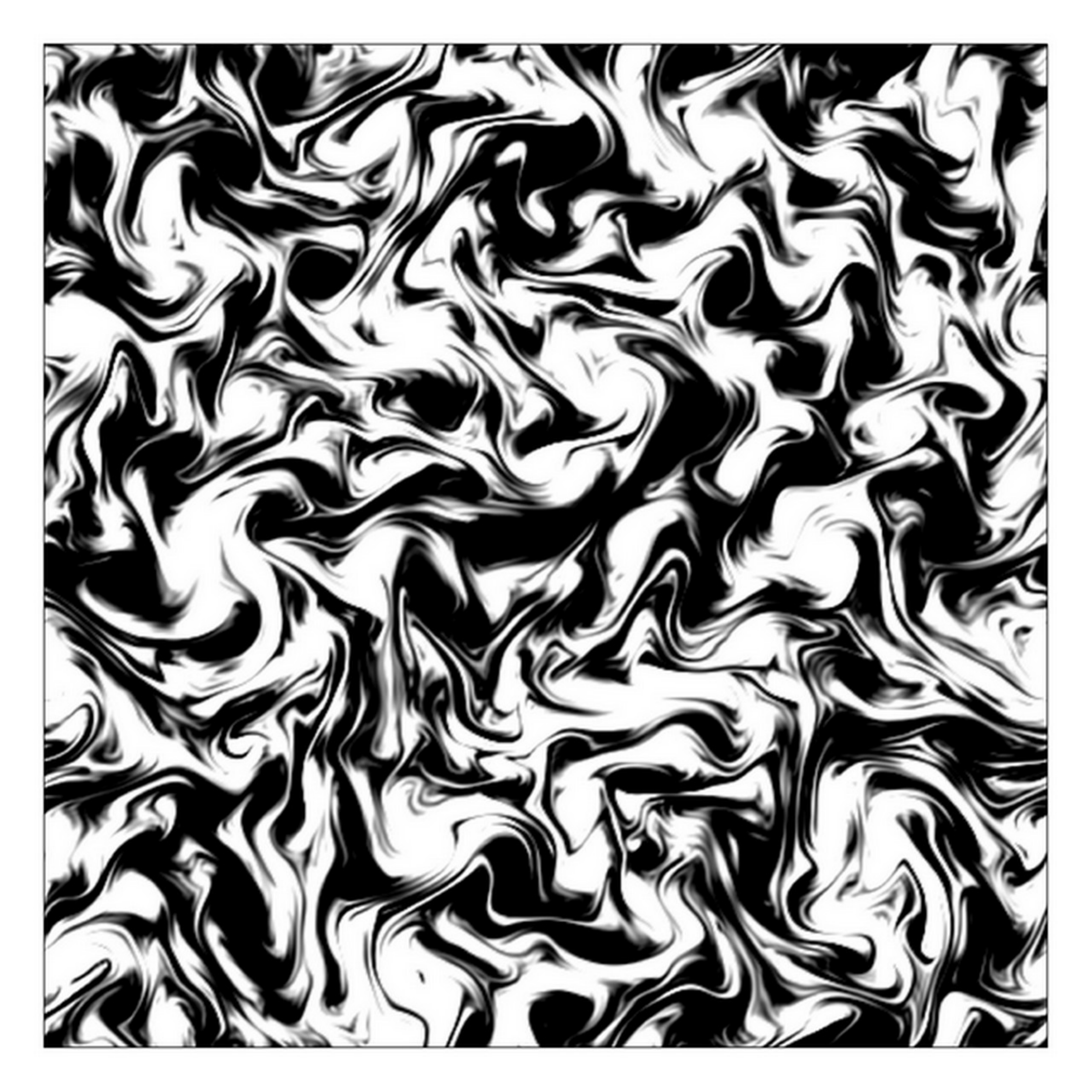}}
}
\caption{Instantaneous vorticity fields at time $t=100$ for varying the forcing bandwidth using the large scale dissipation coefficient of $\lambda=0.05$ and the small scale dissipation coefficients of $\nu=1000$ and $p=8$.}
\label{fig:field_sigma_k10}
\end{figure*}

Figures~\ref{fig:stat_sigma_k10} and~\ref{fig:field_sigma_k10} examine the effects of forcing bandwidth $\sigma$ for the forcing scale $k_f=10$. The effective energy injection occurs in the scales within the interval ($k_f - \sigma$, $k_f + \sigma$). As shown in Figure~\ref{fig:stat_sigma_k10}, the level of energy of the system in statistically steady state increases with increasing the $\sigma$ since we inject more energy due to enhanced forcing zone. In the forward energy range we observe the same scaling for energy spectrum. For $\sigma=8$, it is also shown that the energy spectrum within the forcing zone scales as $E(k)\sim k^{-5/3}$. Comparing the structure functions, we illustrate that $\langle \delta \omega (r)^{2}\rangle$ wiggles more rapidly for smaller values of $\sigma$. It is also interesting to see that the flow pattern shows more vortical filaments for increasing forcing bandwidth.

\begin{figure*}[!t]
\centering
\mbox{
\subfigure[]{\includegraphics[width=0.33\textwidth]{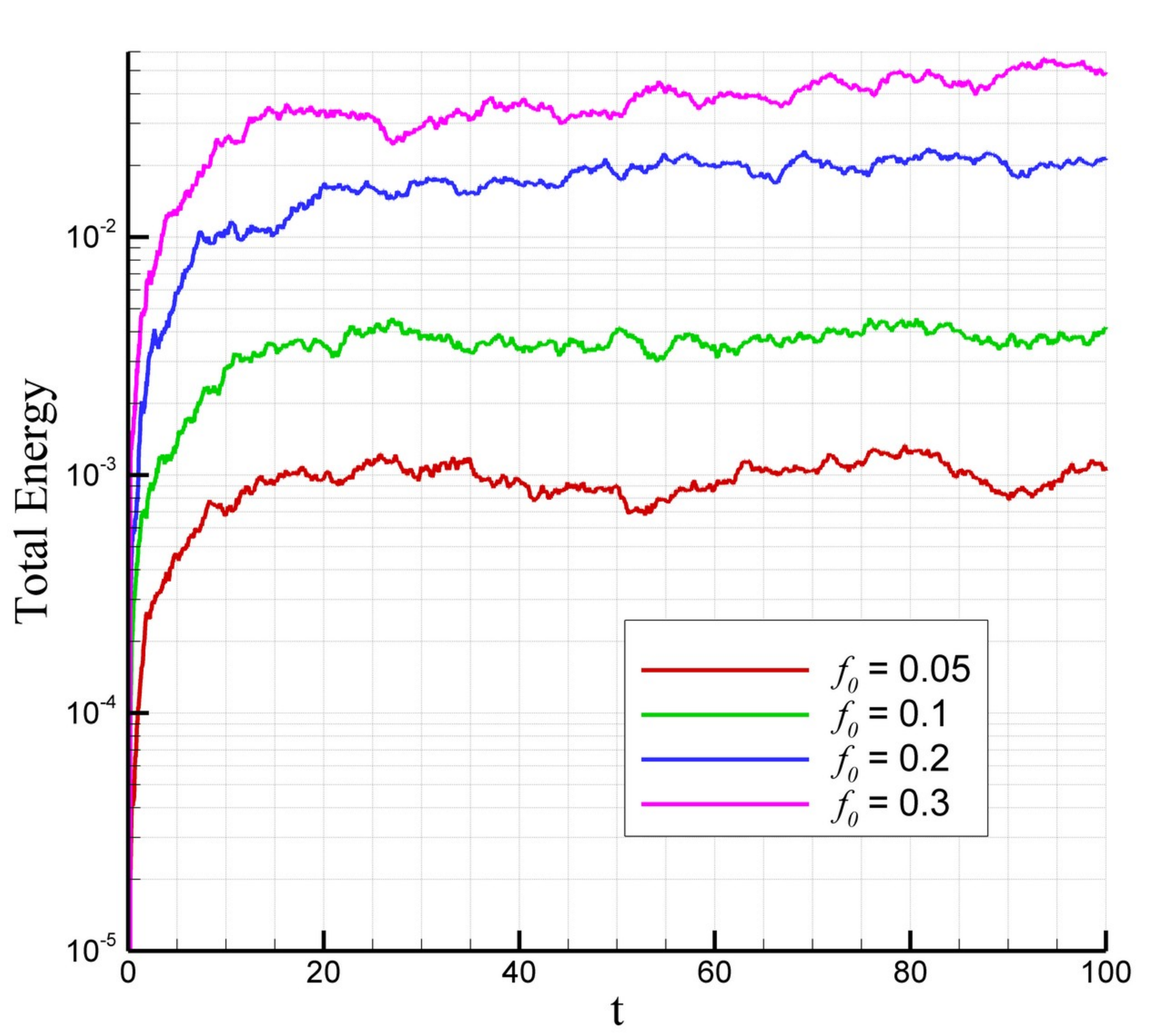}}
\subfigure[]{\includegraphics[width=0.33\textwidth]{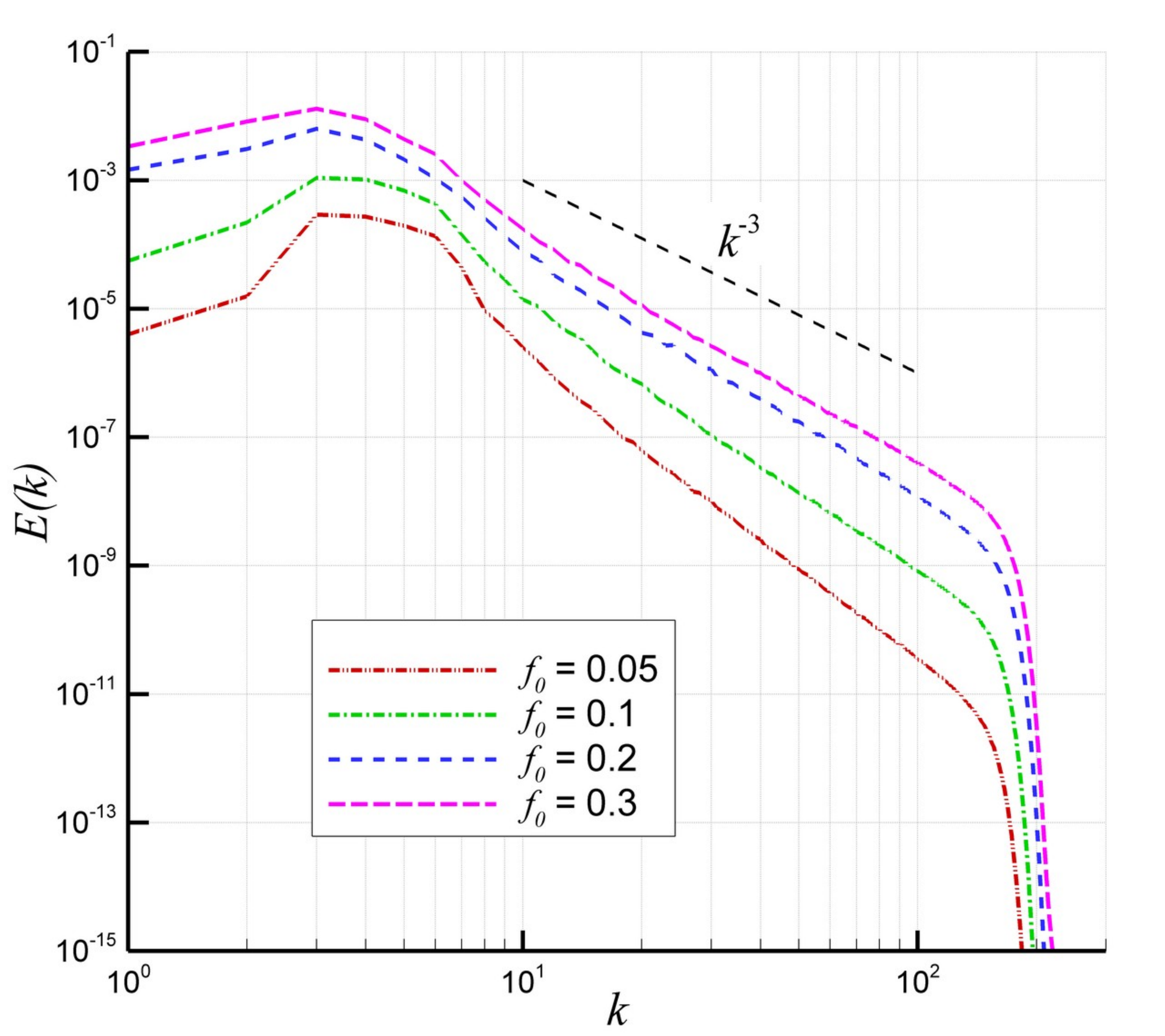}}
\subfigure[]{\includegraphics[width=0.33\textwidth]{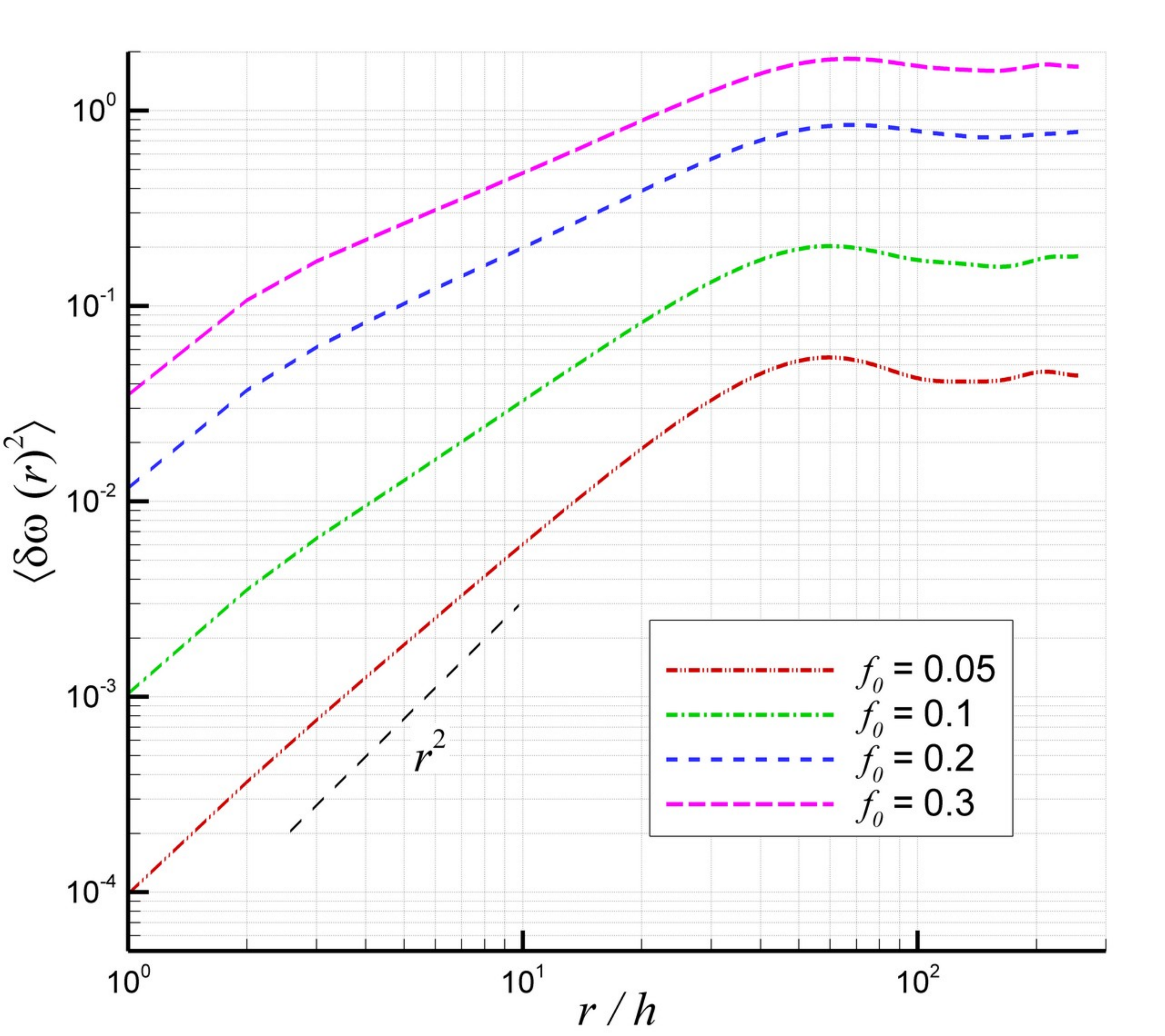}}
}
\caption{The effects of the forcing amplitude on the statistics ($\lambda=0.05$, $k_f=5$, $\rho=0.5$, $\nu=1000$ and $p=8$); (a) time series of total energy, (b) angle averaged energy spectra, and (c) second-order vorticity structure functions.}
\label{fig:stat_f0_k5}
\end{figure*}

\begin{figure*}[!t]
\centering
\mbox{
\subfigure[$f_0=0.05$]{\includegraphics[width=0.245\textwidth]{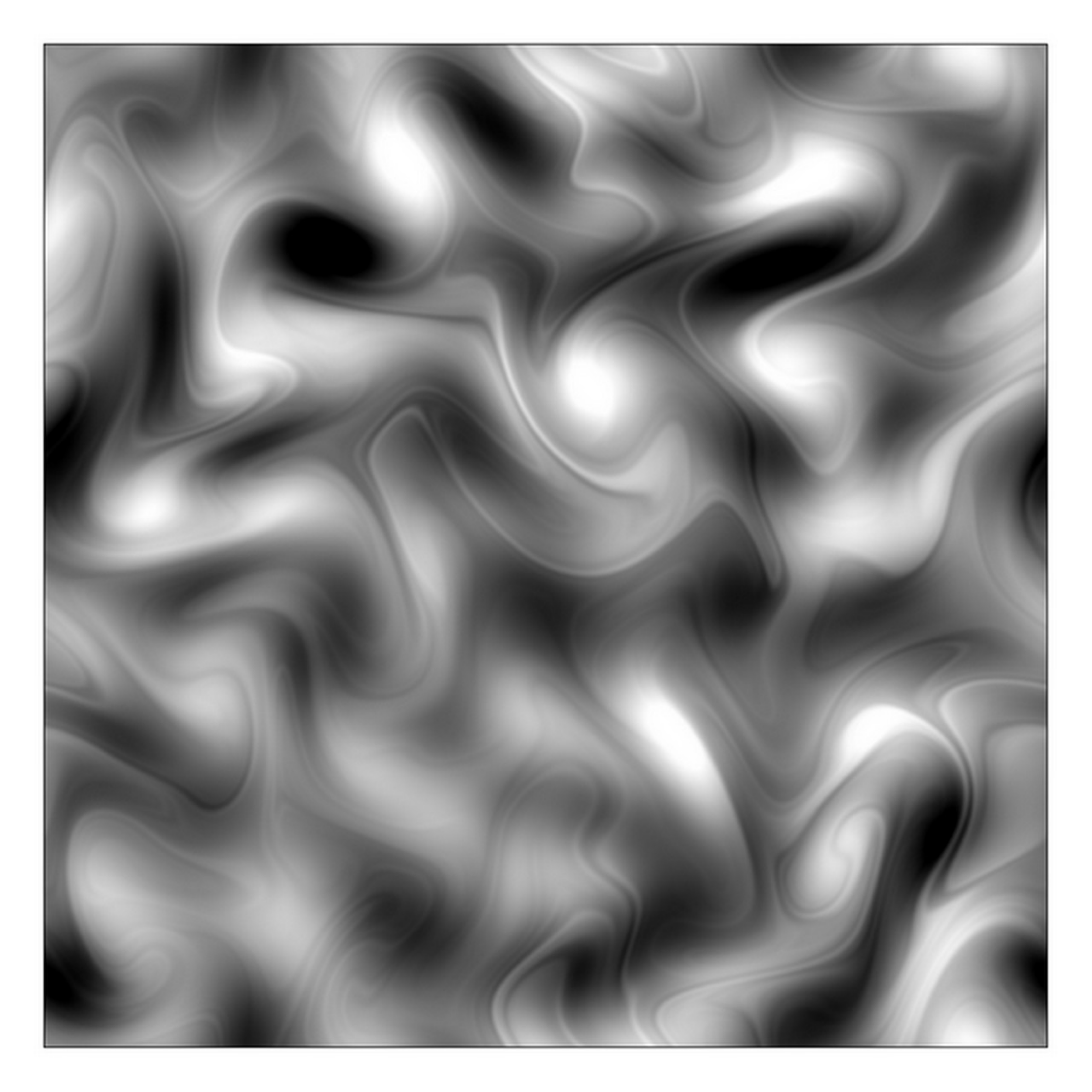}}
\subfigure[$f_0=0.1$]{\includegraphics[width=0.245\textwidth]{64.pdf}}
\subfigure[$f_0=0.2$]{\includegraphics[width=0.245\textwidth]{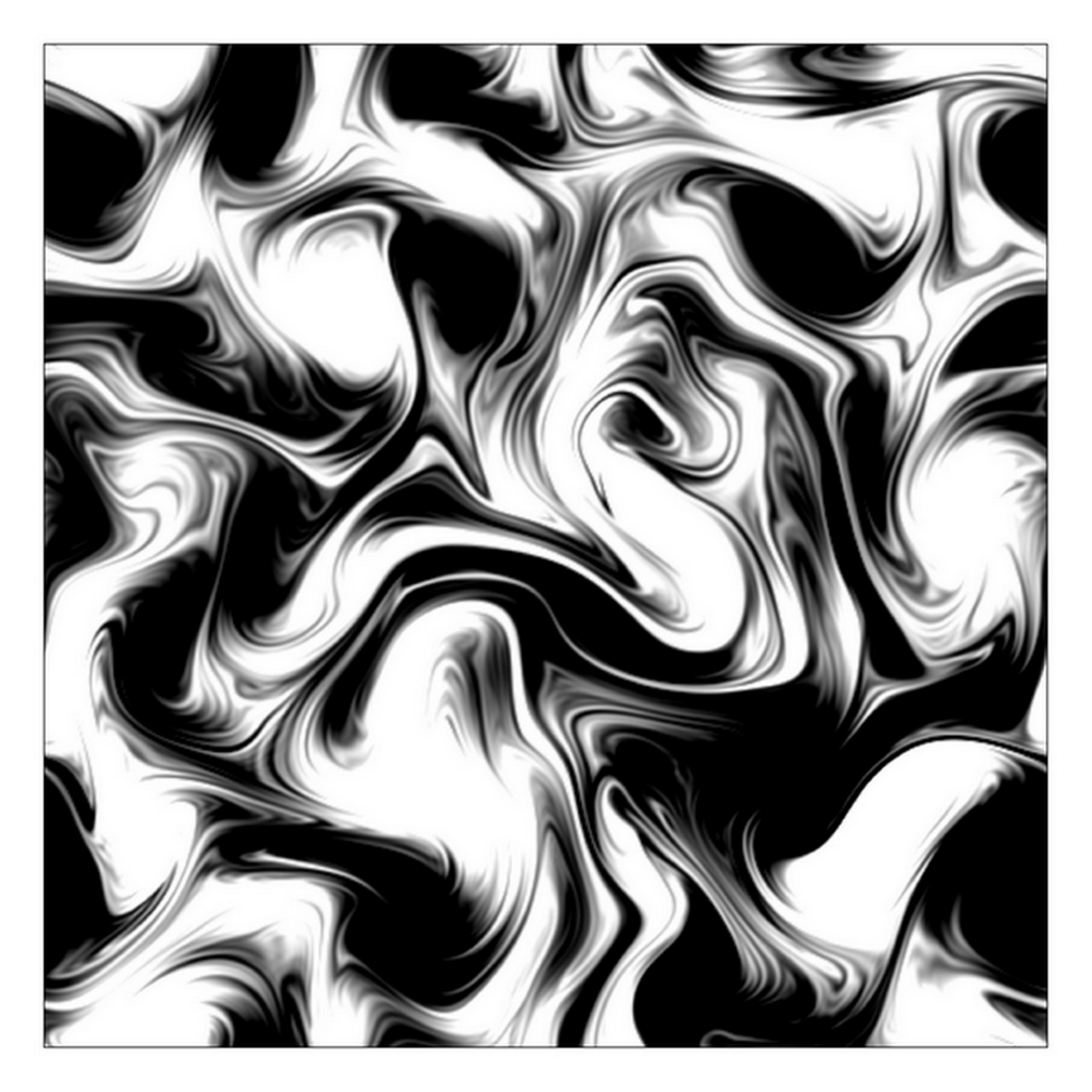}}
\subfigure[$f_0=0.3$]{\includegraphics[width=0.245\textwidth]{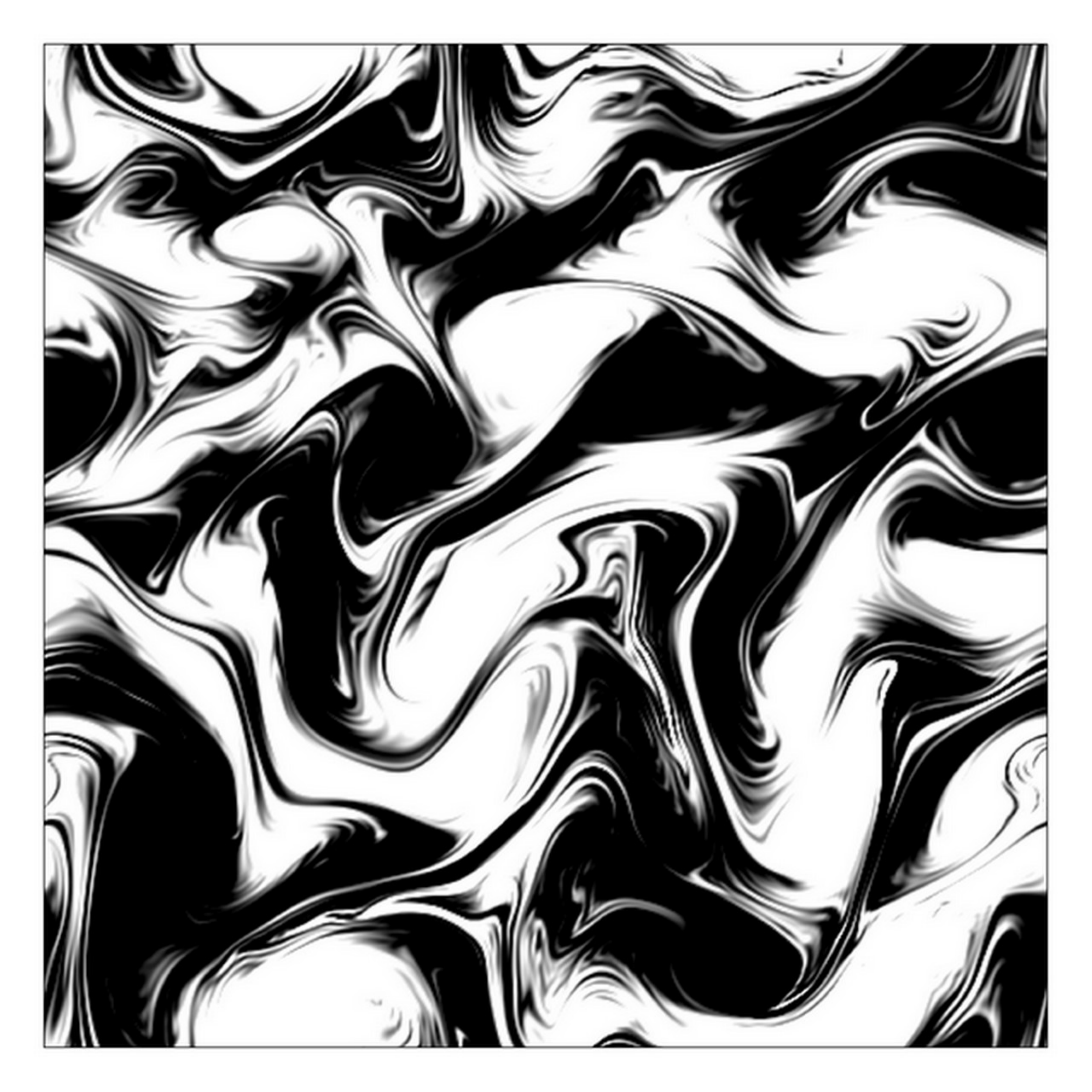}}
}
\caption{Instantaneous vorticity fields at time $t=100$ for varying the forcing amplitude using the large scale dissipation coefficient of $\lambda=0.05$ and the small scale dissipation coefficients of $\nu=1000$ and $p=8$.}
\label{fig:field_f0_k5}
\end{figure*}

Finally, we show the effects of forcing amplitude $f_0$ in Figures~\ref{fig:stat_f0_k5} and~\ref{fig:field_f0_k5}. The linear translation of system's energy level in the statistically steady state can be clearly seen from Figure~\ref{fig:stat_f0_k5}. Here $f_0$ represents the energy injection rate of the external forcing. Therefore, increasing $f_0$ also results in an increase for the amplitude of vorticity which can be seen Figure~\ref{fig:field_f0_k5}. However, the shapes of corresponding energy spectra and structure functions look like similar and there is no significant differences for these statistical quantities.

\section{Summary and conclusions}
\label{sec:conclusions}

Numerical simulations of forced two-dimensional turbulence have been performed by solving the Ekman-Navier-Stokes equations using the Fourier-Galerkin pseudospectral method. Large scale friction, high order small scale dissipation, and Markovian forcing mechanisms have been included in the model. The formulation presented here can be reduced to the classical Navier-Stokes model as a special case by setting the Ekman friction coefficient equal to zero and the order of the viscosity equal to unity. Including the large scale friction mechanism in the model is crucial to be able to obtain a stationary turbulent flow regime in the case of periodic boundary conditions due to the inverse energy cascading phenomenon. Otherwise, the applied forcing mechanism would result in the unbounded growth of the total kinetic energy. The objective of this study was to determine the turbulence statistics for the long-time integration of stationary turbulence simulations, and to investigate the scaling exponents for a large range of physical settings. Specifically, we computed the statistics in terms of the evolution of total kinetic energy, angle averaged energy spectrum, and second-order vorticity structure function.

We first concentrated on the effects of the small scale dissipation mechanism using hyperviscosity, which turns on much more abruptly than the gradual increase of normal viscosity at the small scales. We also studied the Reynolds number dependence when using classical viscosity with the order of Laplacian $p=1$. We showed that the predicted energy spectrum asymptotically converged to the theoretical $k^{-3}$ scaling as the Reynolds number increased, which is predicted by the KBL theory for forward cascading two-dimensional turbulence. We also showed that hyperviscosity effectively eliminates the effects of viscosity at the intermediate scales, thus extending the turbulence inertial range. We demonstrated the shape of the second-order vorticity structure function that is proportional to $r^{2}$ for the smallest separations and flattens when approaching the forcing length scale.

Next, we studied the effects of the large-scale friction mechanism within the turbulence statistics. The linear damping mechanism is utilized in the model, with the friction coefficient $\lambda$ varying from zero to higher values. These computations revealed that the existence of a statistically steady state is maintained by this mechanism and significantly affects the turbulence statistics. We confirmed the classical dual cascade picture of two-dimensional stationary turbulence for $\lambda \rightarrow 0$, which indicates that a direct enstrophy cascade is developed from the forcing range to the dissipation range, and an inverse energy cascade is developed for the scales greater than the effective forcing scale. We found, however, that the large scale damping mechanism is a major source of deviations from the classical $k^{-3}$ scaling in the forward, and $k^{-5/3}$ scaling in the inverse, cascade ranges. We also showed that the tails of corresponding second-order vorticity structure functions scale as $r^{2}$ for large $\lambda$, and reduce to the $r^{2/3}$ scaling as $\lambda$ vanishes.

Finally, we performed a detailed study by systematically varying the parameters associated with Markovian forcing mechanism such as the energy injection scale, the forcing bandwidth, the forcing amplitude, and memory correlation coefficient. We found that these parameters exhibit no significant difference on turbulence statistics, except their translational effects on the total levels of energy in the statistically steady state. We demonstrate that the flow patterns show more vortical filaments for increased forcing bandwidth. We also showed that there is a separation of time scales in the dynamics of the energy spectrum, such that energy is transferred quickly in the forward enstrophy cascade range, while the shape of the spectrum fills out slowly, until, after a long time, the final statistically steady state Kolmogorov spectrum appears in the inverse energy cascade range. This separation of time scales in the energy spectrum might be a useful starting point for developing multiscale computational algorithms for turbulence research, a topic we intend to investigate further in a future study.

\bibliographystyle{model3-num-names}

\bibliography{references}

\end{document}